\documentclass[twocolumn]{article}
\usepackage{latexsym}	      
\usepackage{hyperref}	      
\usepackage{makeidx}
\newcommand{\out}[1]{}        
\newcommand{\ams}[1]{#1}      
\usepackage[preserveurlmacro]{breakurl}
\makeindex

\makeatletter
\def\section{\@startsection {section}{1}{0pt}{-3.25ex plus -1ex minus 
   -.2ex}{1.5ex plus .2ex}{\large\bf}}
\def\subsection{\@startsection {subsection}{2}{0pt}{-2ex plus -1ex minus 
   -.2ex}{1.5ex plus .2ex minus .3ex}{\@setfontsize\large\@xipt{13}\bf}}
\def\paragraph{\@startsection
   {paragraph}{4}{\z@}{2ex plus 1ex minus .2ex}{-1em}{\normalsize\bf}}
\long\def\@makecaption#1#2{
   \vskip 10pt 
   \setbox\@tempboxa\hbox{\sc #1: \it #2}
   \ifdim \wd\@tempboxa >\hsize   
       \sc #1: \it #2\par         
     \else                        
       \hbox to\hsize{\hfil\box\@tempboxa\hfil}  
   \fi}

\makeatother
\newtheorem{defi}{Definition}[section]
\newtheorem{theo}{Theorem}
\newtheorem{prop}{Proposition}[section]
\newtheorem{lemm}{Lemma}
\newtheorem{coro}{Corollary}
\newtheorem{obs}{Observation}[section]
\newtheorem{exam}{Example}

\newenvironment{definition}[1]{\begin{defi} \rm \label{df-#1} }{\end{defi}}
\newenvironment{theorem}[1]{\begin{theo} \rm \label{th-#1} }{\end{theo}}
\newenvironment{proposition}[1]{\begin{prop} \rm \label{pr-#1} }{\end{prop}}
\newenvironment{lemma}[1]{\begin{lemm} \rm \label{lem-#1} }{\end{lemm}}
\newenvironment{corollary}[1]{\begin{coro} \rm \label{cor-#1} }{\end{coro}}
\newenvironment{observation}[1]{\begin{obs} \rm \label{obs-#1} }{\end{obs}}
\newenvironment{example}[1]{\begin{exam} \rm \label{ex-#1} }{\end{exam}}
\newenvironment{proof}{\begin{trivlist} \item[\hspace{\labelsep}\bf
Proof:]}{\hfill $\Box$\end{trivlist}}

\newcommand{\df}[1]{Definition~\ref{df-#1}}
\newcommand{\thm}[1]{Theorem~\ref{th-#1}}
\newcommand{\pr}[1]{Proposition~\ref{pr-#1}}
\newcommand{\lem}[1]{Lemma~\ref{lem-#1}}
\newcommand{\cor}[1]{Corollary~\ref{cor-#1}}
\newcommand{\ob}[1]{Observation~\ref{obs-#1}}
\newcommand{\ex}[1]{Example~\ref{ex-#1}}

\newenvironment{itemise}{\begin{list}{$\bullet$}{\leftmargin 18pt
                        \labelwidth\leftmargini\advance\labelwidth-\labelsep
                        \topsep 4pt \itemsep 2pt \parsep 2pt}}{\end{list}}
\newenvironment{itemise2}{\begin{list}{{\bf --}}{\leftmargin 15pt
                        \labelwidth\leftmargini\advance\labelwidth-\labelsep
                        \topsep 2pt \itemsep 1pt \parsep 1pt}}{\end{list}}
\newenvironment{itemise3}{\begin{list}{*}{\leftmargin 12pt
                        \labelwidth\leftmargini\advance\labelwidth-\labelsep
                        \topsep 0pt \itemsep 0pt \parsep 0pt}}{\end{list}}

\newcommand{\phrase}[1]{\index{#1}{\em #1}}		
\newcommand{\implies}{\Rightarrow}
\newcommand{\turn}{\vdash}                              
\newcommand{\dbigcup}{\bigcup_{\uparrow}}		
\newcommand{\nbigcap}{\bigcap_{\bullet}}		
\newcommand{\bbigcup}{\overline{\bigcup}}		
\newcommand{\bbigcap}{\overline{\bigcap}}		
\newcommand{\nbbigcap}{\bbigcap_{\bullet}}		
\newcommand{\fbbigcup}{\overline{\bigcup}^f}		
\newcommand{\bbbigcup}{\overline{\bigcup}^2}		
\newcommand{\dcup}{~~\makebox[0pt]{\LARGE$\cdot$}\makebox[0pt]{$\cup$}~~}

\newcommand{\bigdvee}{~\makebox[0pt]{\Huge$\cdot$}\makebox[0pt]{$\bigvee$}~}

\newcommand{\dl}[1]{\mbox{\rm I\hspace{-0.75mm}#1}}     
\newcommand{\dc}[1]{\mbox{\rm {\raisebox{.4ex}{\makebox 
        [0pt][l]{\hspace{.2em}\scriptsize $\mid$}}}#1}}
\newcommand{\IZ}{\mbox{\bf Z}}                          
\newcommand{\pow}{{\cal P}}                             
\ams{\newfont{\open}{msbm10}                            
\newcommand{\IT}{\mbox{\open T}}                        
\renewcommand{\IZ}{\mbox{\open Z}}                      
\newfont{\fsc}{eusm10}                                  
\renewcommand{\pow}{\mbox{\fsc P}}}                     
\newcommand{\plat}[1]{\raisebox{0pt}[0pt][0pt]{$#1$}}   
\newcommand{\tuple}[1]{\plat{			 	
	\stackrel{\mbox{\tiny $/$}}
	{\raisebox{-.3ex}[.3ex]{\tiny $\backslash$}}
	\!\!#1\!\!
	\stackrel{\mbox{\tiny $\backslash$}}
	{\raisebox{-.3ex}[.3ex]{\tiny $/$}} }}
\newcommand{\goto}[1]{\stackrel{#1}{\longrightarrow}}   
\newcommand{\gonotto}[1]{\hspace{4pt}\not\hspace{-4pt}  
        \stackrel{#1\ }{\longrightarrow}}

\newcommand{\Con}{{\it Con}}			
\newcommand{\ConGP}{{\it Con}}    		
\newcommand{\fCon}{{\it fCon}}			
\newcommand{\Cn}[1]{{\it Cn}(#1)}		
\newcommand{\fCn}[1]{{\it Cn}_{\it fin}(#1)}	
\newcommand{\bCn}[1]{{\it Cn}_2(#1)}		
\newcommand{\defeq}{:=}	                        
\newcommand{\bsmallcupf}{\overline{\cup}_f}     
\newcommand{\bsmallcap}{\overline{\cap}}        
\newcommand{\bbsmallcupf}{\overline{\cup}_f^2} 
\newcommand{\bbsmallcap}{\overline{\cap}^2}  
\newcommand{\pf}{{\bf Proof:\ }}                

\newcount\PLv\newcount\PLw\newcount\PLx\newcount\PLy\newdimen\PLyy\newdimen\PLX
\newbox\PLdot \setbox\PLdot\hbox{\tiny.} \def\scl{.08} 
\def\PLot#1{\PLx`#1\advance\PLx-42\PLy\PLx\PLv\PLx\divide\PLy9\PLw\PLy\multiply
\PLw9\advance\PLx-\PLw\advance\PLx-4\PLy-\PLy\advance\PLy4\PLX=\the\PLx pt
\advance\PLyy\the\PLy pt\wd\PLdot=\scl\PLX\raise\scl\PLyy\copy\PLdot}
\def\draw#1{\ifx#1\end\let\next=\relax\else\PLot#1\let\next=\draw\fi\next}

\newcommand{\IN}{\dl{N}}                        
\newcommand{\IQ}{\dc{Q}}                        
\newcommand{\fC}{{\cal C}}                      
\newcommand{\fE}{{\cal E}}                      
\newcommand{\fN}{{\cal N}}                      
\newcommand{\fF}{{\cal F}}                      
\newcommand{\fL}{{\cal L}}                      
\newcommand{\fM}{{\cal M}}                      
\newcommand{\fS}{{\cal S}}                      
\newcommand{\fR}{{\cal R}}                      
\newcommand{\eC}{{\rm C}}                       
\newcommand{\eD}{{\rm D}}                       
\newcommand{\eE}{{\rm E}}                       
\newcommand{\eF}{{\rm F}}                       
\newcommand{\eG}{{\rm G}}                       
\newcommand{\eK}{{\rm K}}                       
\newcommand{\eL}{{\rm L}}                       
\newcommand{\eN}{{\rm N}}                       
\newcommand{\eP}{{\rm P}}                       
\newcommand{\eM}{{\rm M}}                       
\newcommand{\eT}{{\rm T}}                       
\newcommand{\fT}{{\cal T}}                      

\begin{document}
\bibliographystyle{plain}

\title{\Large\bf Configuration Structures, Event Structures and Petri Nets%
\thanks{This work was supported by ONR grant N00014-92-J-1974 (when
 both authors were at Stanford), 
 by EPSRC grant GR/S22097 (when both authors were at the
 University of Edinburgh) and by a Royal Society-Wolfson Award.}}
\author{R.J. van Glabbeek\\
 \normalsize NICTA, Sydney, Australia\\[-3pt]
 \normalsize University of New South Wales, Sydney, Australia\\[-3pt]
 \normalsize Stanford University, USA\\[-3pt]
 \normalsize \tt rvg@cs.stanford.edu
\and    G.D. Plotkin\\
 \normalsize Laboratory for Foundations of Computer Science\\[-3pt]
 \normalsize School of Informatics, University of Edinburgh, UK\\[-3pt]
 \normalsize Stanford University, USA\\[-3pt]
 \normalsize \tt gdp@inf.ed.ac.uk}
\date{}
\maketitle
{\small\noindent In this paper the correspondence between safe Petri
nets and event structures, due to Nielsen, Plotkin and Winskel, is
extended to arbitrary nets without self-loops, under the collective
token interpretation.  To this end we propose a more general form of event
structure, matching the expressive power of such nets.  These new event
structures and nets are connected by relating both notions with
\phrase{configuration structures}, which can be regarded as
representations of either event structures or nets that capture their
behaviour in terms of action occurrences and the causal relationships
between them, but abstract from any auxiliary structure.

A configuration structure can also be considered logically, as a class
of propositional models, or---equivalently---as a propositional theory
in disjunctive normal from.  Converting this theory to conjunctive
normal form is the key idea in the translation of such a structure
into a net.

For a variety of classes of event structures we characterise the
associated classes of configuration structures in terms of their
closure properties, as well as in terms of the axiomatisability of the
associated propositional theories by formulae of simple prescribed
forms, and in terms of structural properties of the associated Petri
nets.}
\section*{Introduction}\label{introduction}

The aim of this paper is to connect several models of concurrency, by
providing behaviour preserving translations between them.

\begin{figure}[h]\vspace{-2ex}
\caption{Behaviour preserving translations in \cite{NPW81}}
\input{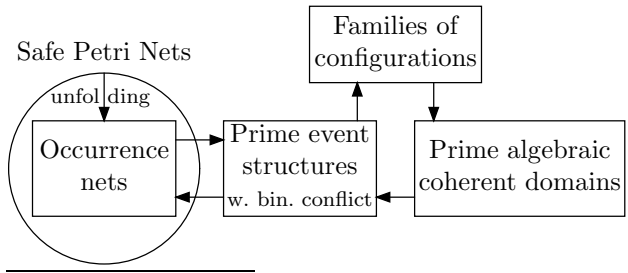}
\centerline{\raise 1ex\box\graph}\vspace{-2ex}
\end{figure}

In {\sc Nielsen, Plotkin \& Winskel} \cite{NPW81} \phrase{event
structures} were introduced as a stepping stone between \phrase{Petri
nets} and \phrase{Scott domains}. It was established that every
\phrase{safe} Petri net can be unfolded into an \phrase{occurrence
net}; the occurrence nets are then in correspondence with event
structures; and they in turn are in correspondence with
\phrase{prime algebraic coherent Scott domains}.  In {\sc Winskel}
\cite{Wi87a} a more general notion of event structure was proposed,
corresponding to a more general kind of Scott domain.  The event
structures from \cite{NPW81} are now called \phrase{prime event
structures with binary conflict}.

The translation from event structures to domains passes through a
stage of \phrase{families of configurations of event structures}.
{\sc Winskel} \cite{Wi82} and {\sc Van Glabbeek {\small\&}$\!$ Goltz} \cite{GG90}
found it convenient to use such families as a model of concurrency in
its own right. In this context the families were called
\phrase{configuration structures} \cite{GG90}.\vspace{-2ex}

\begin{figure}[htb]
\caption{Our main contribution: behaviour preserving translations
between four models of concurrency}
\input{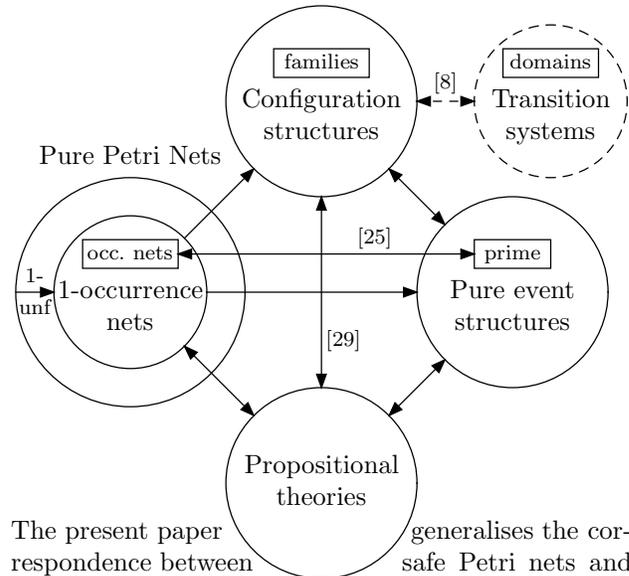}
\centerline{\box\graph}\vspace{-3.6em}
\end{figure}
\vfill

\noindent
The present paper \hfill generalises the cor-\\ respondence between
\hfill safe \,Petri \,nets \,and\\ configuration structures to
(possibly unsafe) nets without self-loops (the \phrase{pure} nets).
\pagebreak
For this purpose we use a more general kind of
configuration structure than in \cite{GG90}, the \phrase{set
systems}. These have an attractive alternative presentation as
\phrase{propositional theories} \cite{Pr94a}, which is exploited in
their translation to nets. We also generalise the event structures of
\cite{Wi87a}, so that, again, our configuration structures arise as
their families of configurations.  The connection between
configuration structures and Scott domains is generalised in {\sc Van
Glabbeek} \cite{vG95c}, who proposes \phrase{transition systems} as
alternative presentations of domains; we do not consider these matters
further in the present paper.

The relationship between configuration structures, infinitary
propositional theories, event structures and Petri nets is described
in Section~\ref{four models}. We \emph{1-unfold} pure nets
into \emph{pure 1-occurrence nets}, which generalise the occurrence
nets of \cite{NPW81}, and argue that this 1-unfolding preserves the
causal and branching time behaviour of the represented system. This
allows us to restrict attention to pure 1-occurrence nets in the rest
of the paper.  Moreover, we give translations showing that
configuration structures, propositional theories and event structures
are equivalent up to so-called \phrase{configuration equivalence}
(which is defined as being mapped to the same configuration
structures) and that, with a slight restriction, all four models are
equivalent up to \emph{finitary equivalence}.

Section~\ref{computational} introduces a computational interpretation
of configuration structures, Petri nets and event structures in terms
of associated transition relations;
restricted to pure Petri nets and pure event structures, these
transition relations can be derived from the relevant sets of
configurations, but not in general.  With that,
Section~\ref{equivalence} provides definitions of notions
of \emph{reachable} and \index{secured configurations}\emph{secured}
(reachable in the limit) configurations and considers corresponding
notions of equivalence by restricting to reachable or secured, and
possibly finite, configurations.

With the general framework thus provided, Section~\ref{brands} considers
the various brands of event structures introduced by Winskel and his
co-workers. They are shown to correspond to natural restrictions on
the general notion of event structure, adapting the comparisons, on the
one hand, to the original notion of configuration and, on the other
hand, to the relevant one from the general theory. These comparisons
are summarised in Table~\ref{7 classes}.

It is then natural to enquire how the event structure restrictions are
reflected in corresponding restrictions on configurations structures
and so on; this is the subject of Section~\ref{ComparingModels}.
Sections~\ref{EvsC} and~\ref{theories} provide such comparisons,
summarised in Table~\ref{correspondence}, for configuration structures
and propositional theories up to configuration equivalence.  The
restrictions on configuration structures are natural closure
properties, and those on propositional theories concern the form of
the formulae occurring in an axiomatisation. Section~\ref{EvsC-secured}
does the same, see Table~\ref{correspondence secured}, but now with
the comparison based on the secured configurations.

Section~\ref{finitary comparisons} concerns the finitary case, with
general comparisons being summarised in Table~\ref{correspondence_finite}
and the restriction to the finite reachable configurations summarised in
Table~\ref{correspondence_finite_reachable}. Section~\ref{Tie-nets}
ties in Petri nets, providing corresponding structurally defined
subclasses; however we were not successful in doing this in all cases. The
main mathematical work is done in Sections~\ref{EvsC} and~\ref{theories},
with the rest of the section adapting this work to the various cases at hand.

Section~\ref{related work} contains a discussion of related work and
presents some possibilities for future research. Finally, there is an
index for the many technical terms introduced in the course of the paper.

The papers \cite{GP95} and \cite{GP04} contain extended abstracts of
parts of this work, together with additional material.$\!$

\section{Four models of concurrency}\label{four models}

In this section we present the four models of concurrency mentioned in
the introduction, and provide translations between them.

\subsection{Configuration structures}

\begin{definition}{set system}\index{set systems}
A \emph{set system} is a pair $\eC=\tuple{E,C}$ with $E$ a
set and $C \subseteq \pow(E)$ a collection of subsets.
\end{definition}
When a set system is used to represent a concurrent system, we call it
a \phrase{pure configuration structure}\index{configuration
structures}, but generally drop the word ``pure''. (We envision
introducing a broader class of configuration structures in the future,
matching the expressive power of impure nets.) The elements of $E$ are
then called \phrase{events} and the elements of $C$
\phrase{configurations}. An event represents an occurrence of an
action the system may perform; a configuration $x$ represents a state
of the system, namely the state in which the events in $x$ have
occurred.

\subsection{Propositional theories}

A set system can also be considered from a logical point of view: $E$
is thought of as a collection of \phrase{propositions} and $C$ as the
collection of \phrase{models}. Connecting with the computational point
of view, we associate with an event the proposition that it has
happened.  This point of view is due to {\sc Pratt}~\cite{GP93a,Pr94a}.
We can now represent a set system by the valid sentences, those holding
in all models; these are the \phrase{laws of $\eC$}.

To make this precise, we choose a language: infinitary propositional
logic.  Given a set $E$ of \emph{(propositional) variables}, the
\phrase{formulae over $E$} form the least class including $E$ and
closed under $\neg$ (negation) and $\bigwedge$ (conjunction of sets of
formulae). We make free use of other standard connectives such as
$\implies, \bigvee, \perp, \top$: they are all definable from $\neg$
and $\bigwedge$.  As usual, an \phrase{interpretation} of $E$ is just
a subset of $E$ and one defines in the standard way when an
interpretation makes a formula true.

\begin{definition}{theory}\index{propositional theories}
An \emph{(infinitary)} \emph{propositional theory} is a pair $\eT =
\tuple{E,T}$ with $E$ a set of propositional variables and $T$ a class
of infinitary propositional formulae over $E$.
\end{definition}
A formula $\varphi$ over $E$ is \phrase{valid} in a set system $\eC =
\tuple{E,C}$ iff it is true in all elements of $C$; the \emph{theory
associated to}\index{propositional theory associated to} C
is $\fT(\eC) := \tuple{E,T(C)}$, where $T(\eC)$
denotes the class of formulae valid in $\eC$.  Equally, given a
propositional theory ${\eT}=\tuple{E,T}$, its \phrase{associated set
system} is $\fM(\eT) := \tuple{E,{M}(T)}$, where ${M}(T)$ is the set of
\phrase{models} of $T$, those interpretations of $E$ making every formula
in $T$ true.  We say that T \phrase{axiomatises} $\fM(\eT)$.
A formula $\varphi$ over $E$ is a \phrase{logical consequence} of a
theory $\eT$ if $\varphi$ is true in any model of $\eT$; a formula
$\psi$ over $E$ \phrase{implies} $\varphi$ iff the latter is a logical
consequence of the theory $\tuple{E,\{\psi\}}$.  Two propositional
theories T and $\eT'$ are \phrase{logically equivalent} if $\fM(\eT) =
\fM(\eT')$, which is easily seen to be the case iff they have the same
logical consequences.

\begin{theorem}{CtoTtoC}
Let $\eC=\tuple{E,C}$ be a set system. Then  $\fT(\eC)$ axiomatises
$\eC$, i.e., $\fM(\fT(\eC)) = \eC$.
\end{theorem}

\begin{proof}
The single formula $\bigvee_{X \in C} (\bigwedge X \wedge \bigwedge
\neg (E-X))$ already constitutes an axiomatisation of $\eC$. It is
called the \phrase{disjunctive normal form} of $\fT(\eC)$.
\end{proof}
Thus $\fT$ and $\fM$ provide a bijective correspondence between set
systems and infinitary propositional theories up to logical
equivalence.  For any two subsets $X$,$Y$ of $E$, let the \phrase{clause}
$X \Rightarrow Y$ abbreviate the implication $\bigwedge X \implies
\bigvee Y$; we say that the elements of $X$ are the \phrase{antecedents}
of the clause, and those of $Y$ its \phrase{consequents}.  Then for any
set system $\eC=\tuple{E,C}$, the set of clauses $\{X \implies (E-X)
\mid X \not\in C \}$ constitutes another axiomatisation of $\eC$.  A
theory consisting of a set of clauses is said to be in
\phrase{conjunctive normal form}.

\subsection{Event structures}\label{event structures}

\begin{definition}{event structure}\index{event structures}
An \emph{event structure} is a pair $\eE = \tuple{E,\turn\;}$ with
\begin{itemise}
\item $E$ a set of \phrase{events},
\item $\turn \; \subseteq \pow(E) \times \pow(E)$, the
      \phrase{enabling relation}.
\end{itemise}
\end{definition}
Like a configuration structure, an event structure describes a
concurrent system in which the events represent action occurrences.
In previous notions of event structure \cite{Wi87a,Wi89}, one only had
singleton enablings: $\turn \; \subseteq \pow(E) \times E$. Here we
generalise $\turn$ to a relation between sets of events. As before,
the enabling relation places some restrictions on which events can
happen when. The idea here is that when $X$ is the set of events that
happened so far, an additional set $U$ of events can happen
(concurrently) iff every subset of $X \cup U$ is enabled by a set of
events that happened before, i.e., a subset of $X$.

\begin{example}{ternary conflict}
Let $E=\{d,e,f\}$ and the enabling relation be given by
$\emptyset \turn X$ for any $X \subseteq E$ with $X \neq E$.
In the initial state of the event structure $\eE =\tuple{E,\turn\;}$,
each of the events $d$, $e$ and $f$ can happen, and any two of them
can happen concurrently. However, there is no way all three events can
ever happen, because there is no set of events $X$ with $X \vdash
\{d,e,f\}$. This is a case of \phrase{ternary conflict}.
\end{example}

\begin{example}{resolved conflict}\index{resolvable conflict}
Let $E=\{a,b,c\}$ and the enabling relation be given by $\{c\} \turn
\{a,b\}$ and $\emptyset \turn X$ for any $X \subseteq E$ with $X \neq
\{a,b\}$. Initially, each of the events $a$, $b$ and $c$ can occur,
and the events $a$ and $c$ can even happen concurrently.  The events
$a$ and $b$, on the other hand, can initially not happen concurrently,
for we do not have $\emptyset \turn \{a,b\}$. However, as soon as $c$
occurs, the events $a$ and $b$ can occur in parallel. We say that the
\emph{conflict} between $a$ and $b$ is \emph{resolved} by the
occurrence of $c$.
\end{example}

\begin{example}{asymmetric conflict}
Let $E=\{d,e\}$ and the enabling relation be given by $\{d\}\turn \{d,e\}$
and $\emptyset \turn X$ for any $X \subseteq E$ with $X \neq \{d,e\}$.
Initially, $d$ and $e$ can both occur, but not in parallel.
After $d$ has happened, $e$ may follow, but when $e$ happens first,
$d$ cannot follow. The reason is that we do not have $X \turn
\{d,e,\}$ for some $X \subseteq \{e\}$. This is a case of
\phrase{asymmetric conflict} \cite{Lk92,PP95}.
\end{example}
In Section \ref{brands} we will explain how these event structures
generalise the ones of \cite{NPW81,Wi87a,Wi89}.  In those papers the
behaviour of an event structure is formalised by associating to it a
family of configurations. However, there are several ways to do so
(cf.\ Section~\ref{equivalence}); here we only consider the
simplest variant.

\begin{definition}{EtoC}
Let $\eE = \tuple{E,\turn\;}$ be an event structure.  The set $L(\eE)$
of \phrase{left-closed configurations} of E is given by\vspace{-2ex}
$$X \in L(\eE) ~\Leftrightarrow~ \forall Y\subseteq X.~
\exists Z \subseteq X.~ Z \turn Y.$$
The \phrase{left-closed configuration structure associated to} E is
$\fL(\eE) := \tuple{E,L(\eE)}$.
Two event structures E and F are \phrase{$\fL$-equivalent}
if $\fL(\eE) = \fL(\eF)$.
\end{definition}
In Section~\ref{computational} we provide a computational
interpretation of event structures with the property that the
left-closed configurations of an event structure adequately represent
the behaviour of the represented system for the following class of
``pure'' event structures:

\begin{definition}{pure es}
An event structure is \phrase{pure} if $X \turn Y$ only if $X \cap Y =
\emptyset$.
\end{definition}
The event structures of Examples~\ref{ex-ternary conflict}
and~\ref{ex-resolved conflict} are pure, but the one of
\ex{asymmetric conflict} is not.

We now show that any configuration structure can be obtained as the
left-closed configuration structure associated to a pure event
structure.

\begin{definition}{CtoE}
Let $\eC = \tuple{E,C}$ be a configuration structure.  The \phrase{event
structure associated to} C is $\fE(\eC) := \tuple{E,\turn\;}$, with
$X \turn Y$ iff $X \cap Y \!=\! \emptyset \;\wedge\; X \cup Y \!\in\! C$.
\end{definition}

\begin{theorem}{CtoEtoC}
Let C be a configuration structure. Then $\fE(\eC)$ is pure and
$\fL(\fE(\eC)) = \eC$.
\end{theorem}

\begin{proof}
Let $\eC = \tuple{E,C}$ and $\fE(\eC) = \tuple{E,\turn\;}$. Suppose $x
\in C$. For any $Y \subseteq x$ take $Z := x-Y$. Then $Z \subseteq x$
and $Z \turn Y$. So $x \in \fL(\fE(\eC))$.  Conversely, suppose $x \in
\fL(\fE(\eC))$. Then there is a $Z \subseteq x$ such that $Z \turn
x$. (In fact, $Z=\emptyset$.) By construction, $x = Z \cup x \in C$.
\end{proof}
Hence, $\fE$ and $\fL$ provide a bijective correspondence between
configuration structures and (pure) event structures up to
$\fL$-equivalence.

\subsubsection*{Event structures vs.\ propositional theories}

By combining Theorems~\ref{th-CtoTtoC} and~\ref{th-CtoEtoC} we find
that $\fT \circ \fL$ and $\fE \circ \fM$ constitute a bijective
correspondence between (pure) event structures up to $\fL$-equivalence
and propositional theories up to logical equivalence. Below we provide
direct translations between them.

To any event structure $\eE=\tuple{E,\turn\;}$ we associate the
propositional theory $\fT(\eE) := \tuple{E,T(\eE)}$, where
$$T(\eE) := \left.\left\{ \bigwedge X \Rightarrow \bigvee_{Y\turn X} \bigwedge Y
\,\right|\, X \subseteq E \right\}.$$
This logical view of event structures corresponds exactly with their
left-closed interpretation:

\begin{proposition}{EtoT} $\fM(\fT(\eE)) = \fL(\eE)$ for any event
structure E.
\end{proposition}

\begin{proof}
Immediate from the definitions.
\end{proof}
Similarly, to any propositional theory $\eT=\tuple{E,T}$ in conjunctive
normal form we associate the (not necessarily pure) event structure
$\fE(\eT) := \tuple{E,\turn_\eT\;}$, where
$$X \turn_\eT Y \Leftrightarrow \forall Z.~ ((Y \implies Z) \in T
\Rightarrow X \cap Z \neq \emptyset).$$

\begin{proposition}{TtoE} $\fL(\fE(\eT)) = \fM(\eT)$ for any theory T
in conjunctive normal form.
\end{proposition}

\begin{proof}
Let $x \in \fM(\eT)$. To establish $x\in\fL(\fE(\eT))$ we take $Y
\subseteq x$ and show $x\turn_\eT Y$. Let $Z \subseteq E$ be such that
$(Y \Rightarrow Z) \in T$. As $Y \Rightarrow Z$ is true in $x$ we have
$Z \cap x \neq \emptyset$. It follows that $x \in \fL(\fE(\eT))$.

Now let $x \in \fL(\fE(\eT))$. To establish $x\in\fM(\eT)$ we take $(Y
\Rightarrow Z) \in T$.  We have to show that $Y \Rightarrow Z$ is true
in $x$. So suppose $Y \subseteq x$.  Then there must be a $W \subseteq
x$ with $W \turn_\eT Y$, hence $W \cap Z \neq \emptyset$. It follows
that $x \cap Z \neq \emptyset$, which had to be shown.
\end{proof}
Thus $\fT$ and $\fE$ provide a bijective correspondence between event
structures up to $\fL$-equivalence and propositional theories up to
logical equivalence.

\begin{definition}{pure PT}
A propositional theory in conjunctive normal form is \phrase{pure} if it
only contains clauses $X \implies Y$ with $X \cap Y = \emptyset$.
\end{definition}
Clearly every propositional theory is logically equivalent to a pure one, as
impure clauses are tautologies, i.e., they hold in all interpretations.
In case $\eT=\tuple{E,T}$ is a pure theory, we can define the
associated pure event structure $\fE_p(\eT) := \tuple{E,\turn_p\;}$ by
$X \turn_p Y \Leftrightarrow X \cap Y = \emptyset \wedge X \turn_\eT Y$.
Note that $\fE_p(\eT)$ is pure and $\fL(\fE_p(\eT)) = \fM(\eT)$.

\subsection{Petri nets}\label{PN}

\begin{definition}{petri}\hfill\mbox{}\index{Petri nets}\\
A \emph{Petri net} is a tuple $\eN = \tuple{S,T,F,I}$ with
\begin{itemise}\vspace{-1ex}
\item $S$ and $T$ two disjoint sets of \phrase{places}
and \phrase{transitions} (\emph{Stellen} and \emph{Transitionen} in German),
\vspace{-1ex}\item $F: (S \!\times\! T \,\cup\, T\!\times\! S)\rightarrow \IN$,
the \phrase{flow relation},
\vspace{-1ex}\item and $I:S \rightarrow \IN$, the \phrase{initial marking}.
\end{itemise}\vspace{-1ex}
\end{definition}
Petri nets are pictured by drawing the places as circles and the
transitions as boxes. For $x,y \in S \cup T$ there are $F(x,y)$
\phrase{arcs} from $x$ to $y$.  A net is said to be \emph{without
arcweights}\phrase{arcweights} if the range of $F$ is $\{0,1\}$.

When a Petri net represents a concurrent system, a global state of
such a system is given as a \phrase{marking}, which is a multiset over
$S$, i.e., a function $M \in \IN^S$.
Such a state is depicted by placing $M(s)$ dots (\phrase{tokens})
in each place $s$. The initial state is given by the marking $I$.  In
order to describe the behaviour of a net, we describe the \phrase{step
transition relation} between markings.

\begin{definition}{multiset}
For two multisets $M$ and $N$ over $S$, or more generally for functions
$M,N \in \IZ^S$, write $M \leq N$ if $M(s) \leq N(s)$ for all $s \mathbin\in S$;
$M+N \in \IZ^S$ is the function given by $(M+N)(s):=M(s)+N(s)$, and
\mbox{$0 \mathbin\in \IN^S$} the one with $0(s):=0$ for all $s \mathbin\in S$;
\mbox{$M-N \in \IZ^S$} is given by $(M-N)(s):=M(s)-N(s)$.

A multiset $M$ over $S$ is \phrase{finite} if $\{s \in S \mid M(s)
> 0\}$ is finite.  A multiset $M \in \IN^S$ with $M(s) \leq 1$ for all
$s \in S$ is identified with the set $\{s \in S \mid M(s)=1\}$.  
\end{definition}
Note that for multisets $M$ and $N$, the function $M-N$ need not be a
multiset.

\begin{definition}{firing}
For a finite multiset $U:T \rightarrow \IN$ of transitions in a Petri net, let
$^\bullet U,~ U^\bullet: S \rightarrow \IN$ be the multisets of
\index{preplaces}\emph{pre-} and \phrase{postplaces} of $U$, given by
$$^\bullet U(s):=\sum_{t\in T} F(s,t) U(t)~~\mbox{and}~~
U^\bullet (s):=\sum_{t\in T} U(t) F(t,s)$$ for $s \in S$.
We say that $U$ is \phrase{enabled} under a marking $M$ if $^\bullet U \leq M$.
In that case $U$ can \phrase{fire} under $M$, yielding the marking
$M':=M - \mbox{$^\bullet U$} + U^\bullet$, written \plat{M\goto{U}M'}.

A chain $I \goto{U_1} M_1 \goto{U_2} \cdots \goto{U_n} M_n$ is called
a \phrase{firing sequence}.  A marking $M$ is
\emph{reachable}\index{reachable marking} if there is
such a sequence ending in $M=M_n$.
\end{definition}
If a multiset $U$ of transitions fires, for every transition $t$ in $U$ and
every arc from a place $s$ to $t$, a token moves along that arc
from $s$ to $t$. These tokens are consumed by the firing, but also new
tokens are created, namely one for every outgoing arc of $t$. These end up
in the places at the end of those arcs. If $t$ occurs several times in
$U$, all this happens several times (in parallel) as well.
The firing of $U$ is only possible if there are sufficiently many
tokens in the preplaces of $U$ (the places where the incoming arcs
come from). In Section~\ref{unbounded par} we explain why we consider
the firing of finite multisets only.

\subsubsection*{From Petri nets to configuration structures}

As for event structures, the behaviour of a net can be captured by
associating to it a family of configurations. 

\begin{definition}{configuration} 
\index{configurations!of a Petri net}
A \emph{(finite) configuration} of a Petri net $\eN=\tuple{S,T,F,I}$ is
any finite multiset $X$ of transitions with the property that the
function $M_X:S \rightarrow \IZ$ given by $M_X := I - \mbox{$^\bullet
\!X$} + X^\bullet$ is a marking, i.e., $M_X \geq 0$.
Let $C(\eN)$ denote the set of\out{ finite} configurations of $\eN$.
\end{definition}
Note that $0$ is a configuration and $M_0=I$; note
further that if $x$ is a configuration and \plat{M_x \goto{~U} M'} then
$x+U$ is a configuration and $M'=M_{x+U}$. So if
\plat{I \goto{U_1} M_1 \goto{U_2} \cdots \goto{U_n} M_n} is a
firing sequence, then $x\defeq U_1 + \cdots + U_n$ is a configuration and
$M_n = M_x$. In general, when $x\in C(\eN)$ then $M_x$ is the marking
that would result from firing all transitions in $x$, if possible,
regardless of the order in which they fire.

Next we will determine which nets can be faithfully described in this
way by means of set systems.

\begin{definition}{1-occurrence}
A \phrase{1-occurrence net} is a net in which every configuration is a set.
\end{definition}
This implies that any transition can fire at most once,
i.e., in every firing sequence \plat{M_0 \goto{U_1} \cdots \goto{U_n} M_n}
the multisets $U_1,...,U_n$ are sets and disjoint.
When dealing with a 1-occurrence net, typically presented as a tuple
$\tuple{S,E,F,I}$, we call its transitions \phrase{events}.

\begin{definition}{loops}
A net $\eN=\tuple{S,T,F,I}$ is \phrase{pure} if there is no $s$ in $S$
and $t$ in $T$ with $F(s,t)>0$ and $F(t,s)>0$, i.e., if it is without
\phrase{self-loops}.
\end{definition}
In Section~\ref{computational} we will argue that the configurations
of a 1-occurrence net adequately represent the behaviour of the
represented system only in the case of pure nets.
Therefore we will restrict attention to pure 1-occurrence nets.

\begin{definition}{NtoC}
Let $\eN=\tuple{S,E,F,I}$ be a pure 1-oc\-cur\-rence net. Its
\phrase{associated\out{ (finitary)} configuration structure}
$\fC(\eN)$ is $\tuple{E,C(\eN)}$. Two such nets N and $\eN'$ are
\emph{configuration equivalent}\index{configuration
equivalence}---written $\eN =_\fC \eN'$---if $\fC(\eN) = \fC(\eN')$.
\end{definition}

\subsubsection*{Individual vs.\ collective tokens}

\index{collective token interpretation}
\index{individual token interpretation}
There are two different schools of thought in interpreting the causal
behaviour of Petri nets, which can be described as the \emph{individual} and
\emph{collective token} philosophy \cite{GP95,vG05}.\footnote{The
individual token interpretation of ordinary nets should not be
confused with the concept of \phrase{Petri nets with individual
tokens} \cite{Rei85} such as \phrase{predicate/transition nets} or
\phrase{coloured Petri nets}; there the individuality is hardwired into the
syntax of nets.} The following example illustrates their difference.

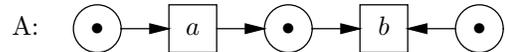
\begin{figure}[htp]
\input{tokens}
\centerline{\raise 1em\box\graph}
\end{figure}
\noindent
In this net, the transitions $a$ and $b$ can fire once each.
After $a$ has fired, there are two tokens in the middle place.
According to the individual token philosophy, it makes a
difference which of these tokens is used in firing $b$. If the token
that was there already is used (which must certainly be the case if
$b$ fires before the token from $a$ arrives), the transitions $a$
and $b$ are causally independent. If the token that was produced by
$a$ is used, $b$ is causally dependent on $a$. Thus, the net A above
has two maximal computations, that can be characterised by partial orders:
\begin{picture}(11,0)
\put(0,0){$a$}
\put(3,1){\vector(1,0){5}}
\put(9,0){$b$}
\end{picture}
and the trivial one
\plat{\begin{array}{@{}c@{}}\,\,a\!\\[-3pt]b\;\,\end{array}}.
According to the collective token philosophy on the other hand, all
that is present in the middle place after the occurrence of $a$ is the
number 2. The preconditions for $b$ to fire do not change, and
consequently $b$ is always causally independent of $a$.

A net is called \phrase{safe} if no reachable marking has multiple
tokens in the same place. For safe nets there is no difference between
the individual and collective token interpretations.

The individual token approach has been formalised by the notion of a
\phrase{process}, described in {\sc Goltz \& Reisig} \cite{GR83}.
A causality-respecting bisimulation relation based on this approach
was proposed by {\sc Best, Devillers, Kiehn \& Pomello} \cite{BDKP91}
under the name \phrase{fully concurrent bisimulation}.
Also the \phrase{unfolding} of non-safe nets into (safe) occurrence nets
proposed by {\sc Engelfriet} \cite{En91} and {\sc Meseguer, Montanari \&
Sassone} \cite{MMS92} embraces the individual token philosophy.

{\sc Best \& Devillers} \cite{BD87} adapted the process concept
of \cite{GR83} to fit the collective token philosophy.  Equivalence
relations on Petri nets based on the collective token interpretation
were proposed by us in \cite{GP95}, and include configuration
equivalence, defined above. There is no unfolding construction that
converts arbitrary non-safe nets into safe nets while preserving their
collective token interpretation, for under the collective token
interpretation non-safe nets are strictly more expressive than safe
ones \cite{vG05}: only the former can express \phrase{resolvable
conflict} \cite{GP04}.

The following example shows that the collective token philosophy
allows the identification of nets that are distinguished under the
individual token philosophy.

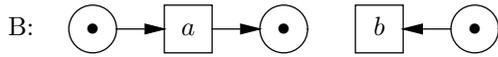
\begin{figure}[htb]
\input{incomparable}
\centerline{\raise 1em\box\graph}
\end{figure}
\noindent
Under the collective token interpretation the precondition of $b$ expressed
by the place in the middle of net A is redundant, and hence A must be
equivalent to B\@. In fact, ${\rm A}=_\fC\mbox{B}$.
However, A and B are not fully concurrent bisimulation equivalent, as
B lacks the computation
\begin{picture}(11,0)
\put(0,0){$a$}
\put(3,1){\vector(1,0){5}}
\put(9,0){$b$}
\end{picture}.

Conversely, the individual token philosophy allows identifications
that are invalid under the collective token philosophy, but these
necessarily involve
\phrase{labelled nets}. A labelled net is a tuple \tuple{S,T,F,I,l} with
\tuple{S,T,F,I} a net and $l:T\rightarrow Act$ a \phrase{labelling function}
over some set of action names $Act$. The labelling enables the presence
of multiple transitions with the same name.
The net A is fully concurrent bisimulation equivalent with
the labelled net C below.
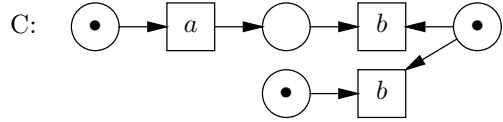
\begin{figure}[htb]
\input{netC}
\centerline{\raise 1em\box\graph}
\end{figure}

\noindent
In fact, C is the occurrence net obtained from A by the
 unfolding of \cite{En91,MMS92}.
In the individual token philosophy, both A and C have the computations
\begin{picture}(11,0)
\put(0,0){$a$}
\put(3,1){\vector(1,0){5}}
\put(9,0){$b$}
\end{picture}
and \plat{\begin{array}{@{}c@{}}\,\,a\!\\[-3pt]b\;\,\end{array}}.
However, in the collective token philosophy A does not have a run
\begin{picture}(11,0)
\put(0,0){$a$}
\put(3,1){\vector(1,0){5}}
\put(9,0){$b$}
\end{picture}
and can therefore not be equivalent to C in any causality preserving way.

Thus, capturing the behaviour of nets by means of our mapping $\fC$ to
configuration structures is compatible with the collective token
interpretation only. In the remainder of this paper, we therefore take
the collective token approach.

\subsubsection*{Rooted structures and finitary equivalence}

The configuration structure associated to a pure 1-occurrence net is
always \phrase{finitary}, meaning that all configurations are finite, and
\phrase{rooted}, meaning that the empty set of events is a configuration.
In order to translate between the models of concurrency seen before
and Petri nets, we therefore restrict attention to rooted structures,
and ignore infinite configurations.

\begin{definition}{rooted}
A configuration structure $\eC\!=\!\tuple{E,C}$ is \phrase{rooted} if
$\emptyset \!\in\! C$.  A propositional theory is \phrase{rooted} if it has
no clause of the form $\emptyset \implies X$ as a logical consequence.
An event structure $\eE=\tuple{E,\turn\;}$ is \phrase{rooted} if
$\emptyset \turn \emptyset$.
\end{definition}

\begin{proposition}{rooted}
If C is rooted, then so are $\fT(\eC)$ and $\fE(\eC)$.
If T is rooted, then so are $\fM(\eT)$ and $\fE(\eT)$.
If E is rooted, then so are $\fL(\eE)$ and $\fT(\eE)$.
\end{proposition}

\begin{proof}
Straightforward.
\end{proof}

\begin{definition}{finitary equivalence}
Given a configuration structure $\eC$, let $\fF(\eC)$ be the
configuration structure with the same events but with only the finite
configurations of $\eC$.
Two configuration structures $\eC$ and $\eD$ are \emph{finitarily
equivalent}\index{finitary equivalence}---written $\eC \simeq_f \eD$---if
$\fF(\eC) = \fF(\eD)$. 
\end{definition}
Instead of considering configuration structures up to finitary
equivalence, we could just as well restrict attention to finitary
configuration structures, thereby taking a normal form in each
equivalence class.  However, on the level of propositional theories
this involves adding clauses $X\implies\emptyset$ for every infinite
set of events $X$, which would needlessly complicate the forthcoming
\pr{NtoT}.  Moreover, the fact that $\fC(\eN)$ is finitary for every
pure 1-occurrence net $\eN$ is more a consequence of not considering
infinite configurations of Petri nets than of there not being any
(cf.\ Section~\ref{unbounded par}).

\out{
 A configuration structure is \phrase{finitary} if all its configurations
 are finite.  A propositional theory is \phrase{finitary} if, for every
 infinite $X \subseteq E$, it has the clause $X \implies \emptyset$ as
 a logical consequence.  An event structure $\tuple{E,\turn\;}$ is \phrase{
 finitary} if $X \turn Y$ only holds for finite sets $Y$.

 \begin{proposition}{finitary}
 If C is finitary, then so are $\fT(\eC)$ and $\fE(\eC)$.
 If T is finitary, then so are $\fM(\eT)$ and $\fE(\eT)$.
 If E is finitary, then so are $\fL(\eE)$ and $\fT(\eE)$.
 \end{proposition}

 \begin{proof}
 Straightforward.
 \end{proof}
}

\subsubsection*{From configuration structures to Petri nets}

We now proceed to show that, up to finitary equivalence, every rooted
configuration structure can be obtained as the image of a pure
1-occurrence net.

\begin{definition}{TtoN}
Let $\eT=\tuple{E,T}$ be a rooted propositional theory in
conjunctive normal form. We define the associated Petri net
$\fN(\eT)$ as follows.  As transitions of the net we take the events
from $E$. For every transition we add one place,
containing one initial token, that has no incoming arcs,
and with its only outgoing arc going to that transition.  These
\phrase{1-occurrence places} make sure that every transition fires at
most once.  For every clause $X \Rightarrow Y$ in $T$ with $X$ finite,
we introduce a place in the net. This place has outgoing arcs to each
of the transitions in $X$, and incoming arcs from each of the
transitions in $Y$. Let $n$ be the cardinality of $X$. As $\eT$ is
rooted, $n \neq 0$.  We finish the construction by putting $n-1$
initial tokens in the created place:
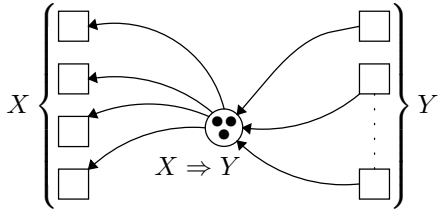
\expandafter\ifx\csname graph\endcsname\relax \csname newbox\endcsname\graph\fi
\expandafter\ifx\csname graphtemp\endcsname\relax \csname newdimen\endcsname\graphtemp\fi
\setbox\graph=\vtop{\vskip 0pt\hbox{%
    \special{pn 8}%
    \special{ar 866 610 98 98 0 6.28319}%
    \graphtemp=.5ex\advance\graphtemp by 0.583in
    \rlap{\kern 0.831in\lower\graphtemp\hbox to 0pt{\hss $\bullet$\hss}}%
    \graphtemp=.5ex\advance\graphtemp by 0.583in
    \rlap{\kern 0.902in\lower\graphtemp\hbox to 0pt{\hss $\bullet$\hss}}%
    \graphtemp=.5ex\advance\graphtemp by 0.650in
    \rlap{\kern 0.866in\lower\graphtemp\hbox to 0pt{\hss $\bullet$\hss}}%
    \special{pa 0 157}%
    \special{pa 157 157}%
    \special{pa 157 0}%
    \special{pa 0 0}%
    \special{pa 0 157}%
    \special{fp}%
    \special{ar 293 654 591 591 -1.802089 -0.242404}%
    \special{sh 1.000}%
    \special{pa 202 94}%
    \special{pa 157 79}%
    \special{pa 190 45}%
    \special{pa 202 94}%
    \special{fp}%
    \special{pa 0 433}%
    \special{pa 157 433}%
    \special{pa 157 276}%
    \special{pa 0 276}%
    \special{pa 0 433}%
    \special{fp}%
    \special{ar 295 1130 787 787 -1.746343 -0.862746}%
    \special{sh 1.000}%
    \special{pa 201 372}%
    \special{pa 157 354}%
    \special{pa 192 323}%
    \special{pa 201 372}%
    \special{fp}%
    \special{pa 0 709}%
    \special{pa 157 709}%
    \special{pa 157 551}%
    \special{pa 0 551}%
    \special{pa 0 709}%
    \special{fp}%
    \special{ar 472 1273 787 787 -1.982313 -1.159279}%
    \special{sh 1.000}%
    \special{pa 204 558}%
    \special{pa 157 551}%
    \special{pa 184 513}%
    \special{pa 204 558}%
    \special{fp}%
    \special{pa 0 984}%
    \special{pa 157 984}%
    \special{pa 157 827}%
    \special{pa 0 827}%
    \special{pa 0 984}%
    \special{fp}%
    \special{ar 703 1395 787 787 -2.335515 -1.488036}%
    \special{sh 1.000}%
    \special{pa 203 818}%
    \special{pa 157 827}%
    \special{pa 169 781}%
    \special{pa 203 818}%
    \special{fp}%
    \special{pa 1575 157}%
    \special{pa 1732 157}%
    \special{pa 1732 0}%
    \special{pa 1575 0}%
    \special{pa 1575 157}%
    \special{fp}%
    \special{pa 936 541}%
    \special{pa 1280 138}%
    \special{pa 1575 79}%
    \special{sp}%
    \special{sh 1.000}%
    \special{pa 980 527}%
    \special{pa 936 541}%
    \special{pa 942 494}%
    \special{pa 980 527}%
    \special{fp}%
    \special{pa 1575 433}%
    \special{pa 1732 433}%
    \special{pa 1732 276}%
    \special{pa 1575 276}%
    \special{pa 1575 433}%
    \special{fp}%
    \special{ar 1069 -169 787 787 0.872902 1.703581}%
    \special{sh 1.000}%
    \special{pa 1000 640}%
    \special{pa 965 610}%
    \special{pa 1007 591}%
    \special{pa 1000 640}%
    \special{fp}%
    \special{pa 1654 433}%
    \special{pa 1654 827}%
    \special{dt 0.079}%
    \special{pa 1575 984}%
    \special{pa 1732 984}%
    \special{pa 1732 827}%
    \special{pa 1575 827}%
    \special{pa 1575 984}%
    \special{fp}%
    \special{ar 1492 122 787 787 1.465367 2.355163}%
    \special{sh 1.000}%
    \special{pa 946 725}%
    \special{pa 936 680}%
    \special{pa 981 690}%
    \special{pa 946 725}%
    \special{fp}%
    \graphtemp=.5ex\advance\graphtemp by 0.827in
    \rlap{\kern 0.728in\lower\graphtemp\hbox to 0pt{\hss $X \implies Y$\hss}}%
    \hbox{\vrule depth0.984in width0pt height 0pt}%
    \kern 1.732in
  }%
}%

$$X \left\{\raisebox{1.3cm}[1.5cm][0pt]{\raise .5 mm \box\graph}\right\}Y$$
\end{definition}
The place belonging to the clause $X \Rightarrow Y$ does not place any
restrictions on the firing of the first \mbox{$n-1$} transitions in
$X$.  However, the last one can only fire after an extra token arrives
in the place.  This can happen only if one of the transitions in $Y$
fires first.  The firing of more transitions in $Y$ has no adverse
effects, as each of the transitions in $X$ can fire only once.  Thus
this place imposes the same restriction on the occurrence of events as
does the corresponding clause.

\begin{theorem}{PtoNtoC}
Let T be a rooted propositional theory in conjunctive normal
form. Then $$\fC(\fN(\eT)) \simeq_f {\cal M}(\eT).$$
\end{theorem}

\begin{proof}
$z \in C(\fN(\eT))$ iff $z$ is finite and $M_z(s) \geq 0$ for any
place $s$. We have $M_z(s) \geq 0$ for all 1-occurrence places $s$
exactly when no transition fires twice in $z$, i.e., when $z$ is a set.
For a place $s$ belonging to the clause $X \Rightarrow Y$ we have
$M_z(s) \geq 0$ iff either one of the transitions in $Y$ has fired, or
not all of the transitions in $X$ have fired, i.e., when $X \implies Y$
holds in $z$, seen as a model of propositional logic. The clauses
$X \implies Y$ of $\eT$ with $X$ infinite surely hold in any finite
configuration $z$.
Thus, $z \in C(\fN(\eT))$ iff $z$ is a finite model of $\eT$.
\end{proof}
The net $\fN(\eT)$ is always without arcweights.
Moreover, in case $\eT$ is pure (cf.\ \df{pure PT}), so is the net
$\fN(\eT)$. As any rooted configuration structure can be axiomatised
by a pure rooted propositional theory in conjunctive normal form,
it follows that
\begin{corollary}{CtoN}
For every rooted configuration structure there exists a pure
1-occurrence net without arcweights with the same finite configurations.
\hfill $\Box$
\end{corollary}
Thus we have established a bijective correspondence between rooted
configuration structures up to finitary equivalence and pure
1-oc\-cur\-rence nets up to configuration equivalence.
Moreover, every pure 1-occurrence net is configuration
equivalent to a pure 1-occurrence net without arcweights.

\begin{example}{ternary-PN}
The event structure with ternary conflict of \ex{ternary conflict} can
be represented by the propositional theory $$(d\wedge e \wedge f)
\Rightarrow \bot\;.$$
The Petri net associated to\\
this theory by \df{TtoN} is:\hfill
\expandafter\ifx\csname graph\endcsname\relax \csname newbox\endcsname\graph\fi
\expandafter\ifx\csname graphtemp\endcsname\relax \csname newdimen\endcsname\graphtemp\fi
\setbox\graph=\vtop{\vskip 0pt\hbox{%
    \special{pn 8}%
    \special{ar 492 98 98 98 0 6.28319}%
    \graphtemp=.5ex\advance\graphtemp by 0.098in
    \rlap{\kern 0.492in\lower\graphtemp\hbox to 0pt{\hss $\bullet\bullet$\hss}}%
    \special{pa 0 591}%
    \special{pa 197 591}%
    \special{pa 197 394}%
    \special{pa 0 394}%
    \special{pa 0 591}%
    \special{fp}%
    \graphtemp=.5ex\advance\graphtemp by 0.492in
    \rlap{\kern 0.098in\lower\graphtemp\hbox to 0pt{\hss $d$\hss}}%
    \special{pa 423 168}%
    \special{pa 197 394}%
    \special{fp}%
    \special{sh 1.000}%
    \special{pa 242 384}%
    \special{pa 197 394}%
    \special{pa 207 348}%
    \special{pa 242 384}%
    \special{fp}%
    \special{pa 394 591}%
    \special{pa 591 591}%
    \special{pa 591 394}%
    \special{pa 394 394}%
    \special{pa 394 591}%
    \special{fp}%
    \graphtemp=.5ex\advance\graphtemp by 0.492in
    \rlap{\kern 0.492in\lower\graphtemp\hbox to 0pt{\hss $e$\hss}}%
    \special{pa 492 197}%
    \special{pa 492 394}%
    \special{fp}%
    \special{sh 1.000}%
    \special{pa 517 354}%
    \special{pa 492 394}%
    \special{pa 467 354}%
    \special{pa 517 354}%
    \special{fp}%
    \special{pa 787 591}%
    \special{pa 984 591}%
    \special{pa 984 394}%
    \special{pa 787 394}%
    \special{pa 787 591}%
    \special{fp}%
    \graphtemp=.5ex\advance\graphtemp by 0.492in
    \rlap{\kern 0.886in\lower\graphtemp\hbox to 0pt{\hss $f$\hss}}%
    \special{pa 562 168}%
    \special{pa 787 394}%
    \special{fp}%
    \special{sh 1.000}%
    \special{pa 777 348}%
    \special{pa 787 394}%
    \special{pa 742 384}%
    \special{pa 777 348}%
    \special{fp}%
    \special{ar 886 846 98 98 0 6.28319}%
    \graphtemp=.5ex\advance\graphtemp by 0.846in
    \rlap{\kern 0.886in\lower\graphtemp\hbox to 0pt{\hss $\bullet$\hss}}%
    \special{pa 886 748}%
    \special{pa 886 591}%
    \special{fp}%
    \special{sh 1.000}%
    \special{pa 861 630}%
    \special{pa 886 591}%
    \special{pa 911 630}%
    \special{pa 861 630}%
    \special{fp}%
    \special{ar 492 846 98 98 0 6.28319}%
    \graphtemp=.5ex\advance\graphtemp by 0.846in
    \rlap{\kern 0.492in\lower\graphtemp\hbox to 0pt{\hss $\bullet$\hss}}%
    \special{pa 492 748}%
    \special{pa 492 591}%
    \special{fp}%
    \special{sh 1.000}%
    \special{pa 467 630}%
    \special{pa 492 591}%
    \special{pa 517 630}%
    \special{pa 467 630}%
    \special{fp}%
    \special{ar 98 846 98 98 0 6.28319}%
    \graphtemp=.5ex\advance\graphtemp by 0.846in
    \rlap{\kern 0.098in\lower\graphtemp\hbox to 0pt{\hss $\bullet$\hss}}%
    \special{pa 98 748}%
    \special{pa 98 591}%
    \special{fp}%
    \special{sh 1.000}%
    \special{pa 73 630}%
    \special{pa 98 591}%
    \special{pa 123 630}%
    \special{pa 73 630}%
    \special{fp}%
    \hbox{\vrule depth0.945in width0pt height 0pt}%
    \kern 0.984in
  }%
}%

\raisebox{1.5cm}[0pt][0pt]{\box\graph}
\vspace{.8cm}
\end{example}

\begin{example}{resolvable}
Below are the event structure with resolvable conflict from \ex{resolved conflict},
its representation as a propositional theory, and the associated Petri
net, as well as its configurations, ordered by inclusion.
\\
\expandafter\ifx\csname graph\endcsname\relax \csname newbox\endcsname\graph\fi
\expandafter\ifx\csname graphtemp\endcsname\relax \csname newdimen\endcsname\graphtemp\fi
\setbox\graph=\vtop{\vskip 0pt\hbox{%
    \special{pn 8}%
    \special{ar 1811 98 98 98 0 6.28319}%
    \graphtemp=.5ex\advance\graphtemp by 0.098in
    \rlap{\kern 1.811in\lower\graphtemp\hbox to 0pt{\hss $\bullet$\hss}}%
    \special{pa 1811 197}%
    \special{pa 1811 394}%
    \special{fp}%
    \special{sh 1.000}%
    \special{pa 1836 354}%
    \special{pa 1811 394}%
    \special{pa 1786 354}%
    \special{pa 1836 354}%
    \special{fp}%
    \special{pa 1713 591}%
    \special{pa 1909 591}%
    \special{pa 1909 394}%
    \special{pa 1713 394}%
    \special{pa 1713 591}%
    \special{fp}%
    \graphtemp=.5ex\advance\graphtemp by 0.492in
    \rlap{\kern 1.811in\lower\graphtemp\hbox to 0pt{\hss $c$\hss}}%
    \special{pa 1811 591}%
    \special{pa 1811 787}%
    \special{fp}%
    \special{sh 1.000}%
    \special{pa 1836 748}%
    \special{pa 1811 787}%
    \special{pa 1786 748}%
    \special{pa 1836 748}%
    \special{fp}%
    \special{ar 1811 886 98 98 0 6.28319}%
    \graphtemp=.5ex\advance\graphtemp by 0.886in
    \rlap{\kern 1.811in\lower\graphtemp\hbox to 0pt{\hss $\bullet$\hss}}%
    \special{ar 1417 886 98 98 0 6.28319}%
    \graphtemp=.5ex\advance\graphtemp by 0.886in
    \rlap{\kern 1.417in\lower\graphtemp\hbox to 0pt{\hss $\bullet$\hss}}%
    \special{pa 1417 984}%
    \special{pa 1417 1181}%
    \special{fp}%
    \special{sh 1.000}%
    \special{pa 1442 1142}%
    \special{pa 1417 1181}%
    \special{pa 1392 1142}%
    \special{pa 1442 1142}%
    \special{fp}%
    \special{pa 1319 1378}%
    \special{pa 1516 1378}%
    \special{pa 1516 1181}%
    \special{pa 1319 1181}%
    \special{pa 1319 1378}%
    \special{fp}%
    \graphtemp=.5ex\advance\graphtemp by 1.280in
    \rlap{\kern 1.417in\lower\graphtemp\hbox to 0pt{\hss $a$\hss}}%
    \special{pa 1741 955}%
    \special{pa 1516 1181}%
    \special{fp}%
    \special{sh 1.000}%
    \special{pa 1561 1171}%
    \special{pa 1516 1181}%
    \special{pa 1526 1136}%
    \special{pa 1561 1171}%
    \special{fp}%
    \special{ar 2205 886 98 98 0 6.28319}%
    \graphtemp=.5ex\advance\graphtemp by 0.886in
    \rlap{\kern 2.205in\lower\graphtemp\hbox to 0pt{\hss $\bullet$\hss}}%
    \special{pa 2205 984}%
    \special{pa 2205 1181}%
    \special{fp}%
    \special{sh 1.000}%
    \special{pa 2230 1142}%
    \special{pa 2205 1181}%
    \special{pa 2180 1142}%
    \special{pa 2230 1142}%
    \special{fp}%
    \special{pa 2106 1378}%
    \special{pa 2303 1378}%
    \special{pa 2303 1181}%
    \special{pa 2106 1181}%
    \special{pa 2106 1378}%
    \special{fp}%
    \graphtemp=.5ex\advance\graphtemp by 1.280in
    \rlap{\kern 2.205in\lower\graphtemp\hbox to 0pt{\hss $b$\hss}}%
    \special{pa 1881 955}%
    \special{pa 2106 1181}%
    \special{fp}%
    \special{sh 1.000}%
    \special{pa 2096 1136}%
    \special{pa 2106 1181}%
    \special{pa 2061 1171}%
    \special{pa 2096 1136}%
    \special{fp}%
    \graphtemp=.5ex\advance\graphtemp by 0.295in
    \rlap{\kern 0.787in\lower\graphtemp\hbox to 0pt{\hss $\begin{array}{c@{~}c@{~}l@{}l}E&=&\{a,b,c\}\!\!\!\!\!\!\!\\ \{c\} &\turn& \{a, b\}\\\emptyset &\turn& X& \mbox{for }X\neq\{a,b\}\end{array}$\hss}}%
    \graphtemp=.5ex\advance\graphtemp by 1.083in
    \rlap{\kern 0.433in\lower\graphtemp\hbox to 0pt{\hss $(a\wedge b)\implies c$\hss}}%
    \hbox{\vrule depth1.378in width0pt height 0pt}%
    \kern 2.303in
  }%
}%

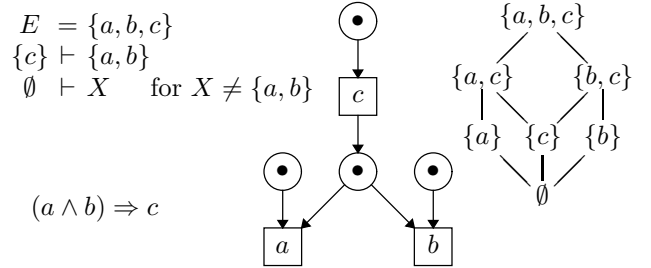
\raisebox{3.8cm}{\box\graph}
\hfill\begin{picture}(25,50)(-10,-14)
\put(0,0){\makebox[0pt]{$\emptyset$}} \put(-2,3){\line(-1,1){5}}
\put(-10,10){\makebox[0pt]{$\{a\}$}}  \put(2,3){\line(1,1){5}}
\put(10,10){\makebox[0pt]{$\{b\}$}}   \put(-7.5,23.5){\line(1,1){5}}
\put(0,10){\makebox[0pt]{$\{c\}$}}    \put(2.5,13.5){\line(1,1){5}}
\put(0,4){\line(0,1){4}}            \put(-2.5,13.5){\line(-1,1){5}}
\put(-10,20){\makebox[0pt]{$\{a,c\}$}}  \put(-10,14){\line(0,1){5}}
\put(10,20){\makebox[0pt]{$\{b,c\}$}}   \put(10,14){\line(0,1){5}}
\put(0,30){\makebox[0pt]{$\{a,b,c\}$}}  \put(7.5,23.5){\line(-1,1){5}}
\end{picture}
\end{example}

\begin{example}{disjunctive}
Below is a propositional theory describing a system in which either
$a$ or $b$ is sufficient to enable the event $c$; this is sometimes
called \phrase{disjunctive causality}.  We also display the associated
Petri net, and its representation as an event structure and a
configuration structure.
\\
\expandafter\ifx\csname graph\endcsname\relax \csname newbox\endcsname\graph\fi
\expandafter\ifx\csname graphtemp\endcsname\relax \csname newdimen\endcsname\graphtemp\fi
\setbox\graph=\vtop{\vskip 0pt\hbox{%
    \special{pn 8}%
    \special{ar 1654 886 98 98 0 6.28319}%
    \special{pa 1654 984}%
    \special{pa 1654 1181}%
    \special{fp}%
    \special{sh 1.000}%
    \special{pa 1679 1142}%
    \special{pa 1654 1181}%
    \special{pa 1629 1142}%
    \special{pa 1679 1142}%
    \special{fp}%
    \special{pa 1555 1378}%
    \special{pa 1752 1378}%
    \special{pa 1752 1181}%
    \special{pa 1555 1181}%
    \special{pa 1555 1378}%
    \special{fp}%
    \graphtemp=.5ex\advance\graphtemp by 1.280in
    \rlap{\kern 1.654in\lower\graphtemp\hbox to 0pt{\hss $c$\hss}}%
    \special{ar 1260 98 98 98 0 6.28319}%
    \graphtemp=.5ex\advance\graphtemp by 0.098in
    \rlap{\kern 1.260in\lower\graphtemp\hbox to 0pt{\hss $\bullet$\hss}}%
    \special{pa 1260 197}%
    \special{pa 1260 394}%
    \special{fp}%
    \special{sh 1.000}%
    \special{pa 1285 354}%
    \special{pa 1260 394}%
    \special{pa 1235 354}%
    \special{pa 1285 354}%
    \special{fp}%
    \special{pa 1161 591}%
    \special{pa 1358 591}%
    \special{pa 1358 394}%
    \special{pa 1161 394}%
    \special{pa 1161 591}%
    \special{fp}%
    \graphtemp=.5ex\advance\graphtemp by 0.492in
    \rlap{\kern 1.260in\lower\graphtemp\hbox to 0pt{\hss $a$\hss}}%
    \special{pa 1358 591}%
    \special{pa 1584 816}%
    \special{fp}%
    \special{sh 1.000}%
    \special{pa 1574 771}%
    \special{pa 1584 816}%
    \special{pa 1538 806}%
    \special{pa 1574 771}%
    \special{fp}%
    \special{ar 2047 98 98 98 0 6.28319}%
    \graphtemp=.5ex\advance\graphtemp by 0.098in
    \rlap{\kern 2.047in\lower\graphtemp\hbox to 0pt{\hss $\bullet$\hss}}%
    \special{pa 2047 197}%
    \special{pa 2047 394}%
    \special{fp}%
    \special{sh 1.000}%
    \special{pa 2072 354}%
    \special{pa 2047 394}%
    \special{pa 2022 354}%
    \special{pa 2072 354}%
    \special{fp}%
    \special{pa 1949 591}%
    \special{pa 2146 591}%
    \special{pa 2146 394}%
    \special{pa 1949 394}%
    \special{pa 1949 591}%
    \special{fp}%
    \graphtemp=.5ex\advance\graphtemp by 0.492in
    \rlap{\kern 2.047in\lower\graphtemp\hbox to 0pt{\hss $b$\hss}}%
    \special{pa 1949 591}%
    \special{pa 1723 816}%
    \special{fp}%
    \special{sh 1.000}%
    \special{pa 1769 806}%
    \special{pa 1723 816}%
    \special{pa 1733 771}%
    \special{pa 1769 806}%
    \special{fp}%
    \special{ar 1260 886 98 98 0 6.28319}%
    \graphtemp=.5ex\advance\graphtemp by 0.886in
    \rlap{\kern 1.260in\lower\graphtemp\hbox to 0pt{\hss $\bullet$\hss}}%
    \special{pa 1329 955}%
    \special{pa 1555 1181}%
    \special{fp}%
    \special{sh 1.000}%
    \special{pa 1545 1136}%
    \special{pa 1555 1181}%
    \special{pa 1510 1171}%
    \special{pa 1545 1136}%
    \special{fp}%
    \graphtemp=.5ex\advance\graphtemp by 1.083in
    \rlap{\kern 0.709in\lower\graphtemp\hbox to 0pt{\hss $\begin{array}{c@{~}c@{~}l@{}l}E&=&\{a,b,c\}\!\!\!\!\!\!\!\\ \{a\} &\turn& \{c\}\\ \{b\} &\turn& \{c\}\\\emptyset &\turn& X& \mbox{for }X\neq\{c\}\end{array}$\hss}}%
    \graphtemp=.5ex\advance\graphtemp by 0.295in
    \rlap{\kern 0.354in\lower\graphtemp\hbox to 0pt{\hss $c \implies (a\vee b)$\hss}}%
    \hbox{\vrule depth1.378in width0pt height 0pt}%
    \kern 2.146in
  }%
}%

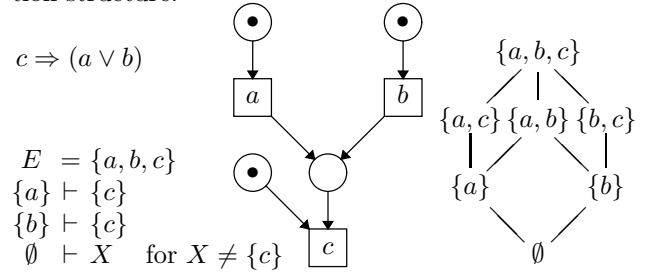
\raisebox{3.8cm}{\box\graph}
\unitlength .9mm
\hfill\begin{picture}(24,20)(-10,-4)
\put(0,0){\makebox[0pt]{$\emptyset$}} \put(-2,3){\line(-1,1){5}}
\put(-10,10){\makebox[0pt]{$\{a\}$}}  \put(2,3){\line(1,1){5}}
\put(10,10){\makebox[0pt]{$\{b\}$}}   \put(-7.5,23.5){\line(1,1){5}}
\put(0,20){\makebox[0pt]{$\{a,b\}$}}  \put(-7.5,13.5){\line(1,1){5}}
\put(0,24){\line(0,1){4}}             \put(7.5,13.5){\line(-1,1){5}}
\put(-10,20){\makebox[0pt]{$\{a,c\}$}}  \put(-10,14){\line(0,1){5}}
\put(10,20){\makebox[0pt]{$\{b,c\}$}}   \put(10,14){\line(0,1){5}}
\put(0,30){\makebox[0pt]{$\{a,b,c\}$}}  \put(7.5,23.5){\line(-1,1){5}}
\end{picture}
In case we modify the event structure by omitting the enabling
$\emptyset\turn\{a,b\}$, the propositional theory gains a clause
$(a\wedge b) \Rightarrow \bot$, the Petri net gains a marked place
with arrows to $a$ and $b$, and the configuration structure loses the
configurations $\{a,b\}$ and  $\{a,b,c\}$.
\end{example}

\subsubsection*{From nets to theories and event structures}
\label{NtoT}

We know already how to translate pure 1-occurrence nets into
propositional theories and event structures, namely through the
intermediate stage of configuration structures. Below we provide
direct translations that might shed more light on the relationships
between these models of concurrency.

Let $\eN=\tuple{S,E,F,I}$ be a 1-occurrence net.
For any place $s \in S$ let $s^\bullet := \{t \in E \mid F(s,t)>0\}$ be
its set of \phrase{posttransitions} and $^\bullet\! s := \{t \in E \mid
F(t,e)>0\}$ its set of \phrase{pretransitions}.
For any finite set $Y \subseteq s^\bullet$ of posttransitions of $s$,
$^\bullet Y(s)$ is the number of tokens needed in place $s$ for
all transitions in $Y$ to fire,\footnote{In case $\eN$ is without
arcweights, $^\bullet Y\!(s)$ is simply $|Y|$ (cf.$\!$\ \cite{GP04}).}
so $^\bullet Y(s)- I(s)$, if positive, is the number of tokens that
have to arrive in $s$ before all transitions in $Y$ can fire.
Furthermore, for $n \!\in\! \IZ$, let \mbox{$^n\!s := \{X \!\subseteq
\mbox{$^\bullet\!s$} \mid X^\bullet(s) \geq n\}$} be the collection of
sets $X$ of pretransitions of $s$, such that if all transitions in $X$
fire, at least $n$ tokens will arrive in $s$.
Write $^Y\!\!s$ for \plat{^{^\bullet Y(s)-I(s)}\!s}. One
of the sets of transitions in $^Y\!\!s$ has to fire entirely before
all transitions in $Y$ can fire.

The formula $\varphi_s^n := \bigvee_{X \in \,^n\!s} \bigwedge X$
expresses which transitions need to fire for $n$ tokens to arrive in
$s$. The formula $\bigwedge Y \Rightarrow \varphi^ {^\bullet
Y(s)-I(s)}_s$ expresses that one of the sets of transitions in
$^Y\!\!s$ has to fire entirely before all transitions in $Y$ can
fire. The \phrase{propositional theory associated to} $\eN$ is defined
as $\fT(\eN):=\tuple{E,T(\eN)}$, where $T(\eN)$ consists of all
formulae $\bigwedge Y \Rightarrow \varphi^{^\bullet Y(s)-I(s)}_s$ with
$s\in S$ and $Y \subseteq_{\it fin} s^\bullet$. It follows that

\begin{proposition}{NtoT}
$\fM(\fT(\eN)) \simeq_f \fC(\eN)$ for any pure 1-occurrence net $\eN$.
\hfill$\Box$
\end{proposition}

\begin{proof}
Let $\eN=\tuple{S,E,F,I}$ be a pure 1-occurrence net and
$X\subseteq_{\it fin} E$ be a finite set of transitions of $\eN$.
Then $X \in \fM(\fT(\eN))$ iff for all $s \in S$ and $Y
\subseteq_{\it fin} s^\bullet$ the formula \plat{\bigwedge Y \Rightarrow
\varphi^{^\bullet Y(s)-I(s)}_s} is true in $X$, which is
the case iff $(Y\subseteq X) \implies \exists Z \subseteq X.~ Z \in\!
\plat{^{^\bullet Y(s)-I(s)}\!s}$, or $$(Y\subseteq X)
\implies \exists Z \subseteq X.~ Z \subseteq \mbox{$^\bullet s$} \wedge
Z^\bullet(s) \geq  \mbox{$^\bullet Y(s)$}-I(s).$$
In the latter formula the clause $Z \subseteq \mbox{$^\bullet s$}$ can just
as well be deleted, as transitions in $Z$ that are not in $^\bullet s$
do not make a contribution to $Z^\bullet(s)$ anyway. Thus this formula
is equivalent to $$(Y\subseteq X) \implies X^\bullet(s) \geq
\mbox{$^\bullet Y(s)$}-I(s).$$
Likewise, requiring this implication to merely hold for sets of
transitions $Y$ with $Y \subseteq_{\it fin} s^\bullet$ is moot.
Hence $$X \in \fM(\fT(\eN))
\begin{array}[t]{@{~}l@{}}
 \Leftrightarrow
\forall s \in S.~ X^\bullet(s) \geq \mbox{$^\bullet X(s)$}-I(s)\\
 \Leftrightarrow
\forall s \in S.~ I(s)- \mbox{$^\bullet X(s)$}+X^\bullet(s) \geq 0\\
 \Leftrightarrow M_X \geq 0\\
 \Leftrightarrow X \in \fC(\eN).
\end{array}\vspace{-1.5em}$$
\end{proof}

For any finite set of transitions $Y \subseteq E$, let $S_Y$ be the set of
places $s$ with $Y \subseteq s^\bullet$ and $^\bullet Y(s)-I(s)>0$.
Now write $X \turn_\eN Y$ whenever $X = \bigcup_{s \in S_Y} X_s$ with
$X_s \in\, ^Y\!\!s$. We also write $\emptyset \turn_\eN Y$ whenever
$Y$ is infinite. The \phrase{event structure associated to} $\eN$ is
defined as $\fE(\eN):=\tuple{E,\turn_\eN\;}$.
Note that if $\eN$ is pure, then so is $\fE(\eN)$.

\begin{proposition}{NtoE}
Let $\eN$ be a pure 1-occurrence net. Then
$\fL(\fE(\eN)) \simeq_f \fC(\eN)$.
\hfill$\Box$
\end{proposition}

\begin{proof}
Let $\eN=\tuple{S,E,F,I}$ be a pure 1-occurrence net and
$X\subseteq_{\it fin} E$ be a finite set of transitions of $\eN$.
Then\vspace{-2em}$$$$
$$\mbox{~~~~~} X \in \fL(\fE(\eN)) \Leftrightarrow \forall Y\subseteq X.~
\exists Z \subseteq X.~ Z \turn_\eN Y$$
$$\Leftrightarrow\forall Y\subseteq X.~\forall s\in S_Y.~\exists Z_s
\subseteq X.~ Z_s\in\mbox{$^Y\!\!s$}\Leftrightarrow$$
$$\forall Y\!\subseteq\! X.~\forall s\!\in\! S_Y.~\exists Z_s
\!\subseteq\! X.~ Z_s \!\subseteq\! \mbox{$^\bullet s$} \wedge
Z_s^\bullet\!(s) \!\geq\! \mbox{$^\bullet Y\!(s)$}-I(s)$$
$$\Leftrightarrow\forall Y\!\subseteq\! X.~\forall s\!\in\! S_Y.~
X^\bullet\!(s) \geq \mbox{$^\bullet Y(s)$}-I(s)$$
$$\Leftrightarrow\forall Y\!\subseteq\! X.~\forall s\!\in\! S.~
(Y\!\subseteq\! s^\bullet \Rightarrow
X^\bullet\!(s) \geq \mbox{$^\bullet Y(s)$}-I(s))$$
$$\Leftrightarrow\forall s\!\in\! S.~\forall Y\!\subseteq\! X.~
X^\bullet\!(s) \geq \mbox{$^\bullet Y(s)$}-I(s)$$
$$\Leftrightarrow\forall s\!\in\! S.~X^\bullet\!(s) \geq
\mbox{$^\bullet\! X(s)$}-I(s)$$
$$\Leftrightarrow M_X \geq 0 \Leftrightarrow X \in \fC(\eN).\vspace{-1.7em}$$
\end{proof}
The size of $\fT(\eN)$ and $\fE(\eN)$ can be reduced by redefining
$^n\!s$ to consist of the \emph{minimal} subsets $X$ of
\mbox{$^\bullet\!s$} with $X^\bullet(s) \geq n$. This does not affect
the truth of Propositions~\ref{pr-NtoT} and~\ref{pr-NtoE}, although it
slightly complicates their proofs. Likewise, in the definition of
$\fT(\eN)$ only those formulae \plat{\bigwedge Y \Rightarrow
\varphi^{^\bullet Y(s)-I(s)}_s}
are needed for which $^\bullet Y(s)-I(s)>0$ (the remaining formulae
being tautologies). This yields the maps of \cite{GP04}.

\out{
 Let $\fT'$ denote the thusly
 modified map from nets to propositional theories. The translations
 $\fN$ and $\fT'$ between propositional theories in conjunctive normal
 form and Petri nets preserve even more information than finite
 models/configurations:

 \begin{proposition}{TtoNtoT}
 Let $\eT$ be a rooted propositional theory WITH FINITE CONFLICT in
 conjunctive normal form. Then $\fT'(\fN(\eT))=\eT$.
 \end{proposition}

 \begin{proof}
 The pairs $(s,Y)$ with $s$ a place in $\fN(\eT)$ and $Y\subseteq_{\it
 fin} s^\bullet$ satisfying $^\bullet Y(s)-I(s)>0$ are exactly those
 with $s$ the place belonging to a clause $Y\implies X$ in $\eT$, and
 for all such pairs we have $^\bullet Y(s)-I(s)=1$. The formula
 generated by $\fT'$ is again $Y\implies X$.
 \end{proof}
}

\subsubsection*{1-Unfolding}

Below we show that the restriction to 1-occurrence nets is not very
crucial; every net can be ``unfolded'' into a 1-occurrence net
without changing its behaviour in any essential way. However, the
unfolding cannot be configuration equivalent to the original, as
the identity of transitions cannot be preserved.

\begin{definition}{1-unfolding}
Let $N=\tuple{S,T,F,I}$ be a Petri net. Its \phrase{1-unfolding}
$N':=\tuple{S',T',F',I'}$ into a 1-occurrence net is given by
(for $s \in S$, $t\in T$, $u,u'\in T'$)
\begin{itemise}
\item $T':=T \times \IN$,
\item $S':=S \cup (T' \times \{*\})$,
\item $F'(s,(t,n)):=F(s,t)$ and $F'((t,n),s):=F(t,s)$,
\item $F'(u,(u,*)) = F'((u,*),u') = F(u',(u,*)) := 0$ and
      $F'((u,*),u):=1$ for $u,u'\in T'$ with $u\neq u'$,
\item $I'(s):=I(s)$ and $I'((u,*)):=1$.
\end{itemise}
\end{definition}
Thus, every transition is replaced by countably many copies, each of
which is connected with its environment (though the flow relation) in
exactly the same way as the original. Furthermore, for every such copy
$u$ an extra place $(u,*)$ is created, containing one initial token, and
having no incoming arcs and only one outgoing arc, going to $u$.
This place guarantees that $u$ can fire only once.

\begin{figure}[htb]
\input{1-unfolding}
\centerline{\raise 1em\box\graph}\vspace{1em}
~~~~\emph{A net fragment \hfill and its 1-unfolding}~~~
\vspace{-.65em}
\end{figure}

We argue that the causal and branching time behaviour of the
represented system is preserved under 1-unfolding. When dealing with
labelled Petri nets, all copies $(t,n)$ of a transition $t$ carry the
same label as $t$. In this setting, common semantic equivalences like
the \phrase{fully concurrent bisimulation} \emph{equivalence} \cite{BDKP91}
or the \emph{(hereditary) history preserving bisimulation
equivalence}\index{history preserving bisimulation} \cite{vG06} under
either the individual or collective token interpretation identify a
net and its 1-unfolding.

Note that the construction above does not introduce self-loops. Thus
unfoldings of pure nets remain pure.
We therefore have translations between arbitrary pure nets, event
structures, configuration structures and propositional theories, as
indicated in the introduction.

\index{self-sequential interpretation of Petri nets}
It is possible to give a slightly different interpretation of nets,
namely by excluding transitions from firing concurrently with
themselves (cf.\ \cite{GR83}).\footnote{This distinction is
independent of the individual--collective token dichotomy, thus
yielding four computational interpretations of nets \cite{vG05}.}
This amounts to simplifying \df{firing} by requiring $U$ to be a set
rather than a multiset. Under this interpretation our unfolding
could introduce concurrency that was not present before. However, for
this purpose \df{1-unfolding} can be adapted by removing the initial
tokens from the places $((t,n),*)$ for $t\in T$ and $n>0$ (but leaving
the token in $((t,0),*)$), and adding an arc from transition $(t,n)$
to place $((t,n+1),*)$ for every $t\in T$ and $n \in \IN$.

\out{
 \subsection{Self-concurrency}

 In older papers on Petri nets a multiset of transitions was allowed to
 fire only if it was a set, i.e., no transition could fire multiple
 times concurrent with itself. The argument for this restriction was
 that a transition could be thought of as a subsystem like a printer,
 that can only print one file at a time. When there are enough tokens in
 its preplaces (representing print-requests and other preconditions for
 printing) to handle two files, these have to be printed one by one.
 This argument has been convincingly rebutted in {\sc Goltz \& Reisig}
 \cite{GR83}, and since then multisets are generally allowed to fire.
 In any case, the behaviour of nets under the \phrase{self-sequential}
 firing rule can easily be encoded into the behaviour of nets
 under the \phrase{self-concurrent} firing rule of \df{firing} by the
 following proposition.

 \begin{proposition}{self-sequential}
 Any net N can be transformed into a net $\eN'$ such that
 \begin{itemise}
 \item under the self-sequential firing rule $\eN'$ behaves the same as N,
 \item in N' we have $M \goto{U} M'$ only if $U$ is a set.
 \end{itemise}
 \end{proposition}

 \begin{proof}
 For any transition $t$ in N add a \phrase{self-loop}, consisting of a place
 $s_t$ with $I(s_t)=F(s_t,t)=F(t,s_t):=1$ and $\forall u\neq t.
 F(s_t,u)=F(u,s_t):=0$. This yields the required net $\eN'$.
 \end{proof}
 Further on we assume the firing rule of \df{firing}, but indicate when
 necessary what has to be changed in case the self-sequential firing
 rule is assumed.
}

\out{
 \subsubsection*{Multiset systems}

 \begin{definition}{multiset system}
 A \phrase{multiset system} is a pair $\tuple{E,M}$ with $E$ a
 set and $M$ a collection of multisets over $E$.
 \end{definition}
 A multiset system is a generalised configuration structure in which
 events may occur multiple times in the same system run. As indicated in
 \df{configuration}, the configuration structure associated to any pure
 Petri net $\eN$ is a multiset system; it is a set system only when
 $\eN$ is a 1-occurrence net.  We have no direct method of translating
 multiset systems back into Petri nets, nor do we have generalisations
 of event structures or propositional theories matching the expressive
 power of multiset systems.  However, every multiset system can be
 ``1-unfolded'' into a set system representing the same behaviour (cf.\
 Section~\ref{computational}), thereby making the extra expressive
 power of multiset systems less relevant.

 \begin{definition}{multi2set}
 Let $\tuple{E,M}$ be a multiset system. The \phrase{1-unfolding}
 $\tuple{E',C}$ of $\tuple{E,M}$ into a set system is given by
 \begin{itemise}
 \item $E':=E \times \IN$,
 \item $x \in C$ iff $\pi(x)\in M$, for $x\subset E'$, where $\pi(x)\in
 \IN^E$ is given by $\pi(x)(e):=|\{n\mid (e,n)\in x\}|$.
 \end{itemise}
 \end{definition}
 The above definitions assume that multisets may have only finitely
 many occurrences of the same event, an assumption that holds for the
 multiset systems associated to Petri nets. However, the generalisation
 to multisets with arbitrary many occurrences of the same event is
 fairly straightforward.

 Now we have two translation from Petri nets into configuration
 structures, as indicated below.
 $$\begin{array}{ccc}
 \mbox{\em Petri nets}&\goto{\fC}&\mbox{\em multiset systems}\\
 \makebox[0pt][r]{\raisebox{5pt}{\footnotesize\em 1-unfolding}}
 \stackrel{|}{\raisebox{0pt}[6pt]{$\downarrow$}}&&
 \stackrel{|}{\raisebox{0pt}[6pt]{$\downarrow$}}
 \makebox[0pt][l]{\raisebox{5pt}{\footnotesize\em 1-unfolding}}\\
 \mbox{\em 1-occurrence nets}&\goto{\fC}&\mbox{\em set systems}
 \end{array}$$
 The following proposition, whose proof is straightforward, says that
 this diagram commutes.
 \begin{proposition}{1-unfolding}
 Let $\eN$ be a Petri net.
 The configuration structure associated to the 1-unfolding of $\eN$
 equals the 1-unfolding of the configuration structure associated to
 $\eN$. \hfill $\Box$
 \end{proposition}
}

\out{
 \subsection{Marking equivalence (to be omitted)}

 One often is interested in the behaviour of nets as far as it can be
 expressed in terms of transition firings. The places etc.\ are then
 seen as just a tool in specifying such behaviour. In this view, one of the
 most discriminating behavioural equivalences we can think of in the
 collective token framework is the following notion of \phrase{marking
 equivalence}:
 \begin{definition}{marking equivalence}
 Two nets $N$ and $N'$ are \phrase{marking equivalent} if $T_N = T_{N'}$
 and there exists a bijection $i$ between their reachable markings,
 such that the initial markings are related, $M \goto{U}_{N} M'
 \Leftrightarrow i(M) \goto{U}_{N'} i(M')$, and $l_N = l_{N'}$.
 \end{definition}
 Note that marking equivalence preserves all causal information
 present in the net representation of a concurrent system. Whether two
 transitions are causally independent is a context-sensitive matter. It
 varies with the markings enabling them both. In such a marking two
 transitions are independent iff they can fire in one step. This
 kind of information is present in the step transition relation \plat{\goto{U}}.
 The nets A and B are marking equivalent. However,
 for many purposes marking equivalence is too fine. 
 It distinguishes for instance the nets P and Q below, as well as M and N\@.

 \begin{figure}[t]
  \input{marking}
  \centerline{\raise 1em\box\graph}
  \end{figure}


 \phrase{Reachable configuration equivalence} is defined similarly and is
 strictly coarser than (reachable) marking equivalence:
 The nets P and Q are (reachable) configuration equivalent.
 However M and N below are not.  The reason is that although in N all
 transitions have a different identity, even though they have the same label.
 \out{
  \begin{figure}[htb]
  \input{identity}
  \centerline{\raise 1em\box\graph}
  \end{figure}
  }
}

\section{Computational interpretation}\label{computational}

In this section we formalise the dynamic behaviour of configuration
structures, Petri nets and event structures, by defining a transition
relation between their configurations. This transition relation tells
how a represented system can evolve from one state to another.  We
prove that on the classes of pure 1-occurrence nets and pure event
structures the translations of Section~\ref{four models} preserve
these transition relations, and show that this result does not extend
to impure 1-occurrence nets or impure event structures.

We indicate that impure nets and event structures may be captured by
considering configuration structures upgraded with an explicit
transition relation between their configurations. However, the
methodology of the present paper is incapable of providing transition
preserving translations between general event structures, 1-occurrence
nets and the upgraded configuration structures.  It is for this reason
that we focus on pure nets and pure event structures.

Our transition relation for Petri nets is derived directly from the
firing rule, which constitutes the standard computational
interpretation of nets. The idea of explicitly defining a transition
relation between the configurations of an event structure may be new,
but we believe that our transition relation is the only natural
candidate that is consistent with the notion of configuration employed
in {\sc Winskel} \cite{Wi87a,Wi89}
(cf.~Sections~\ref{equivalence} and~\ref{brands}).
Our transition relation on configuration structures is chosen so as to
match the ones on nets and on event structures, and formalises a
computational interpretation of configuration structures which we call
the \phrase{asynchronous interpretation}.

We briefly discuss two alternative interpretations of configuration
structures, formalised by alternative transition relations.
The first is the computational interpretation of Chu spaces from {\sc
Gupta \& Pratt} \cite{GP93a,Gup94,Pr94a}.  The second
is a variant of our asynchronous interpretation, based on the
assumption that only finitely many events can happen in a finite time.
This \phrase{finitary asynchronous interpretation} matches the
standard computational interpretation of Petri nets better than does
the asynchronous interpretation, although it falls short in explaining
uncountable configurations of event structures \cite{Wi87a,Wi89}.  We
point out some problems that stand in the way of lifting the
computational interpretation of nets to the infinitary level.

\subsection{The asynchronous interpretation}\label{asynchronous}


\begin{definition}{transitions}
Let $\eC=\tuple{E,C}$ be a configuration structure. For $x,y$ in $C$
write $x \goto{}_\eC y$ if $x \subseteq y$
and $$\forall Z (x \subseteq Z \subseteq y \Rightarrow Z\in C).$$
The relation $\goto{}_\eC$ is called the \phrase{step transition relation}.
\end{definition}
Here $x\goto{}_\eC y$ indicates that the represented system can go from
state $x$ to state $y$ by concurrently performing a number of events
(namely those in $y-x$). The first requirement is unavoidable.
The second one
says that a number of events can be performed concurrently, or
simultaneously, only if they can be performed in any order. This
requirement represents our postulate that different events do not
synchronise in any way; they can happen in one step only if they are
causally independent. Hence our transition relation $\goto{}_\eC$ and
the corresponding computational interpretation of configuration
structures is termed \phrase{asynchronous}.

The \phrase{single-action transition relation} $\goto{}^1_\eC$ on
$C\times C$ is given by $x \goto{}^1_\eC y$ iff $x \subseteq y$ and
$y-x$ is a singleton.  In pictures we omit transitions of the form
$x \goto{}_\eC x$, that exists for every configuration $x$, we indicate
the single-action transition relation by solid arrows, and the rest of
the step transition relation by dashed ones.

\begin{example}{transitions}
These are the transition relations for
$\eD = (\{d,e\},\{ \emptyset, \{d\}, \{e\}, \{d,e\}\})$ and
two structures E and F.
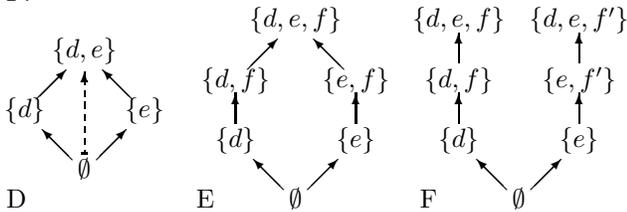
\begin{center}
\begin{picture}(23,25)(-13,-5)
\put(0,0){\makebox[0pt]{$\emptyset$}} \put(-2,3){\vector(-1,1){5}}
\put(-10,10){\makebox[0pt]{$\{d\}$}}  \put(2,3){\vector(1,1){5}}
\put(10,10){\makebox[0pt]{$\{e\}$}}   \put(-8,13){\vector(1,1){5}}
\put(0,20){\makebox[0pt]{$\{d,e\}$}}  \put(8,13){\vector(-1,1){5}}
\put(0,16.6){\vector(0,1){1}}
\put(0,4){\dashbox{1}(0,12){}}
\put(-13,-5){\makebox[0pt][l]{$\eD$}}
\end{picture}
\hfill
\begin{picture}(22,30)(-10,0)
\put(0,0){\makebox[0pt]{$\emptyset$}} \put(-2,3){\vector(-1,1){5}}
\put(-10,10){\makebox[0pt]{$\{d\}$}}  \put(2,3){\vector(1,1){5}}
\put(10,10){\makebox[0pt]{$\{e\}$}}   \put(-8,23){\vector(1,1){5}}
\put(-10,20){\makebox[0pt]{$\{d,f\}$}}  \put(-10,14){\vector(0,1){5}}
\put(10,20){\makebox[0pt]{$\{e,f\}$}}   \put(10,14){\vector(0,1){5}}
\put(0,30){\makebox[0pt]{$\{d,e,f\}$}}  \put(8,23){\vector(-1,1){5}}
\put(-15,0){\makebox[0pt]{$\eE$}}
\end{picture}
\hfill
\begin{picture}(28,30)(-10,0)
\put(0,0){\makebox[0pt]{$\emptyset$}} \put(-2,3){\vector(-1,1){5}}
\put(-10,10){\makebox[0pt]{$\{d\}$}}  \put(2,3){\vector(1,1){5}}
\put(10,10){\makebox[0pt]{$\{e\}$}}   \put(-10,24){\vector(0,1){5}}
\put(-10,20){\makebox[0pt]{$\{d,f\}$}}  \put(-10,14){\vector(0,1){5}}
\put(10,20){\makebox[0pt]{$\{e,f'\}$}}   \put(10,14){\vector(0,1){5}}
\put(-10,30){\makebox[0pt]{$\{d,e,f\}$}} \put(10,24){\vector(0,1){5}}
\put(10,30){\makebox[0pt]{$\{d,e,f'\}$}} 
\put(-15,0){\makebox[0pt]{$\eF$}}
\end{picture}
\end{center}
\end{example}
Such pictures of configuration structures are somewhat misleading
representations, as they suggest a notion of global time, under which
at any time the represented system is in one of its states, moving
from one state to another by following the transitions. Although this
certainly constitutes a valid interpretation, we favour a more truly
concurrent view, in which all events can be performed independently,
unless the absence of certain configurations indicates otherwise.
Under this interpretation, the configurations can be thought of as
\emph{possible} states the system can be in, \emph{from the point of
view of a possible observer}. They are introduced only to indicate (by
their absence) the dependencies between events in the represented system.

In particular, in the structure D above, the events $d$
and $e$ are completely independent, and there is no need to assume
that they are performed either simultaneously or in a particular
order.  The ``diagonal'' in the picture serves merely to remind us of
the independence of these events.  In terms of higher dimensional
automata \cite{Pr91a} it indicates that ``the square is filled in''.

On the other hand, the absence of any ``diagonals'' in E indicates two
distinct linearly ordered computations. In one the event $f$ can only
happen after event $d$, and $e$ in turn has to wait for $f$; the other
has a causal ordering $e<f<d$. There is no way to view $d$ and $e$ as
independent; if there were, there should be a transition
\plat{\emptyset \goto{}_\eC \{d,e\}}.  In labelled versions of
configuration structures, a computationally motivated semantic
equivalence would identify the structures E and F, provided the events
$f$ and $f'$ carry the same label. We do not address such semantic
equivalences in this paper, however.

The configuration structure E is completely axiomatised by the two
clauses $$f \Rightarrow d\vee e$$ \vspace{-2em} $$d\wedge e \Rightarrow f$$
indicating the absence of configurations $\{f\}$ and $\{d,e\}$, respectively.
On the other hand, D has the empty axiomatisation.
An event structure representing D is given by the enabling relation
$\emptyset \turn \emptyset$, $\emptyset \turn \{d\}$,
$\emptyset \turn \{e\}$ and $\emptyset \turn \{d,e\}$, whereas
an enabling relation
for $E$ is $\emptyset \turn \emptyset$, $\emptyset \turn \{d\}$,
$\emptyset \turn \{e\}$, $\{d\} \turn \{f\}$, $\{e\} \turn \{f\}$,
$\emptyset \turn \{d,f\}$, $\emptyset \turn \{e,f\}$,
$\{f\} \turn \{d,e\}$ and $\emptyset \turn \{d,e,f\}$.
Petri net representations of D and E are given below.
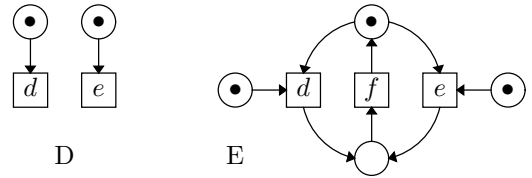
\expandafter\ifx\csname graph\endcsname\relax \csname newbox\endcsname\graph\fi
\expandafter\ifx\csname graphtemp\endcsname\relax \csname newdimen\endcsname\graphtemp\fi
\setbox\graph=\vtop{\vskip 0pt\hbox{%
    \special{pn 8}%
    \special{ar 89 89 89 89 0 6.28319}%
    \graphtemp=.5ex\advance\graphtemp by 0.089in
    \rlap{\kern 0.089in\lower\graphtemp\hbox to 0pt{\hss $\bullet$\hss}}%
    \special{pa 0 536}%
    \special{pa 179 536}%
    \special{pa 179 357}%
    \special{pa 0 357}%
    \special{pa 0 536}%
    \special{fp}%
    \graphtemp=.5ex\advance\graphtemp by 0.446in
    \rlap{\kern 0.089in\lower\graphtemp\hbox to 0pt{\hss $d$\hss}}%
    \special{pa 89 179}%
    \special{pa 89 357}%
    \special{fp}%
    \special{sh 1.000}%
    \special{pa 114 321}%
    \special{pa 89 357}%
    \special{pa 64 321}%
    \special{pa 114 321}%
    \special{fp}%
    \special{ar 446 89 89 89 0 6.28319}%
    \graphtemp=.5ex\advance\graphtemp by 0.089in
    \rlap{\kern 0.446in\lower\graphtemp\hbox to 0pt{\hss $\bullet$\hss}}%
    \special{pa 357 536}%
    \special{pa 536 536}%
    \special{pa 536 357}%
    \special{pa 357 357}%
    \special{pa 357 536}%
    \special{fp}%
    \graphtemp=.5ex\advance\graphtemp by 0.446in
    \rlap{\kern 0.446in\lower\graphtemp\hbox to 0pt{\hss $e$\hss}}%
    \special{pa 446 179}%
    \special{pa 446 357}%
    \special{fp}%
    \special{sh 1.000}%
    \special{pa 471 321}%
    \special{pa 446 357}%
    \special{pa 421 321}%
    \special{pa 471 321}%
    \special{fp}%
    \graphtemp=.5ex\advance\graphtemp by 0.804in
    \rlap{\kern 0.268in\lower\graphtemp\hbox to 0pt{\hss D\hss}}%
    \special{ar 1875 89 89 89 0 6.28319}%
    \graphtemp=.5ex\advance\graphtemp by 0.089in
    \rlap{\kern 1.875in\lower\graphtemp\hbox to 0pt{\hss $\bullet$\hss}}%
    \special{ar 1875 804 89 89 0 6.28319}%
    \special{pa 1429 536}%
    \special{pa 1607 536}%
    \special{pa 1607 357}%
    \special{pa 1429 357}%
    \special{pa 1429 536}%
    \special{fp}%
    \graphtemp=.5ex\advance\graphtemp by 0.446in
    \rlap{\kern 1.518in\lower\graphtemp\hbox to 0pt{\hss $d$\hss}}%
    \special{ar 1866 437 357 357 -2.915184 -1.797205}%
    \special{sh 1.000}%
    \special{pa 1550 328}%
    \special{pa 1518 357}%
    \special{pa 1502 317}%
    \special{pa 1550 328}%
    \special{fp}%
    \special{ar 1866 456 357 357 1.797205 2.915184}%
    \special{sh 1.000}%
    \special{pa 1757 771}%
    \special{pa 1786 804}%
    \special{pa 1745 820}%
    \special{pa 1757 771}%
    \special{fp}%
    \special{ar 1161 446 89 89 0 6.28319}%
    \graphtemp=.5ex\advance\graphtemp by 0.446in
    \rlap{\kern 1.161in\lower\graphtemp\hbox to 0pt{\hss $\bullet$\hss}}%
    \special{pa 1250 446}%
    \special{pa 1429 446}%
    \special{fp}%
    \special{sh 1.000}%
    \special{pa 1393 421}%
    \special{pa 1429 446}%
    \special{pa 1393 471}%
    \special{pa 1393 421}%
    \special{fp}%
    \special{pa 1786 536}%
    \special{pa 1964 536}%
    \special{pa 1964 357}%
    \special{pa 1786 357}%
    \special{pa 1786 536}%
    \special{fp}%
    \graphtemp=.5ex\advance\graphtemp by 0.446in
    \rlap{\kern 1.875in\lower\graphtemp\hbox to 0pt{\hss $f$\hss}}%
    \special{pa 1875 357}%
    \special{pa 1875 179}%
    \special{fp}%
    \special{sh 1.000}%
    \special{pa 1850 214}%
    \special{pa 1875 179}%
    \special{pa 1900 214}%
    \special{pa 1850 214}%
    \special{fp}%
    \special{pa 1875 714}%
    \special{pa 1875 536}%
    \special{fp}%
    \special{sh 1.000}%
    \special{pa 1850 571}%
    \special{pa 1875 536}%
    \special{pa 1900 571}%
    \special{pa 1850 571}%
    \special{fp}%
    \special{pa 2143 536}%
    \special{pa 2321 536}%
    \special{pa 2321 357}%
    \special{pa 2143 357}%
    \special{pa 2143 536}%
    \special{fp}%
    \graphtemp=.5ex\advance\graphtemp by 0.446in
    \rlap{\kern 2.232in\lower\graphtemp\hbox to 0pt{\hss $e$\hss}}%
    \special{ar 1884 437 357 357 -1.344388 -0.226408}%
    \special{sh 1.000}%
    \special{pa 2248 317}%
    \special{pa 2232 357}%
    \special{pa 2200 328}%
    \special{pa 2248 317}%
    \special{fp}%
    \special{ar 1884 456 357 357 0.226408 1.344388}%
    \special{sh 1.000}%
    \special{pa 2005 820}%
    \special{pa 1964 804}%
    \special{pa 1993 771}%
    \special{pa 2005 820}%
    \special{fp}%
    \special{ar 2589 446 89 89 0 6.28319}%
    \graphtemp=.5ex\advance\graphtemp by 0.446in
    \rlap{\kern 2.589in\lower\graphtemp\hbox to 0pt{\hss $\bullet$\hss}}%
    \special{pa 2500 446}%
    \special{pa 2321 446}%
    \special{fp}%
    \special{sh 1.000}%
    \special{pa 2357 471}%
    \special{pa 2321 446}%
    \special{pa 2357 421}%
    \special{pa 2357 471}%
    \special{fp}%
    \graphtemp=.5ex\advance\graphtemp by 0.804in
    \rlap{\kern 1.161in\lower\graphtemp\hbox to 0pt{\hss E\hss}}%
    \hbox{\vrule depth0.893in width0pt height 0pt}%
    \kern 2.679in
  }%
}%

\centerline{\box\graph}

\begin{example}{absence}
Take the system G, represented below as a configuration structure with
a transition relation, a propositional theory, an event structure and
a Petri net.
\begin{figure}[ht]
\begin{picture}(65,40)(-13,-2)
\put(0,0){\makebox[0pt]{$\emptyset$}} \put(-2,3){\vector(-1,1){5}}
\put(-10,10){\makebox[0pt]{$\{d\}$}}  \put(2,3){\vector(1,1){5}}
\put(10,10){\makebox[0pt]{$\{e\}$}}   \put(-8,13){\vector(1,1){5}}
\put(0,20){\makebox[0pt]{$\{d,e\}$}}  \put(8,13){\vector(-1,1){5}}
\put(0,16.6){\vector(0,1){1}}
\put(0,4){\dashbox{1}(0,12){}}
\put(20,20){\makebox[0pt]{$\{e,f\}$}} \put(12,13){\vector(1,1){5}}
\put(10,30){\makebox[0pt]{$\{d,e,f\}$}} \put(2,23){\vector(1,1){5}}
\put(10,26.6){\vector(0,1){1}}
\put(10,14){\dashbox{1}(0,12){}}      \put(18,23){\vector(-1,1){5}}
\put(-13,30){\makebox[0pt][l]{$\eG$}}
\put(34,30){\makebox{$f\implies e$}}
\put(34,7){\makebox{$E=\{d,e,f\}$}}
\put(34,1){\makebox{$\{e\} \turn \{f\}$}}
\put(34,-5){\makebox{$\emptyset \turn X$ for $X\neq\{f\}$}}
\end{picture}~~~~~~~~~~~~~~~~
\input{G}
{\raise 8em\box\graph}
\end{figure}
There is no need to assume, as following the transitions might
suggest, that in any execution of G the event $d$ happens either after
$e$ or before $f$; when actions may have a duration, $d$ may overlap
with both $e$ and $f$. The configuration structure, with its step
transition relation, is not meant to order $d$ with respect to $e$ and
$f$.\linebreak All it does is specify that $f$ comes after $e$, and it
does so by not including configurations $\{f\}$ and $\{d,f\}$. This is
concisely conveyed by the representation of G as a propositional
theory in conjunctive normal form.
\end{example}

\subsection{Petri nets}\label{PN computational}

The firing relation between markings induces a transition relation
between the configurations of a net:
\begin{definition}{transitions-PN}
The \phrase{step transition relation} $\goto{}_\eN$ between the
configurations $x$, $y$ of a net N is given by\vspace{-1ex}
$${x \goto{}_\eN y \Leftrightarrow (x\leq y \wedge M_x \stackrel{y-x}
{\raisebox{0pt}[3pt][0pt]{$-\!\!\!-\!\!\!-\!\!\!\longrightarrow$}}
M_{y})}.$$
\end{definition}
We now show that on pure 1-occurrence nets this step transition
relation matches the one on configuration structures defined above.

\begin{proposition}{transitions-PN}
In a pure net $\eN$ we have
\begin{center} $x \goto{}_\eN y$ iff $x \leq y \wedge
\forall Z (x \leq Z \leq y \Rightarrow Z\in C(\eN))$
\end{center}
for all $x,y$ in $C(\eN)$. (In case N is a pure 1-occurrence net, the
right-hand side can be written as $x \goto{}_{\fC(\eN)}y$.)
\end{proposition}

\begin{proof}``Only if'': Let $x \goto{}_\eN y$ for $x,y \in
C(\eN)$.\\ Then $M_x = I - ^\bullet\!x + x^\bullet \geq 0$ and $y-x$ is
enabled under $M_x$, i.e., $^\bullet (y-x) \leq M_x$. Now let
$x \leq Z \leq y$. Then $^\bullet (Z-x) \leq \mbox{$^\bullet
(y-x)$} \leq M_x$, so\\ $M_Z\,\begin{array}[t]{@{}l@{}}=
 I - ^\bullet\hspace{-2.78995pt}Z + Z^\bullet =  I - \mbox{$^\bullet\!x$} +
 x^\bullet - \mbox{$^\bullet (Z\!-\!x)$} + (Z\!-\!x)^\bullet \\
 =  M_x - \mbox{$^\bullet (Z\!-\!x)$} + (Z\!-\!x)^\bullet \geq 0 +
 (Z\!-\!x)^\bullet  \geq 0,\end{array}$ i.e., $Z$ is a
configuration of $\eN$. Note that for this direction pureness is not needed.

``If'': 
Suppose $x,y\!\in\!C(\eN)$ and $x\!\leq\!y$, but $x \gonotto{}_\eN y$.
Then \mbox{$y-x$} is not enabled under $M_x$, i.e., there is a place
$s\!\in\!S$, such that $^\bullet(y-x)(s) > M_x(s)$. Let $U$ be the multiset of
those transitions $t$ in $y-x$ for which $F(s,t)>0$. Then $^\bullet U(s)
= \mbox{$^\bullet (y-x)$}(s) > M_x(s)$. As N is pure, for all
transitions $t\!\in\!U$ we have $F(t,s)=0$, i.e., $U^\bullet(s)=0$.
Hence $$M_{(x+U)}(s) = M_x(s) - \mbox{$^\bullet U$}(s) + U^\bullet(s) <
0,$$ i.e., $x+U \not\in C(\eN)$. Yet $x \leq (x+U) \leq y$.
\end{proof}
It follows that the step transition relation on a pure net $\eN$ is
\pagebreak[2]
completely determined by the set of configurations of $\eN$, and that
for pure 1-occurrence nets this transition relation exactly matches
the one of \df{transitions}.  This makes $\fC(\eN)$ an acceptable
abstract representation of a pure 1-occurrence net $\eN$.

\out{ Moreover, the transition relation of \df{transitions}
 generalises verbatim to multiset systems (reading $\leq$ for
 $\subseteq$ and $<$ for $\subseteq$), and by \pr{transitions-PN} this
 generalised transition relation exactly matches the one on pure Petri nets.
}

On an impure net $\eN$ the step transition relation is in general not
determined by the set of configurations of $\eN$.  The 1-occurrence
nets P and M below have very different behaviour: in P the transitions
$d$ and $e$ can be done in parallel (there is a transition $\emptyset
\goto{}_\eP \{d,e\}$), whereas in M there is mutual exclusion.
\begin{figure}[htb]
\input{newmutex}
\centerline{\raise 1em\box\graph}
\end{figure}
Yet their configurations are the same: $C(\eP) = C(\eM) =
\{ \emptyset, \{d\}, \{e\}, \{d,e\}\}$.
Therefore it is not a good idea to represent each 1-occurrence net
$\eN=\tuple{S,E,F,I}$ by the configuration structure $\tuple{E,C(\eN)}$.

\subsection{Event structures}\label{ES computational}

\begin{definition}{transitions-ES}
The \phrase{step transition relation} $\goto{}_\eE$ between
configurations $x,y \in L(\eE)$ of an event structure
$\eE=\tuple{E,\turn\;}$ is given by
$${x \goto{}_\eE y \Leftrightarrow (x\subseteq y \wedge
\forall Z \subseteq y.~\exists W \subseteq x.~ W\turn Z)}.$$
\end{definition}
This formalises the intuition provided in Section~\ref{event structures}.
The following proposition says that for pure event structures
this transition relation also exactly matches the one of \df{transitions}.
\begin{proposition}{transitions-ES}
Let $\eE$ be a pure event structure, and
$x,y\mathbin\in L(\eE)$. Then $x \goto{}_\eE y$ iff $x \goto{}_{\fL(\eE)} y$.
\end{proposition}
\begin{proof}
We have to establish that
\begin{center}
$x \goto{}_\eE y$ iff $x\subseteq y \wedge
\forall Z (x \subseteq Z \subseteq y \Rightarrow Z\in L(\eE))$.
\end{center}
``Only if'' follows immediately from the definitions.
For ``if'' let $x\subseteq y$ and $\forall Z (x \subseteq Z \subseteq y
\Rightarrow Z\in L(\eE))$. Let $Z\subseteq y$.
Then $x \subseteq x\cup Z \subseteq y$, so $x\cup Z\in L(\eE)$.
Hence, by \df{EtoC}, $\exists W \subseteq x\cup Z.~W\turn Z$.
As $\eE$ is pure, $W \cap Z = \emptyset$, hence $W \subseteq x$, as
required.
\end{proof}
This makes $\fL(\eE)$ an acceptable abstract representation of a pure
event structure $\eE$.

As for Petri nets, \pr{transitions-ES} does not
generalise to impure event structures, with again the systems $\eP$
and $\eM$ serving as a counterexample. An event structure
representation for $M$ is $\tuple{E,\turn\;}$, with $E=\{d,e\}$ and
$\turn$ given by $\emptyset\turn\emptyset$, $\emptyset\turn\{d\}$,
$\emptyset\turn\{e\}$, $\{d\}\turn\{d,e\}$ and $\{e\}\turn\{d,e\}$.
Another counterexample is the event structure, say H, of \ex{asymmetric
conflict}. The transition relations of P, M and H are
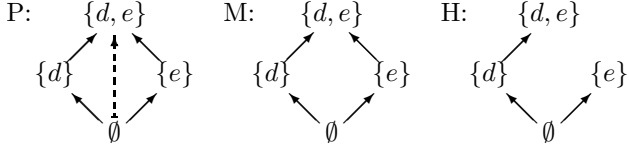
\begin{center}
\begin{picture}(31,22)(-18,0)
\put(0,0){\makebox[0pt]{$\emptyset$}} \put(-2,3){\vector(-1,1){5}}
\put(-10,10){\makebox[0pt]{$\{d\}$}}  \put(2,3){\vector(1,1){5}}
\put(10,10){\makebox[0pt]{$\{e\}$}}   \put(-8,13){\vector(1,1){5}}
\put(0,20){\makebox[0pt]{$\{d,e\}$}}  \put(8,13){\vector(-1,1){5}}
\put(0,16.6){\vector(0,1){1}}
\put(0,4){\dashbox{1}(0,12){}}
\put(-18,20){\makebox[0pt][l]{P:}}
\end{picture}
\hfill
\begin{picture}(31,22)(-18,0)
\put(0,0){\makebox[0pt]{$\emptyset$}} \put(-2,3){\vector(-1,1){5}}
\put(-10,10){\makebox[0pt]{$\{d\}$}}  \put(2,3){\vector(1,1){5}}
\put(10,10){\makebox[0pt]{$\{e\}$}}   \put(-8,13){\vector(1,1){5}}
\put(0,20){\makebox[0pt]{$\{d,e\}$}}  \put(8,13){\vector(-1,1){5}}
\put(-18,20){\makebox[0pt][l]{M:}}
\end{picture}
\hfill
\begin{picture}(31,22)(-18,0)
\put(0,0){\makebox[0pt]{$\emptyset$}} \put(-2,3){\vector(-1,1){5}}
\put(-10,10){\makebox[0pt]{$\{d\}$}}  \put(2,3){\vector(1,1){5}}
\put(10,10){\makebox[0pt]{$\{e\}$}}   \put(-8,13){\vector(1,1){5}}
\put(0,20){\makebox[0pt]{$\{d,e\}$}}  
\put(-18,20){\makebox[0pt][l]{H:}}
\end{picture}
\end{center}

\subsection{The impure case}\label{sec-impure}

In order to provide an adequate abstract representation of impure
1-occurrences nets or impure event structures one could use triples
$\tuple{E,C,\rightarrow\;}$ with $\tuple{E,C}$ a configuration
structure and $\rightarrow\; \subseteq C \times C$ an explicitly
defined transition relation between its configurations. To capture
arbitrary Petri nets one could further allow the configurations to be
multisets of events, rather than sets.

\begin{definition}{multiset-transition system}
A \phrase{multiset transition system} is a tri\-ple
$\tuple{E,C,\rightarrow\;}$ with $E$ a set, $C \subseteq \IN^E$ a
collection of multisets over $E$ and $\rightarrow\; \subseteq C \times C$.

For a configuration structure $\eC=\tuple{E,C}$,
an event structure $\eE=\tuple{E,\turn\;}$ and
a Petri net $\eN=\tuple{S,T,F,I}$,
the \phrase{associated multiset transition system}
is given by $\fC^+(\eC):=\tuple{E,C,\goto{}_\eC\;}$, \hfill
$\fC^+(\eE):=\tuple{E,L(\eE),\goto{}_\eE\;}$ \hfill and
$\fC^+(\eN):=\tuple{T,C(\eN),\goto{}_\eN\;}$, respectively.

Two structures $\eK$ and $\eL$ that may be configuration structures,
event structures and/or Petri nets are \phrase{transition equivalent}
if $\fC^+(\eK)=\fC^+(\eL)$.
\end{definition}
By Propositions~\ref{pr-transitions-PN} and~\ref{pr-transitions-ES},
for pure 1-occurrence nets transition equivalence coincides with
configuration equivalence, and for pure event
structures it coincides with $\fL$-equivalence.

We conjecture that there exist maps between 1-occurrence nets
and event structures that preserve transition equivalence.  However,
the set-up of the present paper, that uses propositional theories up
to logical equivalence as a stepping stone in the translation from
event structures to Petri nets, is insufficient to establish this
beyond the pure case. It is for this reason that we focus on pure nets
and pure event structures.

\subsection{The Gupta-Pratt
interpretation}\label{GP}\index{Gupta-Pratt interpretation}

Our configuration structures are, up to isomorphism, the
\emph{extensional} \phrase{Chu spaces} of {\sc Gupta \& Pratt}
\cite{GP93a,Gup94,Pr94a}. It was in their work that the idea arose of
using the full generality of such structures in modelling
concurrency. It should be noted however that the computational
interpretation they give in \cite{GP93a,Gup94,Pr94a} differs somewhat
from the asynchronous interpretation above;
it can be formalised by means of the step transition relation given by
$x\goto{}_\eC y \Leftrightarrow x \subseteq y$ \cite{GP93a,Pr94a}, thereby
dropping the asynchronicity requirement of \df{transitions}.  This
allows a set of events to occur in one step even if they cannot happen
in any order.

When using the translations between configuration structures and Petri
nets described in Section~\ref{PN}, the Gupta-Pratt interpretation of
configuration structures matches a firing rule on Petri nets
characterised by the possibility of borrowing tokens during the execution
of a multiset of transitions: a multiset $U$ of transitions would be
enabled under a marking $M$ when $M':=M - \mbox{$^\bullet U$} +
U^\bullet \geq 0$. In that case $U$ can fire under $M$, yielding $M'$.
Thus the requirement that $\mbox{$^\bullet U$} \leq M$ is dropped;
tokens that are consumed by the transitions in $U$ may be borrowed
when not available in $M$, as long as they are returned ``to the
bank'' when reproduced by the firing of $U$.

\begin{example}{causality}\index{causality}
The configuration structure $\eC=\tuple{E,C}$ with $E=\{a,b,c\}$ and
$C=\{\emptyset, \{a\}, \{b\}, \{a,b,c\}\}$ models a system in which the
events $a$ and $b$ jointly cause $c$ as their \emph{immediate} effect,
as it is impossible to have done both $a$ and $b$ without doing $c$ also.
Below are the representations of the same system as a propositional
theory, an event structure and a Petri net.
\\
$\begin{array}[t]{c}
\mbox{}\\
c \Leftrightarrow a \wedge b\\
\mbox{}\\[1ex]
\{a,b\} \turn \{c\} \\
\{c\} \turn \{a,b\} \\
\emptyset \turn X \mbox{~for any $X$ with}\\ \{c\} \neq X \neq \{a,b\}
\end{array}$
\hfill\hfill
\expandafter\ifx\csname graph\endcsname\relax \csname newbox\endcsname\graph\fi
\expandafter\ifx\csname graphtemp\endcsname\relax \csname newdimen\endcsname\graphtemp\fi
\setbox\graph=\vtop{\vskip 0pt\hbox{%
    \special{pn 8}%
    \special{ar 89 89 89 89 0 6.28319}%
    \graphtemp=.5ex\advance\graphtemp by 0.089in
    \rlap{\kern 0.089in\lower\graphtemp\hbox to 0pt{\hss $\bullet$\hss}}%
    \special{pa 0 536}%
    \special{pa 179 536}%
    \special{pa 179 357}%
    \special{pa 0 357}%
    \special{pa 0 536}%
    \special{fp}%
    \graphtemp=.5ex\advance\graphtemp by 0.446in
    \rlap{\kern 0.089in\lower\graphtemp\hbox to 0pt{\hss $a$\hss}}%
    \special{pa 89 179}%
    \special{pa 89 357}%
    \special{fp}%
    \special{sh 1.000}%
    \special{pa 114 321}%
    \special{pa 89 357}%
    \special{pa 64 321}%
    \special{pa 114 321}%
    \special{fp}%
    \special{ar 89 804 89 89 0 6.28319}%
    \special{pa 89 536}%
    \special{pa 89 714}%
    \special{fp}%
    \special{sh 1.000}%
    \special{pa 114 679}%
    \special{pa 89 714}%
    \special{pa 64 679}%
    \special{pa 114 679}%
    \special{fp}%
    \special{ar 804 89 89 89 0 6.28319}%
    \graphtemp=.5ex\advance\graphtemp by 0.089in
    \rlap{\kern 0.804in\lower\graphtemp\hbox to 0pt{\hss $\bullet$\hss}}%
    \special{pa 714 536}%
    \special{pa 893 536}%
    \special{pa 893 357}%
    \special{pa 714 357}%
    \special{pa 714 536}%
    \special{fp}%
    \graphtemp=.5ex\advance\graphtemp by 0.446in
    \rlap{\kern 0.804in\lower\graphtemp\hbox to 0pt{\hss $b$\hss}}%
    \special{pa 804 179}%
    \special{pa 804 357}%
    \special{fp}%
    \special{sh 1.000}%
    \special{pa 829 321}%
    \special{pa 804 357}%
    \special{pa 779 321}%
    \special{pa 829 321}%
    \special{fp}%
    \special{ar 804 804 89 89 0 6.28319}%
    \special{pa 804 536}%
    \special{pa 804 714}%
    \special{fp}%
    \special{sh 1.000}%
    \special{pa 829 679}%
    \special{pa 804 714}%
    \special{pa 779 679}%
    \special{pa 829 679}%
    \special{fp}%
    \special{ar 446 446 89 89 0 6.28319}%
    \graphtemp=.5ex\advance\graphtemp by 0.446in
    \rlap{\kern 0.446in\lower\graphtemp\hbox to 0pt{\hss $\bullet$\hss}}%
    \special{pa 357 893}%
    \special{pa 536 893}%
    \special{pa 536 714}%
    \special{pa 357 714}%
    \special{pa 357 893}%
    \special{fp}%
    \graphtemp=.5ex\advance\graphtemp by 0.804in
    \rlap{\kern 0.446in\lower\graphtemp\hbox to 0pt{\hss $c$\hss}}%
    \special{pa 446 714}%
    \special{pa 446 536}%
    \special{fp}%
    \special{sh 1.000}%
    \special{pa 421 571}%
    \special{pa 446 536}%
    \special{pa 471 571}%
    \special{pa 421 571}%
    \special{fp}%
    \special{pa 714 804}%
    \special{pa 536 804}%
    \special{fp}%
    \special{sh 1.000}%
    \special{pa 571 829}%
    \special{pa 536 804}%
    \special{pa 571 779}%
    \special{pa 571 829}%
    \special{fp}%
    \special{pa 179 804}%
    \special{pa 357 804}%
    \special{fp}%
    \special{sh 1.000}%
    \special{pa 321 779}%
    \special{pa 357 804}%
    \special{pa 321 829}%
    \special{pa 321 779}%
    \special{fp}%
    \special{pa 536 446}%
    \special{pa 714 446}%
    \special{fp}%
    \special{sh 1.000}%
    \special{pa 679 421}%
    \special{pa 714 446}%
    \special{pa 679 471}%
    \special{pa 679 421}%
    \special{fp}%
    \special{pa 357 446}%
    \special{pa 179 446}%
    \special{fp}%
    \special{sh 1.000}%
    \special{pa 214 471}%
    \special{pa 179 446}%
    \special{pa 214 421}%
    \special{pa 214 471}%
    \special{fp}%
    \special{ar 446 1161 89 89 0 6.28319}%
    \graphtemp=.5ex\advance\graphtemp by 1.161in
    \rlap{\kern 0.446in\lower\graphtemp\hbox to 0pt{\hss $\bullet$\hss}}%
    \special{pa 446 1071}%
    \special{pa 446 893}%
    \special{fp}%
    \special{sh 1.000}%
    \special{pa 421 929}%
    \special{pa 446 893}%
    \special{pa 471 929}%
    \special{pa 421 929}%
    \special{fp}%
    \hbox{\vrule depth1.250in width0pt height 0pt}%
    \kern 0.893in
  }%
}%

{\box\graph}
\hfill\mbox{}\\[1ex]
It takes the Gupta-Pratt interpretation to obtain the transitions
$\{a\}\goto{}_\eC\{a,b,c\}$ and $\{b\}\goto{}_\eC\{a,b,c\}$, because
under the asynchronous interpretation the configuration $\{a,b,c\}$ is
unreachable.
\end{example}

\subsection{Finite vs.\ infinite steps}\label{unbounded par}

In \cite{GP95} we employed a variant of the transition relation of
\df{transitions}, obtained by additionally requiring, for
$x\goto{}_\eC y$, that $y-x$ be finite.
This transition relation can be motivated computationally by the
assumption that only finitely many events can happen in a finite
amount of time.

In the present paper the asynchronous computational
interpretation of configuration structures given in
Section~\ref{asynchronous} will be our default; we refer to the
interpretation of \cite{GP95} as the \phrase{finitary asynchronous
interpretation}, and denote the associated transition relation by
$\goto{}^f$. 

The step transition relations $\goto{}^f$ and $\goto{}$ on the
configurations of Petri nets coincide; however, this is merely a
spin-off of considering only finite configurations of nets.  It would
be more accurate to recognise the step transition relation of
Section~\ref{PN}, defined between markings, and the inherited step
transition relation between configurations, as finitary ones.

\begin{figure}[htb]
\input{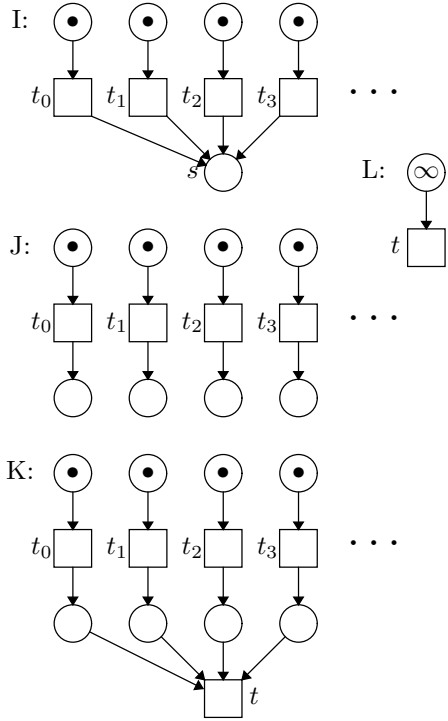}
\centerline{\box\graph}
\caption{Unbounded parallelism\label{unbounded}}\vspace{-3pt}
\end{figure}

It is tempting to generalise the firing rule of \df{firing} to
infinite multisets. The simplest implementation of this idea, however,
yields infinitary markings, as illustrated in Figure~\ref{unbounded}.
After all transitions $t_i$ ($i\mathbin\in \IN$) of net I have fired (in one step)
there are infinitely many tokens in place $s$, contrary to the
definition of a marking. One way to fix this problem is to allow
infinite markings.  This, however, causes the problem illustrated
by the net L: after transition $t$ has fired countably often, are there
tokens left to fire once more? Such problems appear best avoided by
sticking to finitary markings. Another solution is to allow a multiset
of transitions to fire only if by doing so none of its postplaces
receives an infinite amount of tokens. This would enable
any finite multiset over $\{t_i \mid i \in \IN\}$ to fire initially,
but no infinite one. A disadvantage of this solution is that the nets
I and J, which normally would be regarded equivalent, have now a
different behaviour, as in J all transitions $t_i$ can still fire in
one step. As a consequence, the theorem of \cite{GP95} that any net is
step bisimulation equivalent to a safe net, or a prime event
structure, would no longer hold; I constitutes a counterexample.

Therefore we stick in this paper to the convention, formalised by
\df{firing}, that only finitely many transitions can fire in a finite
time. As a consequence, the transition $t$ in net K can never
fire and this net is semantically equivalent to I and J\@.

\section{Four notions of equivalence}\label{equivalence}

In this paper we compare configuration structures, pure event
structures and pure 1-occurrence nets up to four notions of
equivalence, the finest one of which is \phrase{configuration
equivalence}.  Two such structures are configuration equivalent iff
they have the same events and the same configurations (taking
the left-closed configurations of pure event structures).  By
Propositions~\ref{pr-transitions-PN} and~\ref{pr-transitions-ES} this implies
that they also have the same step transition relation between their
configurations. On pure 1-occurrence nets configuration equivalence is
defined in \df{NtoC}, on pure event structures it is defined
in \df{EtoC} under the name \emph{$\fL$-equivalence} and for
configuration structures it is the identity relation. However, we can
also compare nets with event structures, or any other combinations of
models, up to configuration equivalence.  The other three equivalence
relations are obtained by restricting attention to the configurations
that are finite, reachable, or both, as we now see.

\begin{definition}{reachable}
Let $\eC=\tuple{E,C}$ be a configuration structure. A configuration
$x\!\in\! C$ is \emph{reachable}\index{reachable configurations}
if there is a sequence of configurations\vspace{-1ex}
$$\emptyset = x_0 \goto{}_\eC x_1 \goto{}_\eC \ldots \goto{}_\eC x_n =x.$$
Let $R(\eC)$ denote the set of reachable configurations and $F(\eC)$
the set of finite configurations of $\eC$.  The \phrase{reachable
part} of $\eC$ is given by $\fR(\eC):=\tuple{E,R(\eC)}$ and the
\phrase{finite part} by $\fF(\eC):=\tuple{E,F(\eC)}$.

A configuration structure $\eC$ is \phrase{connected} if all its
configurations are reachable, i.e., if $\fR(\eC)=\eC$. It
is \phrase{finitary} if its configurations are finite, i.e., if
$\fF(\eC)=\eC$.

Two configuration structures $\eC$ and $\eD$ are
\begin{itemise3}
\item[$\bullet$] \emph{configuration equivalent}\index{configuration equivalence}
 if $\eC=\eD$;
\item[$\bullet$] \emph{finitarily equivalent}\index{finitary equivalence}
 if $\fF(\eC)=\fF(\eD)$---cf.\ Def.~\ref{df-finitary equivalence};
\item[$\bullet$] \emph{reachably equivalent}\index{reachable equivalence}
if $\fR(\eC)=\fR(\eD)$; and
\item[$\bullet$] \emph{finitarily reachably equivalent}\index{finitary reachable equivalence}
if $\fR(\hspace{-1pt}\fF(\eC)\hspace{-1pt})\mathbin=\fR(\hspace{-1pt}\fF(\eD\!)\hspace{-1pt})$.
\end{itemise3}
For $\eE$ a pure event structure let $\fC(\eE)$ be $\fL(\eE)$, and
for $\eC$ a configuration structure let $\fC(\eC)$ be $\eC$.
Two structures K and L that may be configuration structures, pure event
structures or pure 1-occurrence nets are called \emph{configuration
equivalent} if the configuration structures $\fC(K)$ and $\fC(L)$ are
configuration equivalent. The other three equivalences lift to general
pure structures in the same way.
\end{definition}


\begin{proposition}{reachable transitions}
Let $\eC=\tuple{E,C}$ be a configuration structure, $x\mathbin\in
R(\eC)$ and $Y\mathbin\subseteq E$.
Then $x \goto{}_{\fR(\eC)} Y$ iff $x \goto{}_\eC Y$.
\end{proposition}
\begin{proof}
Let $x \mathbin\in R(\eC)$ and $Y\mathbin\subseteq E$. Then,

$x \goto{}_\eC Y$ iff (by \df{transitions})

$x \subseteq Y \wedge \forall Z (x \subseteq Z
\subseteq Y \Rightarrow Z\in C)$ iff

$x \subseteq Y \wedge \forall Z.~\forall W (x \subseteq W \subseteq Z
\subseteq Y \Rightarrow W,Z \in C)$ iff

$x \subseteq Y \wedge \forall Z (x \subseteq Z
\subseteq Y \Rightarrow (Z\in C \wedge x \goto{}_\eC Z))$ iff

$x \subseteq Y \wedge \forall Z (x \subseteq Z
\subseteq Y \Rightarrow Z\in R(\eC))$.
\end{proof}
In particular (taking $Y\mathbin\in R(\eC)$ above), the step
transition relation on $\fR(\eC)$ is exactly the step transition
relation on $\eC$ restricted to the reachable configurations of $\eC$.
Likewise, the step transition relation on $\fF(\eC)$ is exactly the
step transition relation on $\eC$ restricted to the finite
configurations of $\eC$.

\begin{proposition}{idempotence}
For any configuration structure $\eC$ one has
$\fR(\fR(\eC))=\fR(\eC)$,
$\fF(\fF(\eC))=\fF(\eC)$ and
$\fF(\fR(\eC))=\fR(\fF(\eC))$.
\end{proposition}

\begin{proof}
Straightforward, for the first statement using \pr{reachable transitions}.
\end{proof}
Thus, any reachable equivalence class of configuration structures
contains exactly one connected configuration structure, which can be
obtained as the reachable part of any member of the class.
Likewise, any finitary equivalence class of configuration structures
contains exactly one finitary configuration structure, which can be
obtained as the finite part of any member of the class.
Finally, any finitary reachable equivalence class of configuration
structures contains exactly one configuration structure that is both
finitary and connected; it can be obtained as the reachable part of
the finite part of any member of the class.

By definition, $\fC(\eN)$ is finitary for any pure 1-occurrence net $\eN$.
Hence on pure 1-occurrence nets, finitary equivalence coincides with
configuration equivalence, and finitary reachable equivalence with
reachable equivalence.

If in \df{reachable} we would have defined reachability in terms of
the step transition relation formalising the Gupta-Pratt
interpretation of configuration structures (cf.\ Section~\ref{GP}),
either all or no configurations of a configuration structure $\eC$
would be reachable, depending on whether or not $\eC$ is rooted.
If we would use the step transition relation $\goto{}^f$ formalising
the finitary asynchronous interpretation (cf.\ Section~\ref{unbounded
par}), what we would get as $\fR(\eC)$ is actually $\fR(\fF(\eC))$.

\out{
 The equivalence relations introduced so far on configuration
 structures, event structures and Petri nets are related as follows:
 $$\begin{array}[b]{ccc}
 \simeq	&\Rightarrow	&\simeq_{\fR}   \\
 \Downarrow&		&\Downarrow	\\
 \simeq_f&\Rightarrow	&\simeq_{\fR_f}	\\
 \end{array}.$$
 Here $\simeq$ denotes transition equivalence on nets and event
 structures, and the identity on configuration structures.
}

When dealing with systems that merely take a finite number of
transitions in a finite amount of time, only their reachable parts are
semantically relevant. In this setting it makes sense to study
configuration structures, pure event structures and pure 1-occurrence
nets up to reachable equivalence. When moreover assuming that only
finitely many actions can happen in a finite amount of time, it even
suffices to work up to finitary reachable equivalence.

 Clearly, reachable equivalence is coarser than configuration
 equivalence.  The following example illustrates for event structures
 that this is strictly so.

 \begin{example}{ex-continuous} Let $\eE\!=\!\tuple{E,\turn\;}$ be the
 event structure with as events the set $\IQ$ of rational numbers,
 $\emptyset \turn X$ for any $X$ with $|X|\neq 1$, and $X \turn \{e\}$
 iff \mbox{$X=\{d \in \IQ \mid d<e\}$}.  Then $L(\eE)$ consists of all
 downwards closed subsets of rational numbers and thus contains
 representatives of all reals as well as extra copies of the rationals
 and $\IQ$ itself (infinity); however $R(\fL(\eE))= \{\emptyset\}$. So if
 $\eF$ is $\tuple{\IQ,\{\emptyset\turn\!\emptyset\}}$ then $\eE$ and
 $\eF$ are reachably equivalent, yet $L(\eE) \neq L(\eF)$.
 \end{example}
This example also illustrates that the justification of working up to
reachable equivalence depends on the precise computational
interpretation of event structures. 

We have established the
bijective correspondences between configuration structures,
propositional theories, pure event structures and pure 1-occurrence
nets up to the finest semantic equivalence possible.  This way our
correspondences are compatible, for instance, with the Gupta-Pratt
interpretation of configuration structures.  Under this
interpretation, unreachable configurations may be semantically
relevant, as witnessed by the notions of \phrase{causality} and
\phrase{internal choice} in \cite{GP93a,Pr94a} (see \ex{causality})
and that of \phrase{history preserving bisimulation} in \cite{Gup94}.

\subsection{Hyperreachability}\label{hyperreachability}

Below we consider a class of \phrase{$\fS\fR$-secure} configuration
structures, on which we define the \phrase{hyperconnected} configuration
structures as alternative canonical representatives of reachable
equivalence classes, and we propose a function $\fS$ that transforms
each $\fS\fR$-secure configuration structure into an alternative normal
form: the unique hyperconnected configuration structure inhabiting a
reachable equivalence class of $\fS\fR$-secure configuration structures.
We also show that the function $\fS \circ \fF$ transforms each
configuration function into an alternative canonical representation
of its finitary reachable equivalence class. As we will show in
Section~\ref{brands}, it is the function $\fS$ that generalises the
notion of configuration employed in {\sc Winskel} \cite{Wi87a,Wi89}.

\begin{definition}{hyperreachable}
Let $\eC=\tuple{E,C}$ be a configuration structure.
A set of events $X\subseteq E$ is \phrase{hyperreachable}, or a
\emph{secured configuration}\index{secured configurations} of $\eC$,
if $X=\bigcup_{i=0}^\infty x_i$ for an infinite sequence of configurations
$$\emptyset = x_0 \goto{}_\eC x_1 \goto{}_\eC x_2 \goto{}_\eC \ldots.$$
Let $S(\eC)$ be the set of secured configurations of $\eC$, and write
$\fS(\eC):=\tuple{E,S(\eC)}$.
The structure $\eC$ is \phrase{hyperconnected} if $\fS(\eC)=\eC$.
\end{definition}
The secured configurations include the reachable ones (just take
$x_i=x_n$ for $i>n$). Whereas reachable configurations could be
regarded as modelling possible partial runs of the represented
system,\footnote{The idea of a configuration modelling a possible
partial run is consistent with the idea that it also models a possible
state of the represented system, namely the state obtained after
executing all the events that make up the run.} happening in a finite
amount of time, secured configurations additionally model possible
total runs, happening in an unbounded amount of time.

\begin{proposition}{secured reachable}
Let $\eC$ be a configuration structure.
Then $\fS(\eC)=\fS(\fR(\eC))$ and $R(\eC) \subseteq R(\fS(\eC))$.
\end{proposition}
\begin{proof}
The first statement follows immediately from \pr{reachable transitions};
the second holds because $\mathord{\goto{}_{\fR(\eC)}} \subseteq
\mathord{\goto{}_{\fS(\eC)}}$.
\end{proof}
However, it is not always true that $\fR(\fS(\eC))=\fR(\eC)$.%
\begin{example}{insecure}
Take $E\mathbin{:=}\IN$ and $C\mathbin{:=}\pow_{\it fin}(\IN)$
consisting of all finite
subsets of $\IN$. Then $S(\eC)=\pow(\IN)$ and in the configuration
structure $\fS(\eC)$ one has $\emptyset \goto{}_{\fS(\eC)} X$ for
every $X\subseteq \IN$.
Thus $R(\fS(\eC))=\pow(\IN)$, whereas $R(\eC)=\pow_{\it fin}(\IN)$.
\end{example}
It is also not always the case that $\fS(\fS(\eC))=\fS(\eC)$;
finding a counterexample is left as a puzzle for the reader.
\out{
 do first a number of a's in order (say a1 a2 ... an).
 if n is odd, do all b's in one step followed by a finite number of c's.
 if n is even, do all c's in one step followed by a finite number of b's.
 After applying S we find confs with finitely many a's, and all b's and c's.
 After applying SS we even find a conf with all a's.
}
The problem underlying \ex{insecure} is that the induced step
transition relation on $\fS(\eC)$ may differ from the one on $\eC$,
even when restricting attention to transitions originating from
reachable configurations of $\eC$ (and hence of $\fS(\eC)$).
Thus the map $\fS$ may alter the computational interpretation of
configuration structures as proposed in Section~\ref{computational}.
We now characterise the class of configuration structures for which this
does not happen.

\begin{definition}{secure}
A configuration structure $\eC$ is \phrase{$\fS\fR$-secure} iff $\fR(\fS(\eC))=\fR(\eC)$.
\end{definition}

\begin{observation}{secure}
$\hspace{-1pt}$A configuration structure
$\eC\mathbin=\tuple{E,\hspace{-1pt}C}$ is $\fS\fR$-secure
iff for all $x \mathbin\in R(\eC)$ and all $Y \subseteq E$ one has
$x \goto{}_{\fS(\eC)} Y$ iff $x \goto{}_\eC Y$.
\end{observation}
A configuration structure $\eC\mathbin=\tuple{E,C}$ with $S(\eC)\mathbin\subseteq C$,
i.e., for which all its secured configurations are in fact configurations,
is certainly $\fS\fR$-secure.  \pr{secured reachable} yields:

\begin{proposition}{reachable secured}
Let $\eC$ be an $\fS\fR$-secure configuration structure.
Then $\fS(\fS(\eC))\!=\!\fS(\eC)$, i.e., $\fS(\eC)$ is hyperconnected.
\hfill $\Box$
\end{proposition}
If a configuration structure $\eC$ is $\fS\fR$-secure, then so are
$\fR(\eC)$ and $\fS(\eC)$. However, it is not the case that $\fF(\eC)$
is always $\fS\fR$-secure when $\eC$ is; a counterexample is the $\fS\fR$-secure
configuration structure $\eC:=\tuple{\IN,\pow(\IN)}$: here $\fF(\eC)$
is the $\fS\fR$-insecure configuration structure of \ex{insecure}.

\begin{proposition}{secured equivalence}
Let $\eC$ and $\eD$ be $\fS\fR$-secure configuration structures.
Then $\fR(\eC) \mathbin= \fR(\eD)$ iff $\fS(\eC) \mathbin= \fS(\eD)$.
\end{proposition}
\begin{proof}
It follows from \pr{secured reachable} that $\fS(\eC)$ is completely
determined by $\fR(\eC)$, whereas by \df{secure} $\fR(\eC)$ is completely
determined by $\fS(\eC)$.
\end{proof}
\pr{secured equivalence} says that two $\fS\fR$-secure configuration structures are
reachable equivalent iff they have the same secured configurations.
Thus, in any reachable equivalence class of $\fS\fR$-secure configuration
structures there are two normal forms: a connected configuration
structure that can be obtained as $\fR(\eC)$ for $\eC$ an arbitrary
member of the class, and a hyperconnected configuration structure that
can be obtained as $\fS(\eC)$ for $\eC$ an arbitrary member of the
class. In the sequel we will often use the normal form $\fS$ when
dealing with event structures, as our notion of a secured
configuration of an event structure is the one that generalises the
notion of configuration of \cite{Wi87a,Wi89}.

\ex{insecure} shows that for $\eC$ an $\fS\fR$-insecure configuration
structure, $\fS(\eC)$ need not be reachable equivalent with $\eC$.
Therefore, when working up to reachable equivalence, we will not study
the configuration structures $\fS(\eC)$ for $\fS\fR$-insecure $\eC$.
However, this restriction is not needed when working up to finitary
reachable equivalence, as we will show below.

\begin{proposition}{finitary secured}
Let $\eC$ be a configuration structure.
Then $\fF(\fS(\eC)) = \fF(\fR(\eC))$.
\end{proposition}

\begin{proof}
That any finite secured configuration is reachable follows directly
from \df{hyperreachable}, whereas ``$\supseteq$'' follows from the
earlier observation that $\fS(\eC) \supseteq \fR(\eC)$.
\end{proof}
\out{
 By Propositions~\ref{pr-idempotence},~\ref{pr-secured
 reachable},~\ref{pr-reachable secured} and~\ref{pr-finitary secured},
 the only different configuration structures that can be obtained out
 of a given secure configuration structure $\eC$ by applying the
 functions $\fR$, $\fS$ and $\fF$ are $\fR(\eC)$, $\fS(\eC)$,
 $\fF(\eC)$, $\fR(\fF(\eC))$ and $\fS(\fF(\eC))$.
}

\begin{proposition}{secured finitarily reachably equivalent}
Let $\eC$ be a configuration structure.
Then $\fR(\fF(\fS(\eC))) = \fR(\fF(\eC))$, i.e., $\fS(\eC)$ is finitarily
reachably equivalent with $\eC$.
\end{proposition}

\begin{proof}
$\fR(\fF(\fS(\eC))) \stackrel{\mbox{\footnotesize \ref{pr-finitary secured}}}{=}
\fR(\fF(\fR(\eC))) \stackrel{\mbox{\footnotesize \ref{pr-idempotence}}}{=}
\fR(\fF(\eC))$.
\end{proof}

\begin{proposition}{secured finitary equivalence}
For configuration structures $\eC,\eD$:
$$\fR(\fF(\eC)) = \fR(\fF(\eD)) \mbox{~~iff~~} \fS(\fF(\eC)) = \fS(\fF(\eD)).$$
\end{proposition}
\begin{proof}
It follows from \pr{secured reachable} that $\fS(\fF(\eC))$ is
completely determined by $\fR(\fF(\eC))$, whereas
Propositions~\ref{pr-idempotence} and \ref{pr-finitary
secured} (or~\ref{pr-secured finitarily reachably equivalent}) imply
that $\fR(\fF(\eC))$ is completely determined by $\fS(\fF(\eC))$.
\end{proof}
Thus, in any finitary reachable equivalence class of configuration
structures there are two normal forms: a finitary and connected
configuration structure that can be obtained as $\fR(\fF(\eC))$ for
$\eC$ an arbitrary member of the class, and a configuration structure
that can be obtained as $\fS(\fF(\eC))$ for $\eC$ an arbitrary member
of the class.

\subsection{Petri nets}

In this section we directly define the reachable configurations of a
Petri net, and observe that for pure 1-occurrence nets this definition
agrees with \df{reachable}. Moreover, we infer that finitary, connected
rooted configurations structures are canonical representatives of
equivalence classes of nets that have the same reachable configurations.

\begin{definition}{reachable PN}
The set $R(\eN)$ of \phrase{reachable configurations} of a Petri net
$\eN = \tuple{S,T,F,I}$ consists of the multisets $\Sigma_{i=1}^n U_n$
such that there is a firing sequence
$${I \goto{U_1} M_1 \goto{U_2} \cdots \goto{U_n} M_n}.$$
In case $\eN = \tuple{S,E,F,I}$ is a pure 1-occurrence net, we write
$\fR(\eN) := \tuple{E,R(\eN)}$.
\end{definition}

\begin{proposition}{reachable pure PN}
If $\eN$ is a pure 1-occurrence net, then $\fR(\eN)=\fR(\fC(\eN))$.
\end{proposition}
\begin{proof}
Immediate from Definitions~\ref{df-reachable}
and~\ref{df-transitions-PN}, using \pr{transitions-PN}.
\end{proof}
The configuration structure $\fR(\eN)$ is always rooted, finitary and
connected. Moreover, combining \pr{reachable pure PN} with \cor{CtoN} yields:
\begin{proposition}{CtoN-reachable}
For every rooted, finitary and connected configuration structure $\eC$
there exists a pure 1-occurrence net $\eN$ without arcweights, such that
$\fR(\eN)=\eC$. \hfill $\Box$
\end{proposition}
Thus, we have established a bijective correspondence between pure
1-occurrence nets (with or without arcweights) up to reachable
equivalence and finitary, connected, rooted configuration structures.

\subsection{Event structures}

In this section we define the four notions of configuration $\fS$,
$\fR$, $\fR\circ\fF$ and $\fS\circ\fF$ directly on event structures,
and observe that for pure event structures these definitions agree
with Definitions~\ref{df-reachable} and~\ref{df-hyperreachable}.  In
Section~\ref{brands} we will show that our secured configurations
generalise the configurations of the event structures that appear in
{\sc Winskel} \cite{Wi87a,Wi89}.  As the family of all configurations
of an event structure from \cite{Wi87a,Wi89} is completely determined
by the subfamily of its finite configurations, in \cite{GG90}
attention has been restricted to finite configurations only.  A
generalisation of these finite configurations to the event structures
of this paper are our finite reachable configurations below.

\begin{definition}{reachable ES}
The set $S(\eE)$ of \phrase{secured configurations} of an event
structure $\eE=\tuple{E,C}$ consists of the sets of events
$\bigcup_{i=0}^\infty X_i$ with $X_0=\emptyset$ such that
$$\forall i\in\IN.~ X_i \subseteq X_{i+1} \wedge
\forall Y \subseteq X_{i+1}.~ \exists Z \subseteq X_{i}.~ Z \turn Y.$$
The set $R(\eE)$ of \phrase{reachable configurations} of $\eE$
consists of the sets of events
$\bigcup_{i=0}^n X_i$ with $X_0=\emptyset$ such that
$$\forall i<n.~ X_i \subseteq X_{i+1} \wedge
\forall Y \subseteq X_{i+1}.~ \exists Z \subseteq X_{i}.~ Z \turn Y.$$
The set $R_f(\eE)$ of finite reachable configurations of $\eE$
consists of the sets of events $\{e_1,\ldots,e_n\}$ such that
$$\forall i\leq n.~ \forall Y \subseteq \{e_1,...,e_{i}\}.~
\exists Z \subseteq \{e_1,...,e_{i-1}\}.~ Z \turn Y.$$
Finally, the set $S_f(\eE)$ extends $R_f(\eE)$ with
the infinite sets of events $\{e_1,e_2,\ldots\}$ such that
$$\forall i\in\IN.~ \forall Y \subseteq \{e_1,...,e_{i}\}.~
\exists Z \subseteq \{e_1,...,e_{i-1}\}.~ Z \turn Y.$$
The \phrase{secured configuration structure associated to} E is
$\fS(\eE) := \tuple{E,S(\eE)}$.
Likewise, let $\fR(\eE):=\tuple{E,R(\eE)}$,
$\fR_f(\eE):=\tuple{E,R_f(\eE)}$ and $\fS_f(\eE):=\tuple{E,S_f(\eE)}$.\\
An event structure $\eE$ is \phrase{$\fS\fR$-secure} iff
$\fR(\fS(\eE))\mathbin=\fR(\eE)$.
\end{definition}
Thus $X\in S(\eE)$ iff $X=\bigcup_{i=0}^\infty X_i$ for a sequence
$$\emptyset=X_0\goto{}_\eE X_1\goto{}_\eE X_2\goto{}_\eE\ldots$$
and likewise for $R(\eE)$.
Again, the secured configurations include the reachable ones (just
take $X_i:=X_n$ for $i>n$).  We call a sequence $X_0,X_1,\dots$ as
occurs in the definitions of $S(\eE)$ and $R(\eE)$ a \phrase{stepwise
securing} of $X$; a sequence $e_1,e_2,\dots$ as occurs in the
definitions of $R_f(\eE)$ and $S_f(\eE)$ is an \phrase{eventwise
securing} of $X$.  Computationally, a stepwise securing can be
understood to model a particular run of the represented system by
partitioning time in countably many successive intervals $I_k$ ($k
\geq 1$). The set $X_k-X_{k-1}$ contains the events that occur in the
interval $I_k$.  These events must be enabled by events occurring in
earlier intervals.  The set $X$ contains all events that happen during
such a run.  An eventwise securing can be understood by imposing the
restriction that $|X_k-X_{k-1}| = 1$, i.e., in each interval exactly
one event takes place.  We now show that $R_f(\eE)$ consists of the
finite configurations in $R(\eE)$.

\begin{proposition}{finite reachable ES}
Let E be an event structure. Then $\fR_f(\eE)=\fF(\fR(\eE))=\fF(\fS(\eE))$.
\end{proposition}
\begin{proof}
Given $X\in R_f(\eE)$, let $e_1,\ldots,e_n$ be an eventwise securing
of $X$. Take $X_i := \{e_1,\ldots,e_i\}$ for $i=0,1,\ldots,n$.  Then
$X_0,\ldots,X_n$ is a stepwise securing of $X$. As $X$ is finite we
have $X \in F(\fR(\eE))$.

Given $X \mathbin\in F(\fR(\eE))$, let $X_0,\ldots,X_n$ be a stepwise
securing of $X$. Removing duplicate entries (where $X_{i-1} = X_{i}$
with $1 \leq i \leq n$) from this sequence preserves the property of the
sequence being a stepwise securing. Furthermore, if $X_{i-1} \subset Y
\subset X_{i}$ for some $1 \leq i \leq n$, then adding $Y$ between
$X_{i-1}$ and $X_i$ also preserves the property of the sequence being
a stepwise securing. In this way (using that all $X_i$ are finite) the
stepwise securing $X_0,...,X_n$ can be modified into a stepwise
securing $Y_0,...,Y_m$ with $|Y_i-Y_{i-1}|=1$ for $i=1,...,m$.
The latter can be written as an eventwise securing.

That $\fF(\fR(\eE))=\fF(\fS(\eE))$ is trivial (cf.~Pr.~\ref{pr-finitary secured}).
\end{proof}
The proof above shows that events cannot be ``synchronised'' in event
structures. If a finite number of events takes place simultaneously,
they could just as well have occurred one after the other, in any order.

\begin{proposition}{pure ES commute}
Let E be a pure event structure. Then
$\fR(\eE)=\fR(\fL(\eE))$ and $\fS(\eE)=\fS(\fL(\eE))$.
Moreover, $\fR_f(\eE)=\fR(\fF(\fL(\eE)))$ and
$\fS_f(\eE)=\fS(\fF(\fL(\eE)))$.
\end{proposition}

\begin{proof}
The first two statements follow directly from
Definitions~\ref{df-reachable}, respectively~\ref{df-hyperreachable},
and~\ref{df-transitions-ES}, using \pr{transitions-ES}. The third
statement now follows from \pr{finite reachable ES}, the first
statement, and \pr{idempotence}.

For the last statement, let $X\in S_f(\eE)$. In case $X$ is finite, $X
\in R_f(\eE) = R(\fF(\fL(\eE))) \subseteq S(\fF(\fL((\eE)))$. Otherwise,
let $e_1,e_2,\ldots$ be an eventwise securing of $X$. Let $X_i \mathbin{:=}
\{e_1,\ldots,e_i\}$ for $i\mathbin\geq 0$.  Then $X_i \in F(\fL(\eE))$
for $i\in\IN$ and $\emptyset =X_0 \goto{}_{\fL(\eE)} X_1
\goto{}_{\fL(\eE)} X_2 \goto{}_{\fL(\eE)} \cdots$, so
$X = \bigcup_{i=0}^\infty X_i \in S(\fF(\fL(\eE)))$ by \df{hyperreachable}.

Conversely, let $X\in S(\fF(\fL(\eE)))$. Then, by \df{hyperreachable},
$X=\bigcup_{i=0}^\infty x_i$ for $x_i \mathbin\in F(\fL(\eE))$
($i\mathbin\in\IN$) such that $\emptyset = x_0 \goto{}_{\fL(\eE)} x_1
\goto{}_{\fL(\eE)} x_2 \goto{}_{\fL(\eE)} \ldots$.
As in the proof of \pr{finite reachable ES}, this sequence can be
modified into a finite or infinite sequence $y_i \mathbin\in
F(\fL(\eE))$ with $\emptyset = y_0
\goto{}_{\fL(\eE)} y_1 \goto{}_{\fL(\eE)} y_2 \goto{}_{\fL(\eE)} \ldots$
and $|y_i - y_{i-1}|=1$ for relevant all $i>0$. By \pr{transitions-ES} we have
$\emptyset = y_0 \goto{}_\eE y_1 \goto{}_\eE y_2 \goto{}_\eE \ldots$.
Writing $e_i$ for the unique element of $y_i-y_{i-1}$, for $i>0$,
\df{transitions-ES} yields that $e_1,e_2,\ldots$ is an eventwise
securing of $X$. Hence $X=\{e_1,e_2,\ldots\} \in S_f(\eE)$.
\end{proof}

\begin{corollary}{secure}
A pure event structure $\eE$ is $\fS\fR$-secure iff $\fL(E)$
is an $\fS\fR$-secure configuration structure.
\end{corollary}

\begin{proof}
Let $\eE$ be pure.  If $\eE$ is $\fS\fR$-secure then
\vspace{-1.5ex}
$$\fR(\fS(\fL(\eE)))=\fR(\fS(\eE))=\fS(\eE)=\fS(\fL((\eE)).
\vspace{-1.5ex}$$
Conversely, if $\fL(\eE)$ is $\fS\fR$-secure then\\[1ex]
\mbox{~~~~}\hfill$\fR(\fS(\eE))=\fR(\fS(\fL(\eE)))=\fS(\fL((\eE))=\fS(\eE)$.
\end{proof}

\begin{corollary}{hyperconnected}
Let $\eE$ be a pure and $\fS\fR$-secure event structure.
Then $\fS(\eE)$ is hyperconnected.
Conversely, if $\eC$ is a hyperconnected 
configuration structure, then $\fE(\eC)$ is a pure and
$\fS\fR$-secure event structure.\hfill$\Box$
\end{corollary}
Using \thm{CtoEtoC}, \pr{pure ES commute} yields

\begin{proposition}{CtoE-reachable}
Let $\eC$ be a connected configuration structure. Then
$\fR(\fE(\eC))=\eC$.
\end{proposition}
\begin{proof}
$\fR(\fE(\eC))=\fR(\fL(\fE(\eC)))=\fR(\eC)=\eC$.
\end{proof}
Likewise, if $\eC$ is a hyperconnected configuration structure
then $\fS(\fE(\eC))=\eC$; if $\eC$ is a finitary connected
configuration structure then $\fR_f(\fE(\eC))=\eC$; and if $\eC$ is a
configuration structure of the form $\fS(\eD)$ with $\eD$ a finitary
configuration structure then $\fS_f(\fE(\eC))=\eC$.

Thus $\fR$ and $\fE$ provide a bijective correspondence between pure
event structures up to reachable equivalence and connected
configuration structures (using \pr{pure ES commute}, \df{reachable},
\thm{CtoEtoC} and the above).  Likewise, $\fS$ and $\fE$ provide a
bijective correspondence between pure and $\fS\fR$-secure event
structures up to reachable equivalence and hyperconnected configuration
structures (additionally using Corollaries~\ref{cor-hyperconnected}
and \ref{cor-secure} and \pr{secured equivalence});
$\fR_f$ and $\fE$ provide a bijective correspondence between pure
event structures up to finitary reachable equivalence and finitary
connected configuration structures; and $\fS_f$ and $\fE$ provide a
bijective correspondence between pure event structures up to finitary
reachable equivalence and configuration structures of the form
$\fS(\eD)$ with $\eD$ finitary (additionally using \pr{secured
finitary equivalence}).

\subsubsection*{Impure event structures}

\pr{pure ES commute} does not extend to impure event structures. For those,
their reachable configurations are not determined by their left-closed ones.

\begin{example}{impure}
Let $\eE:=\tuple{\{e\},\{\emptyset\turn\!\emptyset,\,\{e\}\turn\!\{e\}\}}$.
Then $\fL(\eE)=\tuple{\{e\},\{\emptyset,\{e\}\}}$,
whereas $\fR(\eE)=\tuple{\{e\},\{\emptyset\}}$.
Both configuration structures are connected.

Let $\eF:=\tuple{\{e\},\{\emptyset\turn\!\emptyset,\,\emptyset\turn\!\{e\}\}}$.
Then we have $\fL(\eE)=\fL(\eF)$ but $\fR(\eE) \neq \fR(\eF)$.
\end{example}
When the step transition relation of \df{transitions-ES} is taken to
be part of the meaning of an event structure, neither the left-closed
nor the reachable configurations capture the meaning of impure event
structures faithfully, as illustrated by the systems P and M mentioned
in Section~\ref{ES computational}. When, on the other hand, the
behaviour of an event structure is deemed to be determined by its
configurations, then on impure event structures $\fL$ and $\fR$
represent mutually inconsistent interpretations.
However, under either interpretation the impure event structures are
redundant: for every event structure there exists a pure one with the
same configurations.  Obviously, which one depends on whether the
left-closed or the reachable configurations are to be preserved.

\begin{proposition}{purification}
For any event structure E there is
a pure event structure $\eE_\fL$ with $\fL(\eE_\fL)=\fL(\eE)$,
and a pure event structure $\eE_\fR$ with $\fR(\eE_R)=\fR(\eE)$.
\end{proposition}

\begin{proof}
One can take $\eE_\fL$ to be $\fE(\fL(\eE))$
and  $\eE_\fR$ to be $\fE(\fR(\eE))$.
\end{proof}
A structure $\eE_\fL = \tuple{E,\turn_\fL\;}$ can also be directly
obtained by putting $\mathord{\turn_\fL} := \{(X-Y,Y) \mid X \turn Y \}$.

\pr{purification} shows that any event structure could be
transformed into a pure one, while preserving its reachable
configurations.  However, there is no way to purify any event
structure while preserving its secured configurations:

\begin{example}{unpurifiable}
Let $\eE=\tuple{E,\turn\;}$ be given by $\eE:=\IN\cup\{e\}$,
$\{n\}\turn\{n+1\}$ and $\{n\}\turn\{e,n\}$ for $n\in\IN$
and \mbox{$\emptyset\turn X$} for $X$ not of the form $\{n+1\}$ or $\{e,n\}$.
Then\linebreak
$R(\eE)=\{\{i\mid i<n\},\{i\mid i<n\}\cup\{e\}\mid n\in\IN\}$,
$S(\eE)=R(\eE)\cup\{\IN\}$ and $L(\eE)=S(\eE)\cup\{\IN\cup\{e\}\}$.
The configuration $\IN\cup\{e\}$ is not secured, because countably many
stages are needed to perform the events in $\IN$, and whenever both
$e$ and $n$ happen, $n$ needs to happen first.
Using \pr{pure ES commute}, $S(\eE)$ cannot be the set of secured
configurations of a pure event structure, because $S(\eE)$ is not of
the form $S(\eC)$: as $R(\eC)$ would contain all sets
$\{i\mid i<n\}\cup\{e\}$ for $n\in\IN$, $S(\eC)$ would also
contain their limit $\IN\cup\{e\}$.
\end{example}
As $\eE$ above is $\fS\fR$-secure, \ex{unpurifiable} also shows that
\cor{hyperconnected} does not extend to impure structures.

\subsubsection*{Reachably pure event structures}\label{reachably pure}

\out{
 As we will see in Section~\ref{brands}, those event structures
 appearing in the literature for which a notion of configuration
 has been studied that corresponds to our concept of a left-closed
 configuration translate to event structures in our sense that are pure.
 The notion of a secured configuration on the other hand has been
 applied to impure event structures as well. The event structures of
 \ex{impure} for instance can easily be cast as event structures in the
 sense of {\sc Winskel} \cite{Wi87a,Wi89}. However, the event structures
 studied before correspond to even structures in our sense that are
 \phrase{singular}, meaning that whenever $X\turn Y$, either $X=\emptyset$
 or $Y$ is a singleton. For such event structures impureness is
 harmless, because, as is straightforward to check, the impure bits of
 the enabling relation can be deleted without changing the set of
 secured configurations or the transition relation between them in any way.
}

For impure event structures, the functions $\fL$, $\fR$, $\fS$,
$\fR_f$ and $\fS_f$ do not reflect the step transition relation
between configurations and hence may translate an event structure into
a configuration structure with a different computational interpretation.
This is illustrated by the event structure M of Section~\ref{ES
computational}, for which we have $\emptyset \gonotto{}_{\eM} \{d,e\}$
but $\emptyset \goto{}_{\fS(\eM)} \{d,e\}$.
We now extend the class of pure event structures to a slightly
larger class of \phrase{reachably pure} event structures, on which the
functions $\fR$, $\fS$, $\fR_f$ and $\fS_f$, but not $\fL$, still
preserve the computational interpretation of event structures.
This extension is necessary in order to cast the event structures of
{\sc Winskel} \cite{Wi87a,Wi89} as special cases of ours, for they
translate into our framework as event structures that are reachably
pure but not pure.

\begin{definition}{reachably pure}
An event structure is \phrase{reachably pure} if $X \turn Y$ only if
either $X \cap Y = \emptyset$ or $Y \subseteq X$.
\end{definition}
The event structure $\eE$ of \ex{impure} for instance is reachably
pure, but not pure.

\begin{proposition}{reachably pure}
For every reachably pure event structure $\eE$ there exists a pure
event structure $\hat\eE$ such that $X\goto{}_{\hat\eE}Y$ iff
$X\goto{}_{\eE}Y$ for all $X\in L(\hat\eE) \subseteq L(\eE)$ and
$Y\subseteq E$ with $X\neq Y$.
Also, if $\eE$ is rooted, so is $\hat\eE$.
\end{proposition}
\begin{proof}
Obtain $\hat\eE$ by omitting all enablings $X \turn Y$ with $\emptyset
\neq Y \subseteq X$. Apply \df{transitions-ES}.
\end{proof}
\begin{corollary}{reachably pure}
For any reachably pure event structure $\eE$ one has
$\fR(\hat\eE)\mathbin=\fR(\eE)$, $\fS(\hat\eE)\mathbin=\fS(\eE)$,
$\fR_f(\hat\eE)\mathbin=\fR_f(\eE)$ and \plat{\fS_f(\hat\eE)\mathbin=\fS_f(\eE)}.
Moreover, $\hat\eE$ is $\fS\fR$-secure iff $\eE$ is.
\end{corollary}
However, in \ex{impure} we have $\fL(\hat\eE)\neq\fL(\eE)$.

With the above results and \pr{pure ES commute}, all results for
configuration structures in this section, namely
Propositions~\ref{pr-reachable transitions}--\ref{pr-secured finitary
equivalence} and \ob{secure}, lift to reachably pure event structures:

\begin{corollary}{reachable transition equivalence ES}
Let $\eE=\tuple{E,\vdash\;}$ be a reachably pure event structure,
$x\in R(\eE)$ and $Y \subseteq E$.
Then $x \goto{}_{\fR(\eE)} Y$ iff $x \goto{}_\eE Y$.
\hfill $\Box$
\end{corollary}

\begin{corollary}{secured transition equivalence ES}
A reachably pure event structure
$\eE=\tuple{E,\vdash\;}$ is $\fS\fR$-secure
iff for all $x \in R(\eE)$ and all $Y \subseteq E$ one has
$x \goto{}_{\fS(\eE)} Y$ iff $x \goto{}_\eE Y$.
\hfill $\Box$
\end{corollary}

\begin{corollary}{ES analogies}
For any reachably pure event structure $\eE$ it holds that
$\fR(\fR(\eE))=\fR(\eE)$, $\fS(\fR(\eE))=\fS(\eE)$,
$R(\eE)\subseteq R(\fS(\eE))$ and $\fF(\fR(\eE))=\fF(\fR(\fS(\eE)))$.
\hfill $\Box$
\end{corollary}

\begin{corollary}{ES analogies secure}
For any reachably pure and $\fS\fR$-secure event structure $\eE$ it holds that
$\fS(\fS(\eE))=\fS(\eE)$, i.e., $\fS(\eE)$ is hyperconnected.
\hfill $\Box$
\end{corollary}

\begin{corollary}{reachable equivalence ES}
Let $\eE$ and $\eF$ be reachably pure and $\fS\fR$-secure event structures.
Then $\fR(\eE) = \fR(\eF)$ iff $\fS(\eE) = \fS(\eF)$.
\hfill $\Box$
\end{corollary}

\begin{corollary}{finitary reachable equivalence ES}
Let $\eE,\eF$ be reachably pure event structures.
Then $\fR_f(\eE) = \fR_f(\eF)$ iff $\fS_f(\eE) = \fS_f(\eF)$.
\hfill $\Box$
\end{corollary}
We call two reachably pure event structures $\eE$ and $\eF$
\emph{reachably equivalent}\index{reachable equivalence} iff $\fR(\eE)=\fR(\eF)$ and
\emph{finitarily reachably equivalent}\index{finitary reachable equivalence}
iff $\fR_f(\eE)=\fR_f(\eF)$.  Restricted to pure event structures
these definitions agree with \df{reachable}.

\subsubsection*{Secure event structures}\label{secure}

When dealing with secured configurations, we will mainly be interested
in event structures $\eE$ that are reachable pure and
$\fS\fR$-secure, and satisfy $S(\eE) \subseteq L(\eE)$.
The third property says that all secured configurations of $\eE$ are in
fact left-closed configurations. Together, these three properties ensure that the
computational behaviour of $\eE$ is adequately represented by $\fS(\eE)$.
An event structure with these properties is called \phrase{secure}.

\begin{proposition}{hyperconnected secure}
If $\eC$ is a hyperconnected configuration structure, then $\fE(\eC)$
is secure.
\end{proposition}

\begin{proof}
Hyperconnected configuration structures are $\fS\fR$-secure, so
that $\fE(\eC)$ is pure and $\fS\fR$-secure follows from \cor{secure}.
Moreover, using \pr{pure ES commute},
$\fS(\fE(\eC)) = \fS(\fL(\fE(\eC))) = \fS(\eC) = \eC = \fL(\fE(\eC))$.
\end{proof}
Thus $\fS$ and $\fE$ provide a bijective correspondence between secure
event structures up to reachable equivalence and hyperconnected
configuration structures.

\out{
 \begin{example}{insecure analogies}
 Take $\eE\mathbin=\tuple{\IN,\vdash\:}$ with $\emptyset\, \mathbin\vdash \{0\}$, $\{n\}
 \mathbin\vdash \{n\mathord+1\}$, $\emptyset \vdash X$ for all finite $X$ with $|X|\neq
 1$, and $X\vdash X$ for $X$ infinite. This event structure is secure.
 However, $\IN \not\in L(\hat\eE)$, so
 $\hat\eE$ does not satisfy $S(\hat\eE)\subseteq L(\hat\eE)$.
 \end{example}
}

\paragraph{Remark}
For reachably pure event structures $\eE$, unlike for configuration
structures, the requirement $S(\eE) \subseteq L(\eE)$ does not imply
$\fS\fR$-security. Moreover, this requirement would be insufficient in
\cor{reachable equivalence ES}.

\begin{example}{insecure ES}
Take $\eE:=\tuple{\IN,\vdash\:}$ with $\emptyset \vdash X$ for
$X$ finite, and $X\vdash X$ otherwise. This event structure is reachably pure
and satisfies $S(\eE)\subseteq L(\eE)$. However, $R(\eE)=\pow_{\it fin}(\IN)$,
yet $R(\fS(\eE))=\pow(\IN)$.

Take $\eF:=\tuple{\IN,\vdash'\;}$ and $\emptyset \vdash' X$ for all
$X$.  The event structures $\eE$ and $F$ have the same secured
configurations, yet are not reachably equivalent.
\end{example}

\section{Other brands of event structures}\label{brands}

Event structures have been introduced in {\sc Nielsen, Plotkin \&
Winskel} \cite{NPW81} as triples $\tuple{E,\leq,\#}$, in {\sc Winskel}
\cite{Wi87a} as triples $\tuple{E,\Con,\turn\;}$ and $\tuple{E,\Con,\leq\;}$,
and in {\sc Winskel} \cite{Wi89} as triples $\tuple{E,\#,\turn\;}$ and
$\tuple{E,\#,\leq\;}$ ---a special case of those in \cite{NPW81}.  Here
we will explain how our event structures generalise these previous proposals.
The components $\#$, $\Con$, $\turn$ and $\leq$ that occur in the
triples mentioned above can be defined in terms of our event
structures as follows.

\begin{definition}{consistency}
Let $\eE=\tuple{E,\turn\;}$ be an event structure.
A set of events $X \subseteq E$ is \phrase{consistent}, written $\ConGP(X)$, if\vspace{-1ex}
        $$\forall Y \subseteq X.~ \exists Z\subseteq E.~ Z \turn Y.$$
The binary \phrase{conflict relation} $\# \in E \times E$ is given by
$d \# e$ iff $d \neq e \wedge \neg\ConGP(\{d,e\})$.
Write $\fCon(X)$ for ``$X$ is finite and consistent''---this is our
rendering of the component $\Con$ in \cite{Wi87a}.
For $X \subseteq_{\it fin} E$ and $e \in E$, write $X \turn_s e$ for
$$\fCon(X) \wedge \exists Y \subseteq X.~Y \turn \{e\}.$$
The \phrase{direct causality relation} $\prec \; \subseteq E \times E$ is given by
$$d \prec e \Leftrightarrow \forall X.~ (X \turn \{e\} \Rightarrow d \in X).$$
We take the \phrase{causality relation}, $\leq$, to be the reflexive
and transitive closure of $\prec$.
\end{definition}
%
The next definition gives various properties of our event structures
which, in suitable combinations, determine subclasses corresponding to
the various event structures in \cite{NPW81,Wi87a,Wi89}.

\begin{definition}{event-properties}
An event structure $\eE=\tuple{E,\turn\;}$ is
\begin {itemise}
\item \phrase{singular} if $X \turn Y \Rightarrow X=\emptyset \vee |Y| = 1$,
\item \phrase{conjunctive} if $X_i \turn Y ~(i \in I \neq \emptyset) \Rightarrow
      \bigcap_{i\in I} X_i \turn Y$,
\item \phrase{locally conjunctive} if $X_i \turn Y ~(\mbox{for } i \in I
      \neq \emptyset) \; \wedge\\ \ConGP(\bigcup_{i\in I}
      X_i \cup Y) \Rightarrow \bigcap_{i\in I} X_i \turn Y$,
\item \phrase{$\fS$-ir\-re\-dun\-dant} if every event occurs in a secured
      configuration, i.e., $E = \bigcup_{x \in S(\eE)} x$,
\item \phrase{$\fL$-irredundant} if every event occurs in a left-closed
      configuration, i.e., $E = \bigcup_{x \in L(\eE)} x$,
\item and \phrase{cycle-free} if there is no chain\\
     \mbox{}\hfill$e_0 \prec e_1 \prec \cdots \prec e_n \prec e_0$\hfill\mbox{}
\end{itemise}
and has
\begin{itemise}
\item \phrase{finite causes} if $X \turn Y \Rightarrow$ $X$ finite,
\item \phrase{finite conflict} if $X$ infinite $\Rightarrow \emptyset \turn X$
\item and \phrase{binary conflict} if $|X|>2 \Rightarrow \emptyset \turn X$.
\pagebreak[3]
\end{itemise}
\end{definition}
As we will explain below,
the event structures of \cite{NPW81,Wi87a,Wi89} all correspond to
event structures in our sense that are rooted, singular and with
finite conflict. The event structures given as triples involving $\#$
even have binary conflict, the ones from \cite{Wi87a,Wi89} have finite
causes, and the ones involving $\leq$ are conjunctive,
$\fL$-irredundant and cycle-free.  The event structures of
\cite{Wi87a,Wi89} that involve $\leq$ are moreover $\fS$-irredundant,
a property that implies $\fL$-irredundancy and cycle-freeness.
The requirement of \emph{stability} in \cite{Wi87a,Wi89} corresponds to
our notion of local conjunctivity.

Each of the correspondences above will be established by means of
evident translations from the class of event structures from
\cite{NPW81,Wi87a,Wi89} under consideration to the class of our event
structures with the mentioned properties, and vice versa. These
translations will preserve the sets of events of related structures as
well as their configurations. However, which configurations will be
preserved varies, as indicated in Table~\ref{7 classes}.
\begin{table}
\mbox{}\hfill\small
\begin{tabular}{@{}|@{~}l@{}r@{~}|@{~}l@{~\,}r@{~}|@{}}
\hline
ev.\,str.\,\cite{Wi87a} & $\Con,\turn$ & rtd, sing, f.causes \& f.conflict& $\fS$\\
stable     \cite{Wi87a} & $\Con,\turn$ & same \& locally conjunctive& $\fS$\\
prime      \cite{Wi87a} & $\Con,\leq $ & same \& conjunctive \& $\fS$-irr.& $\fS,\fL$\\
ev.\,str.\,\cite{Wi89}  & $\#  ,\turn$ &  rtd, sing, f.causes \& bin.conflict& $\fS$\\
stable     \cite{Wi89}  & $\#  ,\turn$ & same \& locally conjunctive& $\fS$\\
prime      \cite{Wi89}  & $\#  ,\leq $ & same \& conjunctive \& $\fS$-irr.& $\fS,\fL$\\
ev.\,str.\,\cite{NPW81} & $\#  ,\leq $ & rtd, sing, b.c., conj, $\fL$-irr \& c.-f.& $\fL$\\
\hline
\end{tabular}\hfill\mbox{}%
\vspace{-1ex}
\caption{7 corresponding classes of event structures}
\label{7 classes}
\vspace{-.7em}
\end{table}
The configurations employed in \cite{Wi87a,Wi89} correspond to our
secured configurations, whereas the configurations employed for event
structures involving $\leq$ correspond to our left-closed
configurations. In the intersection of those two situations, the
secured and left-closed configurations of event structures coincide.

\begin{definition}{manifestly conjunctive}
An event structure is \phrase{manifestly conjunctive} if for every set of
events $Y$ there is at most one set $X$ with $X \turn Y$.
\end{definition}
Every conjunctive event structure can be made manifestly conjunctive
by deleting from $\turn$, for every set $Y$, all but the smallest $X$
for which $X \turn Y$. The property of conjunctivity implies that such
a smallest $X$ exists. This normalisation preserves $\fL$-equivalence
and even transition equivalence (cf.~\df{multiset-transition system})
and all properties of \df{event-properties}.
The event structures in our sense that arise as translations of event
structures from \cite{NPW81,Wi87a,Wi89} that involve $\leq$ are all
manifestly conjunctive.

\begin{observation}{manifestly conjunctive}
Any singular, cycle-free, manifestly conjunctive event structure is pure.
\end{observation}
Hence the translations between the event structures from
\cite{NPW81,Wi87a,Wi89} involving $\leq$ and subclasses of our event
structures will preserve not only $\fL$-equivalence, but even transition
equivalence.

\begin{lemma}{finite conflict secure}
If $\eE$ has finite conflict, then $S(\eE)\subseteq L(\eE)$.
\end{lemma}
\begin{proof}
Let $X \in S(\eE)$ and let $X_0,X_1,\ldots$ be a stepwise securing of $X$.
Let $Y \subseteq X$. Then either $Y$ is infinite and $\emptyset \vdash Y$ or
$Y$ is finite and hence contained in $X_{i+1}$ for some $i\in\IN$.
In the latter case $\exists Z\subseteq X_i \subseteq X$ with $Z \vdash Y$.
\end{proof}

\begin{observation}{singular}~\vspace{-6pt}
\item Any singular event structure is reachably pure.
\end{observation}

\begin{proposition}{finite conflict secure}
Any singular event structure with finite conflict is \hyperref[secure]{secure}.
\end{proposition}
\begin{proof}
Let $\eE$ be a singular event structure with finite conflict.
Then the event structure $\hat\eE$, as defined in the proof of
\pr{reachably pure}, is pure and with finite conflict. \lem{finite
conflict secure} yields $S(\hat\eE) \subseteq L(\hat\eE)$.  Hence,
$R(\fS(\eE)) = R(\fS(\hat\eE)) \subseteq R(\fL(\hat\eE))=R(\hat\eE)=R(\eE)$.
The other direction follows from \cor{ES analogies}.
\end{proof}
As all event structures of \cite{NPW81,Wi87a,Wi89} correspond to event
structures in our sense that are singular and with finite conflict,
they all fall in the scope of Corollaries~\ref{cor-reachable
transition equivalence ES} and~\ref{cor-reachable equivalence ES}, so
reachable equivalence preserves the computational interpretation of
event structures and is characterised by having the same secured
configurations. Hence the translations between the event structures
from \cite{Wi87a,Wi89} and subclasses of our event structures will
preserve reachable equivalence. We will show that they also preserve
$\fL$-equivalence, and even transition equivalence
(cf.~\df{multiset-transition system}); however, this involves {\em
defining} the left-closed configurations and a transition relation on
the structures of \cite{Wi87a,Wi89}.

\subsection{\normalsize Left-closed configurations and transitions}

For singular event structures $\eE$, the enabling relation consists
of two parts: enablings of the form $\emptyset \turn Y$ with $|Y|\neq 1$, and enablings
of the form $X \turn \{e\}$. When $\eE$ has finite conflict, the first
part can be fully expressed in terms of $\fCon$, at least to the extent
to which it determines which sets of events are configurations. When
$\eE$ has finite causes, the second part can similarly be expressed in
terms of $\turn_s$. One obtains the following.

\begin{observation}{Wi87a}
Let $\eE$ be a singular event structure with finite causes and
finite conflict. Then\\[1.5ex]
\mbox{~}\hfill $~~~X \goto{}_\eE Y \Leftrightarrow\left\{\begin{array}{@{}l@{}}
X \subseteq Y \wedge
\forall Z \subseteq_{\it fin} Y.~ \fCon(Z) \, \wedge ~~~\hfill\footnotemark\\
\forall e \in Y.~\exists W \subseteq X.~ W \turn_s e.~~~
\end{array}\right.$\vspace{-2ex}
\footnotetext{Note that $X\in L(\eE)$ iff $X \goto{}_\eE X$. Hence,
characterisations of $\goto{}_\eE$ such as this one entail also
characterisations of $L(\eE)$.}
\end{observation}
It follows that such structures can alternatively be represented
as triples $\tuple{E,\fCon,\turn_s\;}$ with $\fCon \subseteq \pow_{\it fin}(E)$
and $\turn_s\; \subseteq \fCon\times E$, as are the structures of \cite{Wi87a}.

When $\eE$ moreover is rooted and with binary conflict, $\fCon$, when
applied to non-singleton sets, can be fully expressed in terms of $\#$.

\begin{observation}{Wi89}
Let $\eE$ be a rooted, singular event structure with finite causes and
binary conflict. Then\\[1.5ex]
\mbox{~}\hfill $X \goto{}_\eE Y \Leftrightarrow \left\{\begin{array}{@{}l@{}}
X \subseteq Y \wedge \forall d,e \in Y.~ \neg (d \# e) \, \wedge\\
\forall e \in Y.~\exists W \subseteq X.~ W \turn_s e.
\end{array}\right.$\hfill$\begin{array}{@{}c@{}}~\\\end{array}$
\end{observation}
It follows that such event structures can alternatively be represented
as triples $\tuple{E,\#,\turn_s\;}$ with $\# \subseteq E \times E$
symmetric and irreflexive and $\turn_s\; \subseteq \fCon \times E$, as
are the structures of \cite{Wi89}.

When $d \leq e$, any configuration containing $e$ also contains $d$.
When $\eE=\tuple{E,\turn\;}$ is conjunctive and satisfies
$\ConGP(\{e\})$ for all $e\mathbin\in E$, then for any event
$e \mathbin\in E$ there is a smallest set $X \subseteq E$ with $X \turn e$.
In that case, the part of the enabling relation consisting of enablings $X
\turn e$ is in essence completely determined by the causality relation
$\leq$.

\begin{observation}{prime}
Let $\eE=\tuple{E,\turn\;}$ be a singular, conjunctive event structure
with finite conflict, such that $\ConGP(\{e\})$ for all $e\mathbin\in E$. Then
$$X \in L(\eE) \Leftrightarrow \left\{\begin{array}{@{}l@{}}
\forall Y \subseteq_{\it fin} X.~ \fCon(Y) \, \wedge\\
\forall d,e \in E.~ d \leq e \in X  \Rightarrow d \in X.
\end{array}\right.$$
If $\eE$ moreover is rooted and with binary conflict, then\\[1.5ex]
\mbox{}\hfill $X \in L(\eE) \Leftrightarrow \left\{\begin{array}{@{}l@{}}
\forall d,e \in X.~ \neg (d \# e) \, \wedge\\
\forall d,e \in E.~ d \leq e \in X  \Rightarrow d \in X.
\end{array}\right.$\hfill$\begin{array}{@{}c@{}}~\\\end{array}$
\end{observation}
It follows that, up to $\fL$-equivalence, such structures can
alternatively be represented as triples $\tuple{E,\fCon,\leq\;}$ with
$\fCon \subseteq \pow_{\it fin}(E)$ and $\leq\; \subseteq E \times E$,
as are the prime event structures of \cite{Wi87a}, respectively as
triples $\tuple{E,\#,\leq\;}$, as are the prime event structures of
\cite{NPW81,Wi89}.

\out{
 \begin{observation}{prime-bc}
 Let $\eE$ be a rooted, singular, conjunctive, cycle-free event structure
 with binary conflict. Then
 $$X \in L(\eE) \Leftrightarrow \left\{\begin{array}{@{}l@{}}
 \forall d,e \in X.~ \neg (d \# e) \, \wedge\\
 \forall d,e \in E.~ d \leq e \in X  \Rightarrow d \in X
 \end{array}\right.$$
 \end{observation}
 It follows that, up to $\fL$-equivalence, such event structures can
 alternatively be represented as triples $\tuple{E,\#,\leq\;}$ with
 $\# \subseteq E \times E$ and $\leq\; \subseteq E \times E$, as are
 the prime event structures of \cite{NPW81,Wi89}.
}

\subsection{Secured configurations}\label{secured-Winskel}

In this section we augment Observations \ref{obs-Wi87a} to
\ref{obs-prime} with characterisations of the secured configurations.
To this end we first provide a characterisation of the finite
reachable configurations of singular event structures.

\begin{observation}{Winskel-secured}
Let $\eE$ be a singular event structure. Then\vspace{-2ex}
$$X \in R_f(\eE) \Leftrightarrow \left\{\begin{array}{@{}l@{}}
\fCon(X) \, \wedge\\
\exists e_1, \ldots, e_n \in X.~X = \{e_1,...,e_n\}\,
\wedge \\ \forall i\leq n.~ \{e_1,...,e_{i-1}\} \turn_s e_i.
\end{array}\right.$$
If $\eE$ is furthermore rooted and with binary conflict, then\vspace{-1.4ex}
$$X \in R_f(\eE) \Leftrightarrow \left\{\begin{array}{@{}l@{}}
\forall d,e \in X.~ \neg (d \# e) \, \wedge\\
\exists e_1, \ldots, e_n \in X.~X = \{e_1,...,e_n\}\,
\wedge \\ \forall i\leq n.~ \{e_1,...,e_{i-1}\} \turn_s e_i.
\end{array}\right.$$
\out{In case $\eE$ is conjunctive, singular and cycle-free, every
 finite set $X$ in $L(\eE)$ can be seen to be in $R_f(\eE)$:
 $$X \in R_f(\eE) \Leftrightarrow \left\{\begin{array}{@{}l@{}}
 \fCon(X) \, \wedge\\
 \forall d,e \in E.~ d \leq e \in X  \Rightarrow d \in X.
 \end{array}\right.$$
 If $\eE$ moreover is rooted and with binary conflict, then
 $$X \in R_f(\eE) \Leftrightarrow \left\{\begin{array}{@{}l@{}}
 X \mbox{ is finite} \, \wedge
 \forall d,e \in X.~ \neg (d \# e) \, \wedge\\
 \forall d,e \in E.~ d \leq e \in X  \Rightarrow d \in X.
 \end{array}\right.$$
}
\end{observation}
%
The next proposition says that for certain event structures, including
the ones from \cite{Wi87a,Wi89}, the secured configurations are
completely determined by the finite reachable ones. In addition, it
provides the counterpart of \ob{Wi87a} for the secured configurations.

\begin{proposition}{infinite secured configurations}
Let $\eE = \tuple{E,\turn\;}$ be a singular event structure with finite causes
and finite conflict. Then
$$X \in S(\eE) \Leftrightarrow \forall Y \subseteq_{\it fin} X.~
\exists Z \in R_f(\eE).~ Y \subseteq Z \subseteq X,$$
i.e., $\fS(\eE)$ is the set of \phrase{directed unions} over $\fR_f(\eE)$,
and
$$X \in S(\eE) \Leftrightarrow \left\{\begin{array}{@{}l@{}}
\forall Y \subseteq_{\it fin} X.~ \fCon(Y) \, \wedge\\
\forall e \in X.~ \exists e_0, \ldots, e_n \in X.~ e=e_n\,
\wedge \\ \forall i\leq n.~ \{e_0,...,e_{i-1}\} \turn_s e_i.
\end{array}\right.$$
\end{proposition}

\begin{proof}
``$\Rightarrow$, above'': Let $X \in S(\eE)$ and $Y \subseteq_{\it
fin} X$. Let $X_0,X_1,...$ be a stepwise securing of $X$
(cf.~\df{reachable ES}) and choose
$n$ in $\IN$ such that $Y \subseteq X_n$. For $k=n,n-1,n-2,...,0$
choose the finite subset $Y_k$ of $X_k$ recursively as follows.
\plat{Y_n = Y}.  Given $Y_{k}$ with $1 \leq k \leq n$, choose for any
event $e \in Y_{k}$ a set $Z_e \subseteq X_{k-1}$ with $Z_e \turn e$,
and let $Y_{k-1} = \bigcup_{e \in Y_{k}}Z_e$. 
Because E has finite causes, the sets $Z_e$ are finite, and so is
$Y_{k-1}$. As $\eE$ is singular we have
$\emptyset \turn Z$ for any $Z \subseteq X$ with $|Z| \neq 1$.
Therefore the sets $\bigcup_{i=0}^k Y_i$ for $k\leq n$ form a
stepwise securing of the finite set $Z = \bigcup_{i=0}^n Y_i$.\pagebreak[1]
Hence $Z \in R_f(\eE)$. Furthermore we have $Y \subseteq Z \subseteq X$.

``$\Downarrow$'' follows immediately from \ob{Winskel-secured}.

``$\Leftarrow$, below'': Let $X \subseteq E$ be such that the
right-hand side holds.
Take $X_{n+1} := \{e_n \mid \exists e_0, \ldots, e_{n-1} \in X.\linebreak
\forall i\!\leq\! n.~ \{e_0,...,e_{i-1}\} \turn_s e_i\}$ for $n\!\in\!\IN$,
and take $X_0 := \emptyset$. Now $X = \bigcup_{n=0}^\infty X_n$.  As a
sequence $e_0,...,e_n$ as occurs above can be prolonged by repeating
events, we have $X_n \subseteq X_{n+1}$ for all $n \in \IN$.  Let
{$Y \subseteq X_{n+1}$}. It remains to be shown that $\exists Z
\subseteq X_{n}.~ Z \turn Y$. In case $Y$ is infinite, this follows
because $\eE$ has finite conflict. Otherwise, if $|Y|\neq 1$ it
follows because $\eE$ is singular and $\fCon(Y)$. Now suppose $|Y|=1$.
Then $\exists e_0,..., e_{n} \in X.~ Y\!=\!\{e_{n}\} \wedge \forall i\leq n.~
\{e_0,...,e_{i-1}\} \turn_s e_i$.
So in particular $\exists Z \subseteq \{e_0,...,e_{n-1}\}.~ Z \turn Y$.
We have $Z \subseteq \bigcup_{i=1}^n X_i = X_n$.
\end{proof}
Thus, recalling \ob{singular}, \pr{finite conflict secure} and
\cor{reachable equivalence ES}, for singular event structures with
finite causes and finite conflict we have\vspace{-3pt}
$$\fR(\eE)=\fR(\eF) ~\mbox{ iff }~ \fS(\eE)=\fS(\eF) ~\mbox{ iff }~
\fR_f(\eE)=\fR_f(\eF).\vspace{-3pt}$$
In both statements of \pr{infinite secured configurations},
``$\Rightarrow$'' requires singularity and finite causes, and
``$\Leftarrow$'' singularity and finite conflict. That these
conditions cannot be dropped follows from the following counterexamples.
\begin{itemise}
\item Let $E$ be uncountable let $\emptyset \turn X$ for every finite set
$X$ (with no other enablings). This event structure is singular
and has finite causes, but does not have finite conflict, and
``$\Leftarrow$'' fails for uncountable $X$.
\item Let $E$ be uncountable, with $X \turn Y$ iff $X=\emptyset$ and
$Y$ is empty or infinite, or $Y$ is finite and $X$ contains one event
less. This event structure has finite
causes and finite conflict, but is not singular, and ``$\Leftarrow$'' fails
for uncountable $X$ (even though such $X$ are left-closed configurations).
\item Let $E := \IN \cup \{a\}$, $\emptyset \turn X$ for any $X \neq
\{a\}$, and $\IN \turn a$. This event structure is singular and has
finite conflict, but does not have finite causes, and
``$\Rightarrow$'' fails for $X=E$.
\item Let $E := \IN \cup \{a\}$, $\{0\} \turn a$, $\{n+1\} \turn \{a,n\}$
and $\emptyset \turn X$ for any other set $X$. This event structure has
finite causes and finite conflict, but is not singular, and
``$\Rightarrow$'' fails for $X=E$.
\end{itemise}
The following counterpart of \ob{Wi89} is an easy consequence.

\begin{observation}{Wi89-secured}
Let $\eE$ be a rooted, singular event structure
with finite causes and binary conflict. Then\\[1.5ex]
\mbox{~}\hfill $X \in S(\eE) \Leftrightarrow \left\{\begin{array}{@{}l@{}}
\forall d,e \in X.~ \neg (d \# e) \, \wedge\\
\forall e \in X.~ \exists e_0, \ldots, e_n \in X.~ e=e_n\,
\wedge \\ \forall i\leq n.~ \{e_0,...,e_{i-1}\} \turn_s e_i.
\end{array}\right.$\hfill$\begin{array}{@{}c@{}}~\\\\\end{array}$
\end{observation}
For $X$ a left-closed configuration of a singular, conjunctive event
structure and $e_0\in X$ we say that $e_0$ can happen at stage $n$, if
there is no chain $e_n \prec \cdots \prec e_1 \prec e_0$.  Now we have
\pagebreak[2]
$X\in S(\eE)$ iff each event in $X$ can happen at some finite stage.
It follows that:

\begin{observation}{prime-secured}
Let $\eE=\tuple{E,\turn\;}$ be a singular, conjunctive event structure.
Then
\begin{enumerate}
\item $X \mathbin\in L(\eE) \Leftrightarrow \ConGP(X) \wedge  \forall d,\!e
\mathbin\in E.~ d \mathbin\leq e \mathbin\in X  \Rightarrow d \mathbin\in X$.
\item $\eE$ is $\fL$-irredundant iff $\forall e\in E.~\ConGP(\{d\mid d\leq e\})$.
\item $\eE$ is $\fS$-irredundant iff $\eE$ is $\fL$-irredundant and
for every $e\in E$ there is an $n\in\IN$ such that there is no chain
$e_n \prec \cdots \prec e_1 \prec e_0 = e$.
\item In case $\eE$ is cycle-free we have\vspace{-1ex}
$$X \in R_f(\eE) \Leftrightarrow X \in L(\eE) \wedge X \mbox{ is finite.}
\vspace{-1ex}$$
\item If $\eE$ is $\fS$-irredundant then $L(\eE)\subseteq S(\eE)$.
\end{enumerate}
\out{
 $$X \in S(\eE) \Leftrightarrow \left\{\begin{array}{@{}l@{}}
 \forall Y \subseteq_{\it fin} X.~ \fCon(Y) \, \wedge\\
 \forall d,e \in E.~ d \leq e \in X  \Rightarrow d \in X\\
 \wedge \mbox{ there is no infinite chain }\\
  \cdots \prec e_2 \prec e_1 \prec e_0 \in X.
 \end{array}\right.$$
 If $\eE$ moreover is rooted and with binary conflict, then
 $$X \in S(\eE) \Leftrightarrow \left\{\begin{array}{@{}l@{}}
 \forall d,e \in X.~ \neg (d \# e) \, \wedge\\
 \forall d,e \in E.~ d \leq e \in X  \Rightarrow d \in X.
 \end{array}\right.$$
 Let $\eE$ be a singular, conjunctive event structure.
 In case $\eE$ is cycle-free, every
 finite set $X$ in $L(\eE)$ can be seen to be in $R_f(\eE)$:
 $$X \in R_f(\eE) \Leftrightarrow \left\{\begin{array}{@{}l@{}}
 \fCon(X) \, \wedge\\
 \forall d,e \in E.~ d \leq e \in X  \Rightarrow d \in X.
 \end{array}\right.$$
 If $\eE$ moreover is rooted and with binary conflict, then
 $$X \in R_f(\eE) \Leftrightarrow \left\{\begin{array}{@{}l@{}}
 X \mbox{ is finite} \, \wedge
 \forall d,e \in X.~ \neg (d \# e) \, \wedge\\
 \forall d,e \in E.~ d \leq e \in X  \Rightarrow d \in X.
 \end{array}\right.$$
}
\end{observation}
Together with \lem{finite conflict secure} this yields
\begin{corollary}{prime-secured}
Let $\eE=\tuple{E,\turn\;}$ be a singular, conjunctive,
$\fS$-irredundant event structure with finite conflict.
Then $\fS(\eE)=\fL(\eE)$.\hfill$\Box$
\end{corollary}

\out{
 \begin{proposition}{infinitary conjunctive configurations}
 Let $\eE$ be a singular, conjunctive, $\fS$-irredundant event
 structure with finite conflict. Then $S(\eE) = L(\eE)$.
 \end{proposition}

 \begin{proof}
 ``$\subseteq$'' follows immediately from \df{secured
 configurations}, using that $\eE$ has finite conflict.

 ``$\supseteq$'': For any $e \in E$, let
 $\downarrow\! e$ be $\{d \in E \mid d \leq e\}$. Any secured or left-closed
 configuration containing $e$ must contain $\downarrow\! e$. As $\eE$ is
 $\fS$-irredundant, $e$ must occur in a secured configuration. Hence
 $\downarrow\! e$ is finite, and $\fCon(\downarrow\! e)$. Thus $\downarrow\! e
 \in R_f(\eE)$ by \ob{prime}.

 Now suppose $X \in L(\eE)$. For any $Y \subseteq_{\it fin} X$, let
 $\downarrow\! Y$ be $\bigcup_{e \in Y}\downarrow\! e$. It must be that
 $\downarrow\! Y \subseteq_{\it fin} X$. Hence $\fCon(\downarrow\! Y)$.
 Now \lem{consistent unions} implies that $\downarrow\! Y \in R_f(\eE)$.
 Moreover, $Y \subseteq\; \mbox{$\downarrow\! Y$} \subseteq X$, so
 direction ``$\Leftarrow$'' of \pr{infinite secured configurations}
 (not using the requirement of finite causes) implies that $X \in
 S(\eE)$.
 \end{proof}
}

\subsection[The event structures of Winskel 1987]{The event structures of {\sc Winskel} \cite{Wi87a}}
\label{Wi87a}

These are defined as triples $\eE = \tuple{E,Con,\turn\;}$ where
\begin{itemise}
\item $E$ is a set of \phrase{events},
\item $Con \subseteq \pow_{\it fin}(E)$ is a nonempty \phrase{consistency
      predicate} such that:
      $Y \subseteq X \in Con \Rightarrow Y\in Con$,
\item and $\turn\, \subseteq Con \times E$ is the \phrase{enabling
      relation}, which satisfies $X \turn e \wedge X \subseteq Y\in
      Con \Rightarrow Y\turn e$.
\end{itemise}
Such an event structure is \phrase{stable} if it satisfies
$$X \turn e \wedge Y \turn e \wedge \Con(X \cup Y \cup \{e\})
\Rightarrow X\cap Y \turn e.$$
\index{families of configurations of event structures}%
The family $S(\eE)$ of configurations of such an event
structure (written ${\cal F}(\eE)$ in \cite{Wi87a}) consists of those
$X \subseteq E$ which are
\begin{itemise}
\item \phrase{consistent}: every finite subset of $X$ is in $Con$,
\item and \phrase{secured}: $\forall e \in X.~\exists e_0, \ldots,e_n \in X.~
e_n=e \wedge \forall i\leq n.~ \{e_0,...,e_{i-1}\}\turn e_{i}$,
\end{itemise}
just as in \pr{infinite secured configurations}.
In addition, we define $L(\eE)$ and $\goto{}_\eE$ exactly as in
Observation~\ref{obs-Wi87a}, but reading $\Con$ for $\fCon$ and
$\turn$ for $\turn_s$. Again, we write $\fS(\eE)$ for
$\tuple{E,S(\eE)}$, and $\fC^+(\eE)$ for $\tuple{E,L(E),\goto{}_\eE\;}$.

Here we will show that up to reachable equivalence and even
transition equivalence (cf.~\df{multiset-transition system}) these
event structures are exactly the ones in our sense which are rooted,
singular, with finite causes and with finite conflict; and the stable
event structures of \cite{Wi87a} are the ones which are moreover
locally conjunctive.

For $\eE = \tuple{E_W,\Con_W,\turn_W\;}$ an event structure as in
\cite{Wi87a}, let the event structure $\fE(\eE) := \tuple{E_W, \turn\;}$
be given by\vspace{-1em}
$$X \turn Y \mbox {~iff~} \left\{\begin{array}{@{}l@{}}
\mbox{either $Y=\{e\}$, $\Con_W(\{e\})$ and $X \turn_W e$} \\
\mbox{or $|Y| \neq 1$, $X=\emptyset$ and $\Con_W(Y)$} \\
\mbox{or $Y$ is infinite and $X=\emptyset$.} \end{array}\right.$$
Now, for $X \subseteq_{\it fin} E_W$,
\begin{equation}\label{Con}
\fCon(X) \Leftrightarrow \Con_W(X) \wedge \forall e\mathbin\in X.~ \exists
Y\mathbin\subseteq E_W.~Y \turn_W e
\end{equation}
and whenever $\fCon(X)$ we have
\begin{equation}\label{enabling}
X\turn_s e \Leftrightarrow \Con_W(\{e\}) \wedge X\turn_W e.
\end{equation}

\begin{proposition}{Wi87a to ours}
Let $\eE$ be an event structure as in \cite{Wi87a}. Then
$\fE(\eE)$ is rooted, singular and with finite causes and finite
conflict. If $\eE$ is stable then $\fE(\eE)$ is locally conjunctive.
Moreover, $\fS(\fE(\eE)) = \fS(\eE)$ and $\fC^+(\fE(\eE)) = \fC^+(\eE)$.
\end{proposition}

\begin{proof} Let $\eE=\tuple{E_W,\Con_W,\turn_W\;}$ be an event
structure as in \cite{Wi87a}. As $\Con_W$ is nonempty and
subset-closed we have $\emptyset \in \Con_W$. Thus
$\emptyset \turn \emptyset$, i.e., $\fE(\eE)$ is rooted. By
construction, $\fE(\eE)$ is singular and with finite causes and finite
conflict.  That the stability of $\eE$ implies the local conjunctivity
of $\fE(\eE)$ follows because $\fCon(X) \Rightarrow \Con_W(X)$ and
every collection of finite sets has a finite subcollection with the
same intersection. With \pr{infinite secured configurations} and
\ob{Wi87a}, respectively, using (\ref{Con}) and (\ref{enabling}), one
easily checks that $\fS(\fE(\eE)) = \fS(\eE)$ and $\fC^+(\fE(\eE)) =
\fC^+(\eE)$.
\end{proof}
For $\eE=\tuple{E,\turn\;}$ a rooted event structure, the structure ${\cal
W}(\eE):=\tuple{E,\fCon,\turn_s\;}$, where $\fCon$ and $\turn_s$ are given by
\df{consistency}, is clearly an event structure in the sense of
\cite{Wi87a}.

\begin{proposition}{ours to Wi87a}
Let $\eE$ be a rooted, singular event structure with finite causes and
finite conflict. Then $\fS({\cal W}(\eE)) = \fS(\eE)$
and $\fC^+({\cal W}(\eE)) = \fC^+(\eE)$.
Moreover, ${\cal W}(\eE)$ is stable if $\eE$ is locally conjunctive.
\end{proposition}

\begin{proof}
Trivial, with \pr{infinite secured configurations} and \ob{Wi87a}.
\end{proof}

\subsection[The event structures of Winskel 1989]{The event structures of {\sc Winskel} \cite{Wi89}}
\label{Wi89}

These are defined as triples $\eE = \tuple{E,\#,\turn\;}$ where
\begin{itemise}
\item $E$ is a set of \phrase{events},
\item $\# \subseteq E \times E$ is a symmetric, irreflexive
      \phrase{conflict relation}. Write $\Con$ for the set of finite,
      conflict-free subsets of $E$, i.e., those finite subsets
      $X\subseteq E$ for which $$\forall e,e' \in X.~\neg (e \# e'),$$
\item and $\turn\, \subseteq Con \times E$ is the \phrase{enabling
      relation}, which satisfies $X \turn e \wedge X \subseteq Y\in
      Con \Rightarrow Y\turn e$.
\end{itemise}
Such an event structure is \phrase{stable} if it satisfies
$$X \turn e \wedge Y \turn e \wedge \Con(X \cup Y \cup \{e\})
\Rightarrow X\cap Y \turn e.$$
The family $S(\eE)$ of configurations of such an event
structure (written ${\cal F}(\eE)$ in \cite{Wi89}) consists of those
$X \subseteq E$ which are
\begin{itemise}
\item \phrase{conflict-free}: $\forall e,e' \in X.~\neg (e \# e')$,
\item and \phrase{secured}: $\forall e \in X.~\exists e_0, \ldots,e_n \in X.~
e_n=e \wedge \forall i\leq n.~ \{e_0,...,e_{i-1}\}\turn e_{i}$,
\end{itemise}
just as in \ob{Wi89-secured}.  Note that a set of events $X$ is
conflict-free iff every finite subset of $X$ is in $\Con$. In
addition, we define $L(\eE)$ and $\goto{}_\eE$ exactly as in
Observation~\ref{obs-Wi89}, reading $\Con$ for $\fCon$ and $\turn$
for $\turn_s$.

Say that an event structure $\tuple{E,\Con,\turn\;}$ in the sense of
\cite{Wi87a} has \phrase{binary conflict} if for any $X \subseteq_{\it
fin} E$: $$\Con(X) \Leftrightarrow \forall Y \subseteq X.~(|Y|=2
\Rightarrow \Con(Y)).$$
Clearly, the event structures of \cite{Wi89} are just a reformulation
of the event structures of \cite{Wi87a} that have binary conflict.
A small variation of the arguments from the previous section shows
that, up to reachable equivalence and even transition equivalence, the event
structures of \cite{Wi89} are exactly the ones in our sense which are
rooted, singular, with finite causes and with binary conflict; and the
stable event structures of \cite{Wi89} are the ones which are moreover
locally conjunctive:

For $\eE \mathbin= \tuple{E_W,\#_W,\turn_W\;}$ an event structure as in
\cite{Wi89}, let the event structure $\fE(\eE) := \tuple{E_W, \turn\;}$
be given by $$X \turn Y \mbox {~iff} \left\{\begin{array}{@{}l@{}}
\mbox{either $Y=\{e\}$ and $X \turn_W e$} \\
\mbox{or $|Y| \hspace{-.8pt}=\hspace{-.8pt} \{d,e\}$, $d \!\neq\! e$,
$X \!=\! \emptyset$ and $\neg(d \#_W e)$} \\
\mbox{or $Y=X=\emptyset$}\\
\mbox{or $|Y|>2$ and $X=\emptyset$.} \end{array}\right.$$
Write $\Con_W(X)$ for $|X|<\infty\wedge\forall e,e' \in X.~\neg (e
\#_W e')$.
Then equations (\ref{Con}) and (\ref{enabling}) of Section~\ref{Wi87a} hold again.

\begin{proposition}{Wi89 to ours}
Let $\eE$ be an event structure as in \cite{Wi89}. Then
$\fE(\eE)$ is rooted, singular and with finite causes and binary
conflict. If $\eE$ is stable then $\fE(\eE)$ is locally conjunctive.
Moreover, $\fS(\fE(\eE)) = \fS(\eE)$ and $\fC^+(\fE(\eE)) = \fC^+(\eE)$.
\end{proposition}

\begin{proof} Let $\eE=\tuple{E_W,\#_W,\turn_W\;}$ be an event
structure as in \cite{Wi89}. By construction, $\fE(\eE)$ is rooted,
singular and with finite causes and binary conflict.  That the
stability of $\eE$ implies the local conjunctivity of $\fE(\eE)$
follows exactly as in the proof of \pr{Wi87a to ours}.  With
Observations~\ref{obs-Wi89-secured} and \ref{obs-Wi89}, respectively,
one obtains $\fS(\fE(\eE)) = \fS(\eE)$ and $\fC^+(\fE(\eE)) = \fC^+(\eE)$.
\end{proof}
For $\eE=\tuple{E,\turn\;}$ a rooted event structure with binary
conflict, the structure ${\cal W}^\#(\eE):=\tuple{E,\#,\turn_s^\#\;}$,
where $\#$ is given by \df{consistency} and
 $X \turn_s^\# e$ iff
$$(\fCon(X) \vee |X|=1) \wedge \exists Y \subseteq X.~Y \turn \{e\},$$
is clearly an event structure in the sense of \cite{Wi89}.
\begin{proposition}{ours to Wi89}
Let $\eE$ be a rooted, singular event structure with finite causes and
binary conflict. Then $\fS({\cal W}^\#(\eE)) = \fS(\eE)$
and $\fC^+({\cal W}^\#(\eE)) = \fC^+(\eE)$.
Moreover, ${\cal W}^\#(\eE)$ is stable if $\eE$ is locally conjunctive.
\end{proposition}

\begin{proof}
The first two statements are trivial, with
Observations~\ref{obs-Wi89-secured} and \ref{obs-Wi89}.  Now assume
$\eE$ is locally conjunctive; we show that ${\cal W}^\#(\eE)$ is
stable. So assume $X \turn^\#_s e$, $Y \turn^\#_s e$ and for $d,f\in
X\cup Y \cup\{e\}$ it holds that $\neg(d \# f)$. The latter means that
either $d=f$ or $\Con(\{d,f\})$. We have to show that $X \cap Y \turn_s^\# e$.
\\[1ex]
{\sc Claim}: $\Con(X \cup Y \cup \{e\})$.
\\[1ex]
{\sc proof:}
Let $W \subseteq X \cup Y \cup \{e\}$. We have to find a $Z$ with $Z \turn W$.
In case $W=\emptyset$ or $|W|>2$ we can take $Z=\emptyset$, because E is
rooted and with binary conflict.

In case $W = \{d,f\}$ with $d\mathbin{\neq} f$, we have $\Con(\{d,f\})$.

In case $W=\{e\}$, we use $X \turn^\#_s e$ to
infer that there is an $X' \subseteq X$ with $X'\turn \{e\}$.

In case $W=\{d\}$ with $d\mathbin{\neq} e$, then $\Con(\{d,e\})$ and hence $\Con(\{d\})$.
\\[1ex]
{\sc Application of the Claim}:
Since $X \turn^\#_s e$, there is an $X' \subseteq X$ with $X'\turn e$.
Likewise, there is an $Y' \subseteq Y$ with $Y'\turn e$.
Now $\Con(X' \cup Y' \cup \{e\})$, so the local
conjunctivity of $\eE$ yields $X' \cap Y' \turn \{e\}$.
Furthermore, $X$ and $Y$ must be finite, by definition of $\turn_s^\#$,
so the claim also yields $\fCon(X \cap Y)$. As $X'\cap Y' \subseteq
X\cap Y$ we obtain $X \cap Y \turn_s^\# e$.
\end{proof}

\subsection[The prime event structures of Winskel 1987]{The prime event structures of \cite{Wi87a}}
\label{prime}

These are defined as triples $\eE = \tuple{E,\Con,\leq\;}$ where
\begin{itemise}
\item $E$ is a set of \phrase{events},
\item $Con \subseteq \pow_{\it fin}(E)$ is a nonempty \phrase{consistency
      predicate} such that:
      $Y \subseteq X \in Con \Rightarrow Y\in Con$,
      and $\{e\} \in \Con$ for all $e \in E$,
\item and $\leq \;\subseteq E \times E$ is a partial order,
      the \phrase{causality relation}, satisfying
\begin{itemise2}
\item $d \leq e \in X \in \Con \Rightarrow X \cup \{d\} \in \Con$
\item and $\downarrow\!e = \{d \in E \mid d \leq e\}$ is finite for
      all $e \in E$.
\end{itemise2}
\end{itemise}
The set $L(\eE)$ of configurations of such an event structure
consists of those $X \subseteq E$ which are
\begin{itemise}
\item \phrase{consistent}: every finite subset of $X$ is in $Con$,
\item and \phrase{left-closed}: $\forall d,e \in E.~ d \leq e \in X
      \Rightarrow d \in X$,
\end{itemise}
just as in \ob{prime}. Write $\fL(\eE)$ for $\tuple{E,L(\eE)}$.\pagebreak[2]

Here we will show that up to $\fL$-equivalence these event structures
are exactly the ones in our sense which are rooted, singular,
(manifestly) conjunctive, (pure,) $\fS$-irredundant and with finite
causes and finite conflict.  On this class of event structures,
\cor{prime-secured} says that $\fS$ coincides with $\fL$.  Thus each of
$S$ and $L$ can be understood as generalisation of the notion of
configuration for prime event structures from \cite{Wi87a}.

For $\eE = \tuple{E_W,\Con_W,\leq_W\;}$ a prime event structure as in \cite{Wi87a},
let the event structure $\fE(\eE) := \tuple{E_W,\turn\;}$ be given by\vspace{-11pt}
$$X \turn Y \mbox {~iff~} \left\{\begin{array}{@{}l@{}}
\mbox{$Y=\{e\}$ and $X = \{d \mid d <_W e\}$} \\
\mbox{or $|Y| \neq 1$, $X=\emptyset$ and $\Con_W(Y)$} \\
\mbox{or $Y$ is infinite and $X=\emptyset$.} \end{array}\right.$$
Now $\fCon = \Con_W$ and $\mathord{\leq} = \mathord{\leq_W}$.

\begin{proposition}{prime to ours}
Let $\eE$ be a prime event structure as in \cite{Wi87a}. Then
$\fE(\eE)$ is rooted, singular, manifestly conjunctive,
pure, $\fS$-irredundant and with finite causes and finite
conflict. Moreover, $\fL(\fE(\eE)) = \fL(\eE) = \fS(\eE)$.
\end{proposition}

\begin{proof} Let $\eE=\tuple{E_W,\Con_W,\leq_W\;}$ be a prime event structure as in
\cite{Wi87a}. As $\Con_W$ is nonempty and subset-closed we have
$\emptyset \in \Con_W$. Thus $\emptyset \turn \emptyset$, i.e.,
$\fE(\eE)$ is rooted. By construction, $\fE(\eE)$ is singular,
manifestly conjunctive, pure,
and with finite causes and finite conflict. By \ob{prime-secured} $\fE(\eE)$ is
$\fS$-irredundant, and \ob{prime} and~\cor{prime-secured} yield $\fL(\fE(\eE)) = \fL(\eE) = \fS(\eE)$.
\end{proof}
For $\eE=\tuple{E,\turn\;}$ a rooted, $\fS$-irredundant event
structure with finite causes, the
structure ${\cal W'}(\eE) \mathbin{:=} \tuple{E,\ConGP',\leq\;}$, where
$\ConGP'(X)$ iff $\fCon(\{d \!\in\! E \mid \exists e \!\in\! X.~ d \leq
e\})$, and $\fCon$ and $\leq$ are given by \df{consistency}, is
a prime event structure in the sense of \cite{Wi87a}.
In particular, by $\fS$-irredundancy, for any $e \in E$
there is an $n\in\IN$ such that there is no chain $e_n \prec \cdots
\prec e_1 \prec e$. As $\eE$ has finite causes, for any $e\in E$ there
are only finitely many $d\in E$ with $d\prec e$; thus the set
\mbox{$\downarrow\!e$} is finite. As \mbox{$\downarrow\!e$} must be part of any
configuration containing $e$, $\fCon(\downarrow\!\!e)$, and hence
$\{e\}\mathbin\in\ConGP'$ for any $e\mathbin\in E$.  As $\fS$-irredundancy implies
cycle-freeness, $\leq$ must be a partial order.

\begin{proposition}{ours to prime}
Let $\eE$ be a rooted, singular, conjunctive, $\fS$-irredundant event
structure with finite causes and finite conflict.  Then 
$\fL({\cal W}'(\eE)) = \fL(\eE) = \fS(\eE)$.
\end{proposition}

\begin{proof}
Trivial, with Obs.~\ref{obs-prime} and~\cor{prime-secured}.
\end{proof}

\subsection[The event structures of Nielsen, Plotkin \& Winskel 1981]{The event structures of \cite{NPW81}}
\label{prime-bc}

These are triples $\eE = \tuple{E,\leq,\#}$ where
\begin{itemise}
\item $E$ is a set of \phrase{events},
\item $\leq \;\subseteq E \times E$ is a partial order,
      the \phrase{causality relation},
\item $\# \subseteq E\times E$ is an irreflexive, symmetric relation,
      the \phrase{conflict relation}, satisfying
      $$\forall d,e,f \in E.~ d \leq e \wedge d \# f \Rightarrow e \# f,$$
      the \phrase{principle of conflict heredity}.
\end{itemise}
The set $L(\eE)$ of configurations of such an event structure
consists of those $X \subseteq E$ which are
\begin{itemise}
\item \phrase{conflict-free}: $\# \cap (X \times X) = \emptyset$,
\item and \phrase{left-closed}: $\forall d,e \in E.~ d \leq e \in X
      \Rightarrow d \in X$,
\end{itemise}
just as in \ob{prime}. In addition, we define $R_f(\eE)$ as
the set of finite configurations in $L(\eE)$.

The prime event structures of \cite{Wi89} are defined likewise, but
additionally requiring
$$\{d \in E \mid d \leq e\} \mbox{ is finite for all } e \in E,$$
the \phrase{principle of finite causes}.

Here we will show that up to $\fL$-equivalence these event structures
are exactly the ones in our sense which are (pure,) rooted, singular,
(manifestly) conjunctive, $\fL$-irredundant, cycle-free and with
binary conflict, and for the ones from \cite{Wi89} also
$\fS$-irredundant and with finite causes.

For $\eE = \tuple{E_{N},\leq_{N},\#_{N}}$ a prime event structure as
in \cite{NPW81}, let the event structure $\fE(\eE) :=
\tuple{E_{N},\turn\;}$ be given by\vspace{-11pt}
$$X \turn Y \mbox {~iff~} \left\{\begin{array}{@{}l@{}}
\mbox{$Y=\{e\}$ and $X = \{d \mid d <_{N} e\}$} \\
\mbox{or $Y \!=\! \{d,e\}$, $d \!\neq\! e$, $X\!=\!\emptyset$ and $\neg (d \#_{N} e)$} \\
\mbox{or $|Y| \neq 1,2$ and $X=\emptyset$.} \end{array}\right.$$
Now $\mathord{\leq} = \mathord{\leq_{N}}$ and $\mathord\# = \mathord{\#_{N}}$.

\begin{proposition}{prime-bc to ours}
Let $\eE$ be an event structure as in \cite{NPW81}. Then $\fE(\eE)$ is
pure, rooted, singular, manifestly conjunctive, $\fL$-irredundant,
cycle-free and with binary conflict. If $\eE$ satisfies the principle
of finite causes then $\fE(\eE)$ is moreover $\fS$-irredundant and
with finite causes. Furthermore, $\fL(\fE(\eE)) = \fL(\eE)$.
\end{proposition}

\begin{proof} Let $\eE=\tuple{E_N,\leq_N,\#_N}$ be an event structure as in
\cite{NPW81}. By construction, $\fE(\eE)$ is pure, rooted, singular,
manifestly conjunctive and with binary conflict.
The relation $\prec$ coincides with $<$, so $\fE(\eE)$ is cycle-free.
With \ob{prime}, $\fL(\fE(\eE)) = \fL(\eE)$, and by \ob{prime-secured}.4,
$\fR_f(\fE(\eE)) = \fR_f(\eE)$.

For every $e \in E_N$, the set $\downarrow\! e \defeq \{d \in E_N \mid d
\leq e\}$ must be conflict-free, using the principle of conflict heredity
and the irreflexivity of $\#$.  Hence, $e \in \mbox{$\downarrow\! e$} \in
L(\eE) = L(\fE(\eE))$.  Therefore $\fE(\eE)$ is $\fL$-irredundant.
In case $\eE$ satisfies the principle of finite causes,
$\fE(\eE)$ has finite causes, and $e \in\;\downarrow\! e \in R_f(\eE)
= R_f(\fE(\eE)) \subseteq S(\fE(\eE))$. In this case $\fE(\eE)$ is
even $\fS$-irredundant.
\end{proof}
For $\eE=\tuple{E,\turn\;}$ an $\fL$-irredundant, cycle-free event
structure, the structure ${\cal W}_{\it NP}(\eE) := \tuple{E,\leq,\#_h\,}$,
where $d \#_h e$ iff $\exists d' \leq d.~\exists e' \leq e.~d'\#e'$,
and $\leq$ and $\#$ are given by \df{consistency}, is clearly an event
structure in the sense of \cite{NPW81}.  In particular, $\leq$ is a partial
order since $\eE$ is cycle-free, and $\#_h$ is irreflexive since if
$e\#_h e$ then $e$ could not occur in any configuration, contradicting
$\fL$-irredundancy.  In case $\eE$ is moreover $\fS$-irredundant and
with finite causes, then, by the argument in the previous section, the
sets $\downarrow\!e$ have to be finite. In this case ${\cal W}_{\it NP}(\eE)$
is a prime event structure as in \cite{Wi89}.

\begin{proposition}{ours to prime-bc}
Let $\eE$ be a rooted, singular, conjunctive, $\fL$-irredundant
and cycle-free event structure with binary conflict.  Then
$\fL({\cal W}_{\it NP}(\eE)) = \fL(\eE)$.
\end{proposition}

\begin{proof}
Trivial, with \ob{prime}. 
\end{proof}
If $\eE$ is moreover $\fS$-irredundant, then $\fS(\eE)=\fL(\eE)$, by
\cor{prime-secured}.
This does not extend to the structures corresponding to the event
structures of \cite{NPW81} however:

\begin{example}{LnotS}
Let $\eE$ be given by $E=\{e_0, e_1, ...\} \cup
\{e_\infty\}$, $\#\mathbin=\emptyset$ and $e_i \mathbin< e_j$ iff $i \mathbin< j$.
Then $E \mathbin\in L(\eE)$ but $E \mathbin{\not\in} S(\eE)$.
\end{example}

\subsection{Summary and remarks}

The left-closed configurations of an event structure generalise the
left-closed and conflict-free subsets of events considered in {\sc
Nielsen, Plotkin \& Winskel} \cite{NPW81}, as well as the families of
configurations of prime event structures as considered in {\sc
Winskel} \cite{Wi87a,Wi89}.  The secured configurations generalise the
families of configurations of event structures (prime and otherwise)
considered in \cite{Wi87a,Wi89}. The families of configurations of
such event structures are completely determined by their finite
reachable configurations.

As indicated in Table~\ref{7 classes}, for each of the seven classes
of event structures proposed in \cite{NPW81,Wi87a,Wi89} a
corresponding subclass of our event structures has been defined,
together with event and configuration preserving translations in both
directions. Upon defining left-closed configurations and a transition
relation on the event structures of \cite{NPW81,Wi87a,Wi89}, these
translations even preserve transition equivalence.

For the event structures in our sense corresponding to the prime event
structures of \cite{Wi87a,Wi89}, the requirements of
$\fS$-irredundancy and having finite causes can be replaced by the
requirement of \phrase{$\fR_f$-irredundancy\/}: any event should occur
in a finite reachable configuration.

Preserving finitary equivalence---that is, preserving events and finite
configurations---any event structure can be converted
into one with finite causes and finite conflict, namely by adding all
enablings $\emptyset \turn Y$ with $Y$ infinite, and omitting the
enablings $X \turn Y$ with $X$ infinite. This procedure preserves the
other properties of \df{event-properties}, except $\fS$- and
$\fL$-irredundancy. It also preserves $\fR_f$-irredundancy.
Hence, up to finitary equivalence the first 6
correspondences of Table~\ref{7 classes} hold without
finite causes and finite conflict, and using
$\fR_f$-irredundancy instead of $\fS$-irredundancy.

Any event structure $\eE=\tuple{E,\turn\;}$ can be converted into an
$\fS$-irredundant structure, namely by omitting from $E$ all events
that do not occur in any secured configuration, and omitting from
$\turn$ any enablings $X \turn Y$ in which such events occur in $X$ or
$Y$. This clearly preserves $S(\eE)$, as well as the properties
rootedness, singularity, (local) conjunctivity, cycle-freeness and
having finite causes and finite or binary conflict. Thus, up to having
the same secured configurations, the prime event structures of
\cite{Wi87a} (resp.\ \cite{Wi89}) even correspond to the class of our
event structures that are rooted, singular, conjunctive and with
finite causes and finite (resp.\ binary) conflict, i.e., not requiring
$\fS$-irredundancy.  However, it should be noted that this
correspondence does not hold up to $\fS$-equivalence, as the set of
events is not preserved.
The same can be said for $\fL$ and $\fR_f$-irredundancy.

\out{
 Finally, also preserving $\fR_f$-equivalence, every event structure can
 be converted into a finitary one, namely by omitting all enablings of
 the form $X \turn Y$ with $Y$ infinite. This conversion preserves the
 properties of \df{event-properties}, except $\fL$-irredundancy and, of
 course, finite conflict; it preserves binary conflict only in the form
 $$2<|X|<\infty \Rightarrow \emptyset \turn X.$$
}


\section{Comparing Models}\label{ComparingModels}

Having seen the general correspondences between our various models of
computation---event structures, configuration structures,
propositional theories and Petri nets---we now trace the relationships
for various natural subclasses; we are guided in our choice of these
subclasses by the concepts isolated in our exploration of\linebreak[3]
previous notions of event structure in the last section. 

In Sections~\ref{EvsC},~\ref{EvsC-secured} and~\ref{finitary comparisons}
we first of all give properties of configuration structures
corresponding to those of event structures.  We then tackle the
converse \phrase{completeness} problem for collections of properties:
given a configuration structure with a collection of these properties,
is there an event structure satisfying the corresponding properties
which yields the given configuration structure?
Following our general point of view, we
understand the configuration structures to provide our (semantic)
model of behaviour. So we are content to consider the map from event
structures to configuration structures for each of the various
classes, and show that it is onto; we do not seek such properties
of a map or maps in the converse direction.
As map from event structures to configuration structures we take $\fL$
in Section~\ref{EvsC}, $\fS$ in Section~\ref{EvsC-secured} (but only
covering secure event structures) and $\fF\circ\fL$
and $\fR_f$ in Section~\ref{finitary comparisons}.

In Section~\ref{theories} we provide corresponding classes of
propositional theories, described according to the syntactic form of
the allowed formulae.  Finally, in Section~\ref{Tie-nets} we tie in
corresponding classes of Petri nets.

\subsection{Event vs.\ configuration structures}\label{EvsC}

\begin{table}[tb]
\begin{tabular}{@{}|@{~}l@{\,}|@{~}l@{~}|@{~}l@{\,}|@{}}
\hline
Event           & Configuration                 & Propositional         \\
structures      & structures                    & theories              \\
\hline\hline
rooted          & rooted                        & ($>\!0$, any)		\\
singular        & closed under $\bbigcup$       & (1, any), (any, 0)    \\
conjunctive     & closed under $\nbigcap$       & (any, $\leq\!1$)        \\
locally conj.   & closed under $\nbbigcap$      & (any, ddc)            \\
finite conflict & finite conflict               & (finite, any)         \\
binary conflict & binary conflict               & ($\leq\!2$, any)        \\
\hline
\raisebox{0pt}[13pt]{sing}.\ \& fin.\ con.& closed under $\fbbigcup$  & (1, any), (fin., 0)   \\
sing.\ \& bin.\ con.& closed under $\bbbigcup$  & (1, any), ($\leq\!2$, 0)\\
loc.\ conj.\ \& f.c.& closed under $\nbbigcap^f$& (finite, fddc)         \\
loc.\ conj.\ \& b.c.& closed under $\nbbigcap^2$& ($\leq\!2$, bddc)       \\
\hline
\end{tabular}
\caption{Corresponding properties\label{correspondence}}
\end{table}

Table~\ref{correspondence} gives the various corresponding properties. We 
\hyperref[df-event-properties]{have already defined} all those we need for
event structures. For configuration structures we first need some
notions of consistency.

\begin{definition}{consistency-CS}
Let $\eC=\tuple{E,C}$ be a configuration structure. A set of events $X
\subseteq E$ is \phrase{consistent}%
, written $\Cn{X}$, if $\exists z\!\in\!C.~ X \subseteq z$.\\
Further, $X$ is \phrase{finitely consistent}, written $\fCn{X}$, if
$$\forall Y\!\subseteq_{\it fin}\!X.~ \Cn{Y}$$
and \phrase{pairwise consistent}, written $\bCn{X}$, if
$$\forall Y\!\subseteq\!X.~ (|Y|\!\leq\!2 \implies \Cn{Y}).$$
\end{definition}
Now we can define the corresponding properties used in the table.
\begin{definition} {properties}
Let $\eC=\tuple{E,C}$ be a configuration structure. Then:
\begin{enumerate}
\item C is said to be \phrase{consistently complete} \cite{Plo78} or
closed under \phrase{bounded unions} ($\bbigcup$) if
$$A \subseteq C \wedge \Cn{\bigcup A} \Rightarrow \bigcup A \in C$$
\item C is said to be closed under \phrase{nonempty intersections} ($\nbigcap$) 
if
$$\emptyset \neq A \subseteq C \Rightarrow \bigcap A \in C$$
\item C is said to be closed under \phrase{bounded nonempty intersections} 
($\nbbigcap$) if
$$\emptyset \neq A \subseteq C \wedge \Cn{\bigcup A} \Rightarrow \bigcap A \in C$$
\item C has \phrase{finite conflict} if
$$[\forall Y \!\!\subseteq\! X.~(Y \mbox{ finite} \Rightarrow \exists z
\!\in\! C.~ Y \!\!\subseteq\! z \!\subseteq\! X)] \Rightarrow X \!\!\in\! C$$
\item C has \phrase{binary conflict} if
$$[\forall Y \!\!\subseteq\! X.~(|Y| \leq 2 \Rightarrow \exists z
\!\in\! C.~ Y \!\!\subseteq\! z \!\subseteq\! X)] \Rightarrow X \!\!\in\! C$$
\item C is said to be closed under \phrase{finitely consistent unions} 
($\fbbigcup$) if 
$$A \subseteq C \wedge \fCn{\bigcup A} \Rightarrow \bigcup A \in C$$
\item C is said to be closed under \phrase{pairwise consistent unions} 
($\bbbigcup$) if
$$A \subseteq C \wedge \bCn{\bigcup A} \Rightarrow \bigcup A \in C$$
\item C is said to be closed under \phrase{finitely consistent nonempty 
intersections}\ ($\nbbigcap^f$) if
$$\emptyset \neq A \subseteq C \wedge \fCn{\bigcup A} \Rightarrow \bigcap A \in C$$
\item C is said to be closed under \phrase{pairwise consistent nonempty 
intersections}\ ($\nbbigcap^2$) if
$$\emptyset \neq A \subseteq C \wedge \bCn{\bigcup A} \Rightarrow \bigcap A \in C.$$
\end{enumerate}
 \end{definition}
By their definition, these notions are related as follows:\vspace{-1ex}
$$\begin{array}{ccccc}
\bbbigcup\mbox{-closed}&\implies&\fbbigcup\mbox{-closed}&\implies&
\bbigcup\mbox{-closed}\\
\Downarrow && \Downarrow \\
\mbox{binary conflict} &\implies&\mbox{finite conflict}
\end{array}$$
$\hfill\begin{array}[b]{r@{~~~~~~}l}
\bbbigcup\mbox{- and } \nbbigcap\mbox{-closed}
& \fbbigcup\mbox{- and } \nbbigcap\mbox{-closed}\\
\Downarrow~~~&~~~~~~\Downarrow
\end{array}\hfill$\vspace{-1ex}
$\nbigcap\mbox{-closed}   \hfill\implies\hfill
\nbbigcap^2\mbox{-closed} \hfill\implies\hfill
\nbbigcap^f\mbox{-closed} \hfill\implies\hfill
\nbbigcap\mbox{-closed.}$
\vspace{2mm}
\\
We can illustrate these properties with the aid of previously given
examples.
The configuration structure $G$ from \ex{absence} has all properties
of \df{properties}, and indeed its event structure representation has
all the corresponding properties of Table~\ref{correspondence}.

The configuration structure of \ex{resolvable} fails to be
closed under bounded unions, for there is no configuration $\{a\}\cup
\{b\}$, even though its superset $\{a,b,c\}$ is a configuration.
Indeed, the corresponding event structure is not singular.

The configuration structure of \ex{disjunctive} fails to be closed
under bounded nonempty intersections, for there is no configuration $\{c\}$.
Indeed its associated event structure is not locally conjunctive.
The modified event structure of \ex{disjunctive} \emph{is} locally
conjunctive, although not conjunctive. Its associated configuration
structure is closed under bounded nonempty intersections, but not
under (general) nonempty intersections.

The configuration structure of \ex{insecure} fails to have finite
conflict, whereas the event structure from \ex{ternary conflict} has
finite conflict but fails to have binary conflict.

By combining these examples it is not hard to show that for each
selection from the first five properties, respecting the implications
above, there exists a configuration structure with the selected
properties and none of the others.

The first three conditions above are particularly natural as they are
(essentially) couched in terms of the lattice-theoretic structure
the configuration structure inherits from that of the powerset lattice of all events.
The following are natural replacements of this kind for the remaining
six conditions.

\begin{definition}{compatibility}
Let $\eC=\tuple{E,C}$ be a configuration structure.
\begin{enumerate}
\item[4$'$.] C is said to be closed under \phrase{directed unions}
($\dbigcup$) if for every nonempty family
$A$ of configurations:
$$[\forall x,y\!\in\! A.~\exists z\!\in\!A.~x\cup y \subseteq z]\Rightarrow
\bigcup A \in C$$
\item[5$'$.] C is said to be \phrase{weakly coherent} 
iff for every family 
$A \subseteq C$ of configurations:
$$[\forall x,y\!\in\! A.~\exists z\!\in\!C.~x\cup y \subseteq z
\subseteq \bigcup A]\Rightarrow \bigcup A \in C$$
\item[6$'$.] C is said to be \phrase{finitely complete} \cite{Wi87a} or closed
under \phrase{finitely compatible unions} ($\bbigcup^{\it fc}$) if
$$\vspace{-2ex} A \subseteq C \wedge \forall F\subseteq_{\it fin}\!A.~\Cn
{\bigcup F} \Rightarrow \bigcup A \in C$$ 
\item[7$'$.] C is said to be \phrase{coherent} \cite{Plo78,Wi89} or closed
under \phrase{pairwise compatible unions} ($\bbigcup^{2c}$) if
$$\vspace{-2ex} A \subseteq C \wedge \forall x,y\!\in\!A.~\Cn
{x\cup y} \Rightarrow \bigcup A \in C$$ 
\item[8$'$.] C is said to be closed under \phrase{finitely compatible nonempty
intersections} ($\nbbigcap^{\it fc}$) if
$$\vspace{-2ex} \emptyset\neq A \subseteq C \wedge \forall
F\subseteq_{\it fin}\!A.~\Cn{\bigcup F} \Rightarrow \bigcap A \in C$$ 
\item[9$'$.] C is said to be closed under \phrase{pairwise compatible nonempty
intersections} ($\nbbigcap^{2c}$) if
$$\vspace{-2ex} \emptyset\neq A \subseteq C \wedge \forall
x,y\!\in\!A.~\Cn{x\cup y} \Rightarrow \bigcap A \in C.$$
\end{enumerate}
\end{definition}
In all six cases the property of \df{compatibility} is strictly weaker
than the corresponding one of \df{properties}, except that (weak)
coherence also implies rootedness. Strictness is illustrated by
the configuration structure consisting of $\emptyset$ and the
co-singleton sets of natural numbers.
However, in all six cases (trivially in the last two) the two
properties coincide for those configuration structures closed under
non-empty intersections.

\begin{proposition}{directed-unions} Let $\eC = \tuple{E,C}$ be a
configuration structure that is closed under $\nbigcap$. Then
\begin{itemise}
\item $\eC$ is closed under $\dbigcup$ iff it has finite conflict,
\item $\eC$ is weakly coherent iff it is rooted and has binary conflict,
\item $\eC$ is closed under $\bbigcup^{\it fc}$ iff it is closed
under $\bbigcup^f$, and
\item $\eC$ is coherent iff it is rooted and closed under $\bbigcup^2$.
\end{itemise}
\end{proposition}

\begin{proof}
We only prove the first and last statement; the other proofs are similar.

Suppose $\eC$ has finite conflict. Let $\emptyset\neq A \subseteq C$ satisfy
$$\forall x,y\!\in\! A.~\exists z\!\in\!A.~x\cup y \subseteq z.$$
Then every finite subset $Y$ of $\bigcup A$ is contained in the union
of a finite subset of $A$ and hence in an element of $A$.
As $\eC$ has finite conflict it follows that $\bigcup A \in C$.

Now suppose $\eC$ is closed under $\dbigcup$ and $\nbigcap$.
Let $X$ be a set of events satisfying
$$\forall Y \!\!\subseteq\! X.~(Y \mbox{finite} \Rightarrow \exists
z\!\in\! C.~ Y \!\!\subseteq\! z\!\subseteq\! X).$$
As $\eC$ is closed under $\nbigcap$, for every finite subset $Y$ of
$X$ there is a least configuration $z_Y \in C$ satisfying $Y\subseteq
z_Y\subseteq X$. Clearly $z_{Y} \cup z_{Y'}\subseteq z_{Y\cup Y'}$.
Hence $X=\bigcup_{Y \subseteq_{{\it fin}}X}z_Y \,  \in \, C$.

Suppose $\eC$ is rooted and closed under $\bbigcup^2$. Let $A$ satisfy
$$A \subseteq C \wedge \forall x,y\!\in\!A.~\Cn{x\cup y}.$$
Then $\Cn{Y}$ for each $Y \!\subseteq\! \bigcup A$ with $|Y|\!\leq\! 2$, so
$\bigcup A \!\in\! C$.

Now suppose $\eC$ is closed under $\bbigcup^{2c}$ and $\nbigcap$.
Taking $A=\emptyset$ in \df{compatibility}.7 we find that $\eC$ is rooted.
Let $$A \subseteq C \wedge \bCn{\bigcup A}.$$
As $\eC$ is closed under $\nbigcap$, for every $e\!\in\!x\!\in\!A$
there is a least configuration $z_e \!\in\! C$ satisfying $e\in z_e
\subseteq x$. Moreover, for every $d,e\in\bigcup A$ there is a least
\mbox{$z_{d,e}\!\in C$} satisfying $d,e\in z_{d,e}$. Clearly $z_{d}
\cup z_{e}\subseteq z_{d,e}$.
Hence $\bigcup A = \bigcup_{e\in x \in A} z_e \in C$.
\end{proof}

\begin{proposition}{no packaging}
A configuration structure is closed under $\bbigcup^{\it fc}$ iff it
is closed under $\bbigcup$ and $\dbigcup$.  Likewise, it is
coherent iff it is closed under $\bbigcup$ and weakly coherent.
\end{proposition}
\pf
For both claims ``only if'' is straightforward.  So suppose
$\eC=\tuple{E,C}$ is closed under $\bbigcup$ and $\dbigcup$.  Let $$A
\subseteq C \wedge \forall F\subseteq_{\it fin}\!A.~\Cn{\bigcup F}.$$
As $\eC$ is closed under $\bbigcup$ we have $\forall F\subseteq_{\it
fin}\!A.~\bigcup F \in C$. Thus the family consisting of $\bigcup F\in C$
for $F\subseteq_{\it fin}\!A$ is a directed union, and $\bigcup A =
\bigcup_{F\subseteq_{\it fin}A} \bigcup F \in C$.

The last claim follows because in the presence of closure under
$\bbigcup$, both coherence and weak coherence simplify to:\\[1ex]
\mbox{}\hfill$A \subseteq C \wedge \forall x,y\!\in\! A.~ x \cup y \in
C \implies \bigcup A \in C$.\hfill$\Box$\\[1em]
We will now proceed to establish the correspondence between the
properties of event structures in the first column of
Table~\ref{correspondence} and the properties of configuration
structures in the second column.

\begin{theorem}{EtoC} Let $\eE$ be an event structure.
\begin{enumerate}
\item If $\eE$ is singular, then $\fL(\eE)$ is closed under $\bbigcup$.
\item If $\eE$ is conjunctive, then $\fL(\eE)$ is closed under $\nbigcap$.
\item If $\eE$ is locally conjunctive, then $\fL(\eE)$ is closed under
$\nbbigcap$.
\item If $\eE$ has finite conflict, then so does $\fL(\eE)$.
\item If $\eE$ has binary conflict, then so does $\fL(\eE)$.
\item If $\eE$ is singular and with finite conflict, then $\fL(\eE)$ is
closed under $\fbbigcup$.
\item If $\eE$ is singular and with binary conflict, then $\fL(\eE)$ is
closed under $\bbbigcup$.
\item If $\eE$ is locally conjunctive and with finite conflict, then
$\fL(\eE)$ is closed under $\nbbigcap^f$.
\item If $\eE$ is locally conjunctive and with binary conflict, then
$\fL(\eE)$ is closed under $\nbbigcap^2$.
\end{enumerate}
\end{theorem}
\begin{proof} The details are routine and are omitted.
\end{proof}
In the next theorem we will show that none of the nine properties of
configuration structures that figure in \thm{EtoC} can be strengthened.

Something unexpected arises in the last four statements of \thm{EtoC}:
the conjunction of two properties of an event structure gives rise to
a property of configuration structures which does not follow
from the properties associated to the two event structure properties
separately. The following example illustrates this.
\begin{example}{strange}
Consider the configuration structure\\  $\eC=\tuple{E,C}$ where\vspace{-1ex}
 $$E := \{a_i \mid i \geq 1\}  \cup \{b,c\} \cup  \{d_i \mid i \geq 1\} \vspace{-1ex}$$
 and where $C$ contains the sets:\vspace{-1ex}
 $$\emptyset,~ \{a_i \mid i \geq 1\} \cup \{b\},~ \{a_i \mid i \geq 1\}
 \cup \{c\}\vspace{-1ex}$$
 and, for all $n\geq 1$,\vspace{-1ex}
$$ \{a_1, \ldots, a_{n},d_n,b,c \}.\vspace{-1ex}$$
Then $\eC$ is rooted and closed under $\bbigcup$ and $\nbbigcap$, and
has finite and binary conflict. But it is not closed under either
$\fbbigcup$ or $\bbbigcup$ or $\nbbigcap^f$or $\nbbigcap^2$.
\end{example}
We would therefore not, for example, expect to recognise a configuration
structure closed under $\bbigcup$ and with finite conflict as
the configuration structure of a singular event structure with finite conflict.
For that we should also require the configuration structure to be closed under
\raisebox{-1.5pt}[0pt]{$\fbbigcup$}.

So it is natural to define a notion of \phrase{package} of properties of
configuration structures with the intention that packages are the collections of
properties for which corresponding event structures are expected to exist.
We call a set of properties from the second column of
Table~\ref{correspondence} a {package }if
\begin{itemise2}
\itemsep 0pt
\item it contains the property
``closed under $\fbbigcup$'' iff it contains the properties
``closed under $\bbigcup$'' and ``having finite conflict'',
\item it contains the property
``closed under $\bbbigcup$'' iff  it contains the properties
``closed under $\bbigcup$'' and ``having binary conflict'',
\item it contains the property
``closed under $\nbbigcap^f$'' iff it contains the properties
``closed under $\nbbigcap$'' and ``having finite conflict'', and
\item it contains the property
``closed under $\nbbigcap^2$'' iff it contains the properties
``closed under $\nbbigcap$'' and ``having binary conflict.''
\vspace{2pt}
\end{itemise2}

By Propositions~\ref{pr-directed-unions} and~\ref{pr-no packaging}, the phenomenon of \ex{strange}
does not apply to event structures closed under $\nbigcap$.  When
restricting attention to those, packaging would not be needed.

\begin{theorem}{CtoE}
A configuration structure C has any package of properties from the
second column of Table~\ref{correspondence} iff there is a (pure) event
structure E with the corresponding properties such that $\fL(\eE) = \eC$.
\end{theorem}

\begin{proof}
Let $\eC = \tuple{E,C}$ be a configuration structure.  Define $\eE :=
\tuple{E,\turn\;}$ by $X \turn Y$ iff $X \cap Y = \emptyset \wedge X
\cup Y \!\in\! C$. Thus $\eE=\fE(\eC)$. It is straightforward to check that
\begin{itemise2}
\item E is always pure,
\item if C is rooted, then so is $\eE$,
\item if C is closed under $\nbigcap$ then E is conjunctive,
\item and if C is \plat{\nbbigcap}-closed then E is locally
conjunctive.
\end{itemise2}
We show that $\fL(\eE) = \eC$. Suppose $x \in C$. For any $Y \subseteq x$
take $Z := x-Y$. Then $Z \subseteq x$ and $Z \turn Y$. So $x \in L(\eE)$.
Conversely, suppose $x \in L(\eE)$. Then there is a $Z \subseteq x$
such that $Z \turn x$. (In fact, $Z$ must be $\emptyset$.) By
construction, $x = Z \cup x \in C$.

Next let C have finite conflict.  Let $\eE := \tuple{E,\turn \!\cup
\turn^\omega\:}$
with $\turn$ defined as before, and $X \turn^\omega Y$ iff
$X =\emptyset$ and $Y$ infinite.
It is straightforward to check that
\begin{itemise2}
\item E is always pure and with finite conflict,
\item if C is rooted, then so is $\eE$,
\item if C is closed under $\nbigcap$ then E is conjunctive,
\item and if C is \plat{\nbbigcap^f}-closed then E is locally
conjunctive.
\end{itemise2}
We show that $\fL(\eE) = \eC$. That $C \subseteq L(\eE)$ goes exactly
as in the previous case, so suppose $x \in L(\eE)$. For any finite
$Y \subseteq x$ there must be a $Z \subseteq x$ with $Z \turn Y$.
By construction, $Z \cup Y \in C$. As $Y \subseteq Z \cup Y \subseteq
x$, and C has finite conflict, we have $x \in C$.

The case that C has binary conflict goes similarly.

Now assume C is closed under bounded unions
(\plat{\bbigcup}). Let $\eE := \tuple{E,\turn_1 \cup \turn_2\;}$ with
\begin{center}
\begin{tabular}{@{}l@{}}
$X \turn_1 Y$ iff $|Y| = 1$, $X \cap Y = \emptyset$ and $X \cup Y
\in C$,\\
$X \turn_2 Y$ iff $X = \emptyset$, $|Y| \neq 1$ and $\Cn{Y}$.
\end{tabular}
\end{center}
It is straightforward to check that
\begin{itemise2}
\item E is always pure and singular,
\item if C is rooted, then so is $\eE$,
\item if C is closed under $\nbigcap$ then E is conjunctive,
\item and if C is \plat{\nbbigcap}-closed then E is locally
conjunctive.
\end{itemise2}
We show that $\fL(\eE) = \eC$. Suppose $x \in C$. For any $Y \subseteq x$
take $Z := x-Y$ if $|Y|=1$ and $Z:=\emptyset$ otherwise. Then $Z
\subseteq x$ and $Z \turn Y$. So $x \in L(\eE)$.
Conversely, suppose $x \in L(\eE)$. Then there is a $Z \subseteq x$
such that $Z \turn x$. In case $|x|=1$ we have $x = Z \cup x \in
C$. In case $|x| \neq 1$ it must be that $Z=\emptyset$ and $\Cn{x}$.
Moreover, for any $e \in x$ there is a $Z_e \subseteq x$ such that
$Z_e \turn \{e\}$. By construction, $Z_e \cup \{e\} \in C$. As
$\bigcup_{e \in x}(Z_e \cup \{e\}) = x$ and $\Cn{x}$, and C is closed
under bounded unions, $x \in C$.

Next assume C is closed under $\bbbigcup$.\\
Let $\eE := \tuple{E,\turn_1 \cup \turn_2 \cup \turn_3\;}$ with
\begin{center}
\begin{tabular}{@{}l@{}}
$X \turn_1 Y$ iff $|Y| = 1$, $X \cap Y = \emptyset$ and $X \cup Y
\in C$,\\
$X \turn_2 Y$ iff $X = \emptyset$, ($|Y| = 0$ or $|Y| = 2$) and $\Cn{Y}$,\\
$X \turn_3 Y$ iff $X = \emptyset$ and $|Y| > 2$.
\end{tabular}
\end{center}
It is straightforward to check that
\begin{itemise2}
\item E is always pure, singular and with binary conflict,
\item if C is rooted, then so is $\eE$,
\item if C is closed under $\nbigcap$ then E is conjunctive,
\item and if C is \plat{\nbbigcap^2}-closed then E is locally
conjunctive.
\end{itemise2}
We show that $\fL(\eE) = \eC$. That $C \subseteq L(\eE)$ goes exactly
as in the previous case, so suppose $x \in L(\eE)$. In case $|x| = 1$
there again is a $Z \subseteq x$ such that $Z \turn x$, and we have $x = Z
\cup x \in C$. So suppose $|x| \neq 1$. For every $Y \subseteq x$ with
$|Y|=0$ or $|Y|=2$ there is a $Z$ with $Z \turn Y$. It must be that
$Z=\emptyset$ and $\Cn{Y}$. Hence $\bCn{x}$.  Moreover, for any $e
\in x$ there is a $Z_e \subseteq x$ such that $Z_e \turn \{e\}$. By
construction, $Z_e \cup \{e\} \in C$. As $\bigcup_{e \in x}(Z_e \cup
\{e\}) = x$ and $\bCn{x}$, and C is closed under pairwise consistent
unions, $x \in C$.

The case that C is closed under $\fbbigcup\!$ goes likewise.
\end{proof}
A noteworthy consequence of this theorem is that every event structure 
with a given collection of properties from the first column of
Table~\ref{correspondence} is $\fL$-equivalent to a pure one with the
same set of properties.

The property $\fL$-irredundancy of event structures is defined in
terms of associated configuration structures: call a configuration
structure \phrase{irredundant} if every event occurs in a configuration,
then an event structure $\eE$ is $\fL$-irredundant iff $\fL(\eE)$ is
irredundant. Thus Theorems~\ref{th-EtoC} and~\ref{th-CtoE} can be
trivially upgraded by adding $\fL$-irredundancy and irredundancy
to the table.

Likewise, call a configuration structure \phrase{$\fS$-ir\-re\-dun\-dant}
if every event occurs in a secured configuration. Using that $\fS(\eE)
\subseteq \fS(\fL(\eE))$, even for impure event structures $\eE$,
whenever $\eE$ is an $\fS$-irredundant event structure then $\fL(\eE)$
is an $\fS$-irredundant configuration structure.  Conversely, if $\eC$
is an $\fS$-ir\-re\-dun\-dant configuration structure, then any pure
event structure $\eE$ with $\fL(\eE)=\eC$ is $\fS$-irredundant. Thus
Theorems~\ref{th-EtoC} and~\ref{th-CtoE} can be upgraded by adding
$\fS$-irredundancy to the first two columns of the table.

Cycle-freeness, as defined in Section~\ref{brands}, is a meaningful
concept only for singular conjunctive $\fL$-irredundant event
structures; there it matches the concept of coincidence-freeness on
configuration structures.

\begin{definition}{coincidence-freeness}
A configuration structure is \phrase{co\-in\-ci\-dence-free} if for every
two distinct events occurring in a configuration there is a
subconfiguration containing one but not the other.
\end{definition}

\begin{proposition}{coincidence-freeness}
A singular, conjunctive, $\fL$-ir\-re\-dun\-dant event structure
$\eE\!=\!\tuple{E,C}$ is cycle-free iff $\fL(\eE)$ is coincidence-free.
\end{proposition}

\begin{proof}
For all $e$ in $E$, the set $\downarrow(e) := \{d\in E \mid d \leq e\}$
is the least left-closed configuration of $\eE$ containing $e$.  This
implies that a failure of coincidence-freeness in $\fL(\eE)$ occurs if
and only if there are two distinct events $d$ and $e$ with
$d \leq e \leq d$, i.e., in case of a cycle in $\eE$.
\end{proof}
We can now characterise the configuration structures associated to the
event structures of \cite{NPW81}.

\begin{corollary}{characterisation-prime-binary}
A configuration structure arises as the family of configurations of an
event structure of \cite{NPW81} iff it is rooted, irredundant,
coincidence-free and closed under $\bbbigcup$ and $\nbigcap$,
or, equivalently, irredundant, coincidence-free,
coherent and closed under $\nbigcap$.
\hfill $\Box$
\end{corollary}
Although \cite{NPW81} contains a characterisation of the
\emph{domains} induced by families of configurations of event
structures of \cite{NPW81}, ordered by inclusion, a characterisation as
above seems not to have appeared before.

The (secured) configuration structures that arise as the families of
configurations of the various event structure of {\sc Winskel}
\cite{Wi87a,Wi89} will be characterised in Section~\ref{EvsC-secured}.
We do not have a characterisation of the left-closed configuration
structures associated to event structures with finite causes, and
consequently no characterisation of the left-closed configuration
structures associated to the general and stable event event structures
of \cite{Wi87a,Wi89}.

\subsection{Propositional theories}\label{theories}

\begin{figure}
\input{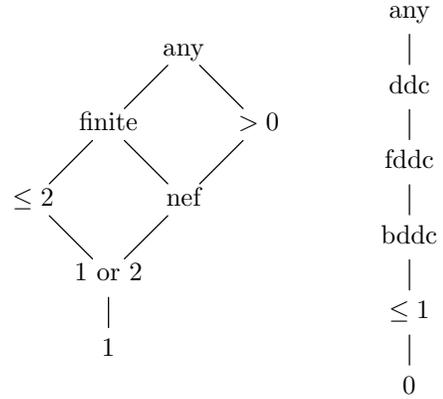}
\mbox{}\vspace{-1.4em}\\
\vspace{-1em}%
\centerline{\raise 1em\box\graph}
\caption{Form Lattices \label{form-lattices}}
\end{figure}

We now consider a variety of \phrase{forms} of formulae, written as $(L,R)$
where $L$ is taken from the lattice on the left of Figure~\ref{form-lattices} and 
$R$ is taken from
the lattice on the right.
Other than the case where $R$ is ``bddc'' these formulae are always
implications, and then they are always clauses except when $R$ is
``ddc'' or ``fddc''. If they are clauses then $L$ and $R$ indicate in
an evident way how many variables there are on each side of the
implication; for example the form $(\mbox{any}, \leq\!1)$ indicates a
clause $X \Rightarrow Y$ such that $Y$ has at most one element and
with no restriction on $X$. In the left-hand lattice, ``nef'' stands
for ``finite and non-empty''.

Formulae of the form $(L,\mbox{ddc})$ 
are implications where the hypothesis is a conjunction of 
 variables whose size is specified by $L$ and whose conclusion is a formula in
``ddc'' form, a \phrase{disjoint disjunction of clauses},
viz.\ a formula \plat{\bigdvee_{j\in J}\bigwedge Y_{j}} where the
$Y_{j}$ are sets of variables, and we write \plat{\bigdvee\Phi} for
$(\bigvee \Phi) \wedge \bigwedge \{ \neg(\varphi \wedge\varphi')
\mid \varphi,\varphi'\mathbin\in \Phi,~ \varphi \neq \varphi'\}$,
the \phrase{disjoint disjunction} of $\Phi$.

Sometimes such inconsistencies are signalled by finitary or even binary means.
Formulae of the form $(L,\mbox{fddc})$ are again 
implications where the hypothesis is a conjunction of 
 variables whose size is specified by $L$ but now the conclusion is a formula in
``fddc'' form, a \phrase{finitely disjoint disjunction of clauses}, viz.\vspace{-1.4ex}
$$(\bigvee_{j\in J}\bigwedge Y_{j})\mbox{ }\wedge \bigwedge_{j,k\in J,~j\neq k} 
         \neg (\bigwedge Z_{j,k} \wedge \bigwedge Z_{k,j})$$
where the $Y_{j}$ are sets of variables and the $Z_{j,k}$ are finite
subsets of $Y_{j}$.
Finally we say that a formula has the
$(L, \mbox{bddc})$ form if it has the form 
$$(\bigwedge X \implies \bigvee_{j\in J}\bigwedge Y_{j})
 \wedge \bigwedge_{j,k\in J,~j\neq k} 
         \neg (e_{j,k} \wedge e_{k,j})$$
where $X$ is a set of variables with size specified by $L$,
the $Y_{j}$ are sets of variables and the $e_{j,k}$ are in $Y_{j}$.

Formulae of any of the above forms are called \phrase{pure} if no
variable occurs at both sides of the implication.

\begin{theorem}{TtoC} Let $\eT$ be a propositional theory
all of whose formulae have one of the forms given in a row of
Table~\ref{correspondence}. Then $\fM(T)$ has the corresponding
property, as given in the table.
\end{theorem}
\begin{proof} We consider only the cases (any, ddc) and\linebreak[3]
(finite, fddc), leaving the others to the reader.  To this end we
first of all show that the collection of models of a family $\Phi$ of
implications of the form \plat{\bigwedge X \implies \bigdvee_{j\in
J}\bigwedge Y_{j}} is closed under bounded non-empty
intersections. Suppose that $\{ m_i\mid i \in I\}$ is a set of models
of $\Phi$, with upper bound $m'$. Let $m$ be the intersection of the
$m_i$; we must show it is a model. To this end, choose one implication
\plat{\bigwedge X \implies\bigdvee_{j\in J}\bigwedge Y_{j}} in $\Phi$ and
suppose that $m$ includes its premise $X$. Then, for $i\in I$, so does
$m_i$ and hence there is a unique $j(i)$ in $J$ such that $Y_{j(i)}
\subseteq m_i$. We claim that, for $i\in I$, all $j(i)$ are the same.
For otherwise $m'$ would contain $X \cup Y_j \cup Y_{k}$ for $j,k \mathbin\in J$,
$j\neq k$, contradicting the fact that $m'$ satisfies \plat{\bigwedge
X \implies\bigdvee_{j\in J}\bigwedge Y_{j}}.  Hence there is a unique
$j$ in $J$ such that $Y_{j}\subseteq m_i$ for all $i\in I$.  So
$Y_{j}\subseteq m$, and this must be the unique $j$ with this property
as $m \subseteq m'$, since $I$ is non-empty.

We deal with the case (finite, fddc) by showing that the set
of models of a family $\Phi$ of formulae of the form $\bigwedge X
\implies (\bigvee_{j\in J}\bigwedge Y_{j})\mbox{ }\wedge
\bigwedge_{j,k\in J,~j\neq k} \neg (\bigwedge Z_{j,k} \wedge \bigwedge
Z_{k,j}),$ with $X$ finite and the $Z_{j,k}$ finite subsets of $Y_j$,
is closed under finitely consistent non-empty intersections.
Suppose that $\{ m_i\mid i \in I\}$ is a set of models of $\Phi$, with
union $m'$ and intersection $m$, such that $\fCn{m'}$.  We must show
that $m$ is a model. To this end, choose one implication
$X\implies (\bigvee_{j\in J}\bigwedge Y_{j})\mbox{ }\wedge
\bigwedge_{j,k\in J,~j\neq k}\neg (\bigwedge Z_{j,k} \wedge \bigwedge Z_{k,j})$
in $\Phi$ and suppose that $m$ includes its premise $X$. Then, for
$i\in I$, so does $m_i$ and hence there is a unique $j(i)$ in $J$ such
that $Y_{j(i)} \subseteq m_i$. We claim that, for $i\in I$, all $j(i)$
are the same.  For otherwise $m'$ would contain the finite set $X \cup
Z_{j,k} \cup Z_{k,j}$ for $j,k \in J$, $j\neq k$. As $\fCn{m'}$ this set
would be included in a model of $\Phi$, which is a contradiction.
Hence there is a unique $j$ in $J$ such that $Y_{j}\subseteq m_i$ for
all $i\in I$.  So $Y_{j}\subseteq m$, and this must be the unique $j$
with this property as $m \subseteq m'$, since $I$ is non-empty.
\end{proof}
For the converse direction we go from event structures to
propositional theories.  Given any event structure $\eE$ satisfying
a collection of properties of Table~\ref{correspondence} we seek to
axiomatise its associated configuration structure $\fL(\eE)$ by
formulae whose form is one of the combinations of the forms found on
the corresponding lines of the table.  
In combining forms $(L_i,R_i)$ $(i \in I)$ into a form $(L,R)$ we
obtain $L$ and $R$ as the meets in the form lattices
given in Figure~\ref{form-lattices}.
For example, for singular conjunctive event
structures the axiomatisation will be by formulae of one
of the two forms $(1,\leq\hspace{-3pt}1)$ and $(\mbox{any},0)$.

\begin{theorem}{EtoT}
Let $\eE$ be a (pure) event structure satisfying any collection of
properties of Table~\ref{correspondence}. Then $\fL(\eE)$ can be
axiomatised by (pure) formulae whose forms are one of the combinations of the
forms found on the corresponding lines of the table.
\end{theorem}

\begin{proof}
We first consider collections of properties not involving
(local) conjunctivity. By \pr{EtoT}, for any event structure $\eE$,
$\fL(\eE)$ can be axiomatised by the set of formulae 
$$\varphi_X := (\bigwedge X \Rightarrow \bigvee_{Y\turn X} \bigwedge Y)$$ 
for $X \subseteq E$;
expanding out $\varphi_X$ to conjunctive normal form yields a set of
clauses $\Phi_X$, and $\bigcup_{X\subseteq E}\Phi_X$ axiomatises $\fL(\eE)$.

If $\eE$ is rooted, i.e., $\emptyset\turn\emptyset$, then $\varphi_{\emptyset}$
is a tautology and $\Phi_{\emptyset}$ is empty. Hence all clauses in
$\bigcup_{X\subseteq E}\Phi_X$ have the form ($>\!0$, any).

If $\eE$ is singular, whenever $Y \turn X$ then either $Y$ is empty or
$X$ is a singleton. If $|X|=1$ then $\Phi_X$ is a set of formulae
of the form (1, any). If $|X|\neq 1$, there either is a relation
$Y \turn X$ or not: if there is then $\varphi_X$ is a tautology and
$\Phi_X$ is empty, and if not we obtain the single clause $(X,\emptyset)$.

If $\eE$ has finite conflict then for any infinite $X$, $\varphi_X$ is
a tautology and $\Phi_X$ empty. For finite $X$, $\Phi_X$ consists of
clauses of the form (finite, any).  The case of binary conflict is
similar.

We now turn to collections of properties including (local) conjunctivity.
Note that conjunctivity implies local conjunctivity.
Let $\eE$ be a locally conjunctive event structure.  One easily sees
that if $Y \turn X$ and $\ConGP(Y \cup X)$ then there is a least $Z
\subseteq Y$ such that $Z \turn X$ and, further, if $Z$ and $W$ are
two minimal sets with $Z\turn X$ and $W\turn X$, then either $Z = W$
or else it is not true that $\ConGP(Z\cup W \cup X)$.  If we now keep
only those pairs $Y \turn X$ such that $\ConGP(Y \cup X)$, and such that $Y$ is a
minimal set with $Y \turn X$, then we obtain an event structure
$\eE' = \tuple{E,\turn'\;}$ with the same collection of configurations
as $\eE$ and such that if $Z \turn' X$ and $W \turn' X$ then either $Z
= W$ or else it is not true that $\ConGP'(Z \cup W \cup X)$.  Further if
$\eE$ was pure, rooted, singular, conjunctive, or with finite or binary
conflict then so is $\eE'$.

First consider the case that conjunctivity is among the considered
properties of $\eE$. Exactly as above we find that $\fL(\eE')$ is
axiomatised by the set of formulae $\varphi'_X := (\bigwedge X \Rightarrow
\bigvee_{Y\turn' X} \bigwedge Y)$ for $X \subseteq E$; and by its
conjunctive normal form $\bigcup_{X\subseteq E}\Phi'_X$.
As $\eE'$ is conjunctive, for each set of events $X$ there is at most
one set $Y$ with $Y\turn' X$. If such a $Y$ exists, $\Phi'_X$ consists
of the formulae $(X,\{e\})$ for $e \in Y$; otherwise it consists of the
single clause $(X,\emptyset)$. Thus all formulae have the form (any,
$\leq\!1$). The arguments for the other properties of $\eE$ are exactly
as before.

We proceed with collections of properties including local
connectivity, but excluding connectivity. We see again that
$\fL(\eE')$ is axiomatised by the set of formulae 
$\varphi'_X := (\bigwedge X \Rightarrow \bigvee_{Y\turn' X} \bigwedge Y)$
for $X \subseteq E$. But since it is false that $\ConGP'(Z \cup W \cup X)$
when $Z \turn' X$, $W \turn' X$ and $Z \neq W$, no configuration can
include $Z \cup W \cup X$. Hence the set of formulae 
$${\dot{\varphi}_X := (\bigwedge X \Rightarrow
\bigvee^{\makebox{\raisebox{-11pt}[0pt][0pt]{\Huge$\cdot$}}}_{Y\turn' X}
\bigwedge Y)}$$
for $X \subseteq E$ axiomatises $\fL(\eE')$, and thus $\fL(\eE)$, as
$\dot\varphi_X$ implies $\varphi_X$ and holds in all interpretations
in $\fL(\eE')$. These formulae have the form (any, ddc).

If $\eE$, and hence $\eE'$, is rooted then $\dot\varphi_{\emptyset}$ is a tautology
and can be omitted. All remaining formulae have the form ($>\!0$, ddc).
 
Suppose next that $\eE$, and hence $\eE'$, is both locally conjunctive
and singular. Then, much as above, for nonsingular $X$,
$\dot{\varphi}_X$ is either a tautology or equivalent to a formula of
the form (any, 0), and for singular $X$ it has the form (1, ddc).

Suppose now that $\eE'$ is both locally conjunctive and with finite
conflict. Then for infinite $X$, $\dot{\varphi}_X$ is a tautology. For
finite $X$ we know that if $Z \turn' X$, $W \turn' X$ and $Z\neq W$
then it is not true that $\ConGP'(Z \cup W \cup X)$. So as $\eE'$ has
finite conflict, it follows that there are finite subsets $Z_1$ and
$W_1$ of, respectively, $Z$ and $W$ such that for no $Y$ is it the
case that $Y \turn' Z_1 \cup W_1 \cup X$. It follows that $Z_1 \cup
W_1 \cup X$ is a subset of no configuration. Since this works for any
such $Z$ and $W$ there is a (finite, fddc) formula that implies
$\dot{\varphi}_X$ and that holds in all interpretations in $\fL(\eE')$,
and so we have the required axiomatisation.

The case where $\eE'$ is locally conjunctive, singular and with finite
conflict is an easy combination of the previous two cases.  When
$\eE'$ is rooted, $\emptyset \turn' \emptyset$ and so we need only then
consider $\dot{\varphi}_X$ for nonempty $X$.

Suppose now that $\eE'$ is both locally conjunctive and with binary
conflict. Then for $X$ with $|X|>2$, $\dot{\varphi}_X$ is a
tautology. For $X$ with $|X|\leq 2$ we know that if $Z \turn'
X$, $W \turn' X$ and $Z\neq W$ then it is not true that $\ConGP'(Z \cup
W \cup X)$. So as $\eE'$ has binary conflict, and $\ConGP'(Z \cup X)$
and $\ConGP'(W \cup X)$, it follows that there are elements $e$ and $e'$
of, respectively, $Z$ and $W$ such that for no $Y$ is it the case that
$Y \turn' \{e,e'\}$. It follows that $ \{e,e'\}$ is a subset of no
configuration. Since this works for any such $Z$ and $W$ there is a
($\leq\!2$, bddc) formula that implies $\dot{\varphi}_X$ and that
holds in all interpretations in $\fL(\eE')$, and so we have the required
axiomatisation.

The cases where $\eE'$ is locally conjunctive and has binary conflict,
and is one or both of singular or rooted are dealt with as before.

Finally we remark that, in the above,  in all cases the 
axiomatisation obtained is pure if $\eE'$ is. 
\end{proof}
An immediate consequence of the above work
(Theorems~\ref{th-TtoC},~\ref{th-CtoE} and~\ref{th-EtoT} and
\pr{directed-unions}) is that a
configuration structure is axiomatisable by formulae of the form 
(finite, $\leq\!1$) iff it is closed under nonempty intersections and directed
unions; this result is essentially due to Larsen and
Winskel~\cite{LW91} as axiomatisations of the form
$(\mbox{finite},\mbox{$\leq\!1$})$ correspond to Scott information
systems. There are two related cases of logical interest:
\phrase{Horn clauses} where there are finitely many
antecedents and one consequent, and \phrase{Scott clauses} where,
more generally, there may be finitely many  consequents~\cite{Ga81,Sc74}.
\begin{proposition}{Scott}
A configuration structure  $\tuple{E,C}$ is Horn clause
axiomatisable iff it  is closed under arbitrary intersections and
directed unions. It is Scott clause axiomatisable  iff $C$ is
closed in the product topology on $2^E$.
\end{proposition}
\begin{proof}
For the implication from left to right in the first statement,
we have just established closure under directed unions and
non-empty intersections. Closure under the empty intersection is
immediate, as $E$ is a model of any set of Horn clauses. For
the converse, we
have an axiomatisation by clauses of the form $X \Rightarrow Y$
where $X$ is finite and $Y$ is empty or a singleton. But the first
case cannot obtain, as here $E$ is a model.

For the second statement, the product topology on $2^E$ is the
$E$-fold power of the discrete topology on the two-point
set. Identifying $2^E$ with ${\cal P}(E)$, we see that the space has
as basis all sets of the form
$${\cal U}_{x,y} = \{m \subseteq E \mid x \subseteq m, (m \cap y)  =\emptyset\}$$
where $x$,$y$ are finite subsets of $E$. The statement now follows,
noting that the complement of ${\cal U}_{x,y}$ is the set of models of
$x \Rightarrow y$.
\end{proof}

\subsection{Secured configuration structures}\label{EvsC-secured}

In Section~\ref{EvsC} we characterised the left-closed configuration
structures associated to various classes of event structures.
Here we do the same for the secured configuration structures of
\hyperref[secure]{secure} event structures.
Our results are indicated in Table~\ref{correspondence secured}.
\begin{table}[htb]
\begin{tabular}{@{}|@{~}l@{\,}|@{~}l@{\,}|@{}}
\hline
Event			& Configuration             \\
structures		& structures                \\
\hline\hline
rooted			& rooted                    \\
singular		& closed under $\bbigcup$   \\
conjunctive		& closed under $\nbigcap$   \\
locally conjunctive	& closed under $\nbbigcap$  \\
finite conflict 	& hyperreachable finite conflict \\
binary conflict		& hyperreachable binary conflict \\
\hline
\raisebox{0pt}[13pt]{singular} \& fin.\ con.	& closed under $\fbbigcup$  \\
singular \& bin.\ con.	& closed under $\bbbigcup$  \\
loc.\ conj.\ \& fin.\ con.& closed under $\nbbigcap^f$\\
loc.\ conj.\ \& bin.\ con.& closed under $\nbbigcap^2$\\
\hline
\end{tabular}
\caption{Corresponding properties, secured case\label{correspondence secured}}
\end{table}
\out{
 The two question marks indicate that we are not sure about the
 implications from left to right, i.e., whether the secured configuration
 structure associated to a (locally) conjunctive event structure $\eE$
 is always closed under (bounded) nonempty intersections. We are sure
 about these implications in case $S(\eE) \subseteq L(\eE)$, which
 is surely the case if $\eE$ has finite conflict.
}

Unlike in \thm{EtoC} it is not always the case that the secured
configuration structure associated to a secure event
structure with finite (resp.\ binary) conflict has finite (resp.\
binary) conflict.

\begin{example}{no finite conflict}
Let $\eE=\tuple{E,\turn\;}$ be given by $$E:=\{a_i, b_i \mid i\in\IN\}
\cup \{c\},$$ $\{a_i\}\turn\{b_i\}$, $\{b_i\}\turn\{a_{i+1}\}$,
$\{b_i\}\turn\{c,a_i\}$ ($i\in\IN$) and $\emptyset \turn X$ for any
$X\subseteq E$ unequal to $\{a_{i+1}\}$, $\{b_i\}$ or $\{c,a_i\}$ ($i\in\IN$).
Then $$R_f(\eE) = \left.\left\{\begin{array}{@{}l@{}}
\{a_i,b_i\mid i<n\}\\
\{a_i,b_i\mid i<n\}\cup\{a_n\}\\
\{a_i,b_i\mid i<n\}\cup\{c\}
\end{array}\;\right|\; n\in\IN \right\},$$
$$S(\eE)=R_f(\eE)\cup\{\{a_i,b_i\mid i\in\IN\}\}$$
and $$L(\eE)=S(\eE)\cup\{E\}.$$
The configuration $E$ is not secured because once $c$ happens only
finitely many of the $a_i$ and $b_i$'s can have happened, and no further
$a_i$ and $b_i$'s can happen, because such an $a_i$ needs to be
preceded by $b_i$ and vice versa. Nevertheless, each finite
subset of $E$ is contained in a secured configuration.
It follows that $\fS(\eE)$ does not have finite (or binary) conflict,
even though $\eE$ does have finite (even binary) conflict.
\end{example}
The event structure of \ex{no finite conflict} is pure, secure, rooted
and conjunctive. By \thm{EtoC-secured}.6 below there can be no
such example with a singular event structure. \ex{no finite conflict}
shows in fact that the requirement of being hyperconnected, which by
\cor{ES analogies secure} holds for configuration structures of the form
$\fS(\eE)$ with $\eE$ secure, can prevent the presence of
configurations required by \df{properties}.4. Hence it is appropriate
to weaken the requirement of \df{properties}.4.

\begin{definition}{conflict closure}
Let $\eC=\tuple{E,C}$ be a configuration structure. Its \phrase{closure
under finite conflict}, $\eC^f:=\tuple{E,C^f}$, is the configuration
structure with the same set of events, and as configurations those
sets $X$ satisfying\vspace{-1.4ex} $$~~~~\forall Y \subseteq X.~ (Y
\mbox{ finite} \implies \exists z\!\in\!C.~ Y \subseteq  z \subseteq  X).$$
Likewise, its \phrase{closure under binary conflict}, $\eC^b := \tuple{E,C^b}$,
has as configurations those sets $X$ satisfying
$$\forall Y \subseteq X.~
(|Y| \leq 2 \implies \exists z\!\in\!C.~ Y \subseteq  z \subseteq  X).$$
\end{definition}
One always has $C \subseteq C^f \subseteq C^b$ (in the first inclusion
take $z$ to be $X$).
Note that $\eC$ has finite conflict iff $\eC = \eC^f$ and
$\eC$ has binary conflict iff $\eC = \eC^b$. Hence the following
appear to be suitable replacements of these notions for hyperconnected
configuration structures.

\begin{definition}{hyperreachable conflict}
A configuration structure $\eC=\tuple{E,C}$ has \phrase{hyperreachable
finite conflict} if $\eC = \fS(\eC^f)$. It has
\phrase{hyperreachable binary conflict} if $\eC = \fS(\eC^b)$.
\end{definition}
In other words, a configuration structure $\eC=\tuple{E,C}$ has
hyperreachable finite (resp.\ binary) conflict iff $X\in C$ exactly
when $X$ can be written as $\bigcup_{i=0}^\infty X_i$ such that
$X_0=\emptyset$ and, for all $i\in\IN$, $X_i \subseteq X_{i+1}$ and
for all $X$ with $X_i \subseteq X \subseteq X_{i+1}$ one has $$\forall
Y\!\subseteq\!X.~Y\mbox{ finite (resp.\ }|Y|\!\leq\!2) \implies \exists
z\!\in\!C.~ Y \subseteq z \subseteq X.$$
We proceed to show that a configuration structure has hyperreachable
finite (resp.\ binary) conflict iff it has the form $\fS(\eC)$ for
$\eC$ a configuration structure with finite (resp.\ binary) conflict.

\begin{lemma}{conflict closure}
Let $\eC$ be a configuration structure.
Then $(\eC^f)^f = \eC^f$ and $(\eC^b)^b = \eC^b$.
\end{lemma}

\begin{proof}
Suppose $X$ is a set of events satisfying
$$\forall Y \subseteq X.~ Y \mbox{ finite} \implies \exists
z\!\in\!C^f.~ Y \subseteq  z \subseteq  X$$
and let $Y \subseteq X$ be finite. Then there is a $z\!\in\!C^f$ with
$Y \subseteq  z \subseteq  X$. Hence there is a $w\!\in\!C$ with
\mbox{$Y \subseteq  w \subseteq z \subseteq  X$}. Thus $X\in C^f$.

That $(\eC^b)^b = \eC^b$ follows likewise.
\end{proof}

\begin{corollary}{conflict closure}
Let $\eC$ be a configuration structure.\\
Then $\eC^f$ has finite conflict and $\eC^b$ binary conflict.
\hfill$\Box$
\end{corollary}

\begin{proposition}{hyperreachable conflict}
A configuration structure has hyperreachable finite conflict iff it
has the form $\fS(\eC)$ for $\eC$ a configuration structure with
finite conflict.

Likewise, a configuration structure has hyperreachable binary conflict
iff it has the form $\fS(\eC)$ for $\eC$ a configuration structure
with binary conflict.
\end{proposition}

\begin{proof}
``Only if'' follows immediately from \df{hyperreachable conflict} and
\cor{conflict closure}. For ``if'' suppose that $\eC=\tuple{E,C}$ has
finite conflict, i.e., $\eC = \eC^f$. By \thm{CtoE} there is a pure
 event structure with finite conflict such that $\eC = \fL(\eE)$.
So $S(\eC)=S(\fL(\eE))=S(\eE)\subseteq L(\eE) = C$
by \pr{pure ES commute} and \lem{finite conflict secure}.
Hence $\eC$ is $\fS\fR$-secure, and \pr{reachable secured} yields
$S(\fS(\eC))=S(\eC)$. Using the monotonicity w.r.t.\
the inclusion ordering of the operators $(\cdot)^f$ and $\fS$ we find
$$S((\fS(\eC))^f)\hspace{-.55pt}\subseteq\hspace{-.55pt} S(\eC^f)
\hspace{-.55pt}=\hspace{-.55pt} S(\eC) \hspace{-.55pt}=\hspace{-.55pt}
S(\fS(\eC)) \hspace{-.55pt}\subseteq\hspace{-.55pt} S((\fS(\eC))^f).$$
Thus $\fS(\eC) = \fS((\fS(\eC))^f)$, i.e.,
$\fS(\eC)$ has hyperreachable finite conflict, which had to be shown.

The second statement is obtained likewise, reading ``binary'' for
``finite'' and $b$ for $f$.
\end{proof}
We are now ready to prove the implications from the left to the right
column of Table~\ref{correspondence secured}.

\begin{theorem}{EtoC-secured}
Let $\eE$ be a secure\footnote{In fact, for Claims 4 and 5 it suffices
to assume that $\eE$ is reachably pure, and for the other claims that
$S(\eE) \subseteq L(\eE)$.} event structure.
\begin{enumerate}\parskip 0pt
\item[0.] If $\eE$ is rooted, then so is $\fS(\eE)$.
\item If $\eE$ is singular, then $\fS(\eE)$ is closed under $\bbigcup$.
\item If $\eE$ is conjunctive, then $\fS(\eE)$ is closed under $\nbigcap$.
\item If $\eE$ is locally conjunctive,
          then $\fS(\eE)$ is closed under $\nbbigcap$.
\item If $\eE$ has finite conflict, then $\fS(\eE)$ has
          hyperreachable finite conflict.
\item If $\eE$ has binary conflict, then $\fS(\eE)$ has
          hyperreachable binary conflict.
\item If $\eE$ is singular and with finite conflict, then
          $\fS(\eE)$ is closed under $\fbbigcup$.
\item If $\eE$ is singular and with binary conflict, then
          $\fS(\eE)$ is closed under $\bbbigcup$.
\item If $\eE$ is locally conjunctive and with finite conflict,
          then $\fS(\eE)$ is closed under $\nbbigcap^f$.
\item If $\eE$ is locally conjunctive and with binary conflict,
          then $\fS(\eE)$ is closed under $\nbbigcap^2$.
\end{enumerate}
\end{theorem}
\begin{trivlist}
\item[\hspace{\labelsep}\bf Proof:]
Claim 0 is immediate from \df{hyperreachable}.
For claims 4 and 5, note that if $\eE$ is reachably pure and with finite (or
binary) conflict, then $\hat\eE$, as constructed in the proof of
\pr{reachably pure}, is pure and with finite (or binary) conflict, and
$\fS(\eE)=\fS(\hat\eE)$. Now the results are immediately from
\thm{EtoC} and Propositions~\ref{pr-pure ES commute} and~\ref{pr-hyperreachable conflict}.
For the remaining claims, let $A \subseteq S(\eE)$.
By ``consistency'' we will mean that $\ConGP(\bigcup A)$.
This follows for Claims 1, 3, 6, 7, 8 and 9 (but not 2) because
$S(\eE) \subseteq L(\eE)$ and either $\Cn{\bigcup A}$, or 
$\fCn{\bigcup A}$ and $\eE$ has finite conflict, or 
$\bCn{\bigcup A}$ and $\eE$ has binary conflict.
Applying \df{reachable ES}, for each $x\in
A$ let $x = \bigcup_{n=0}^\infty x_n$ with $x_0 = \emptyset$ and
$$\forall n\in \IN.~ x_n \subseteq x_{n+1} \wedge \forall Y \subseteq
x_{n+1}.~ \exists Z \subseteq x_{n}.~ Z \turn Y.$$
\begin{list}{-}{\labelwidth\leftmargini\advance\labelwidth-\labelsep
                \topsep 4pt \itemsep 2pt \parsep 2pt}
\item[Ad 1, 6 and 7.~] Let $X_n \defeq \bigcup_{x\in A} x_n$ for $n\in \IN$.
Then \plat{X \defeq \bigcup A = \bigcup_{n=0}^\infty X_n}.
Moreover, $X_0=\emptyset$ and $X_n \subseteq X_{n+1}$ for $n\in \IN$.
Let $e\in X_{n+1}$ for some $n\in\IN$. Then $e\in x_{n+1}$ for
certain $x\in A$. Thus $\exists Z \subseteq x_{n} \subseteq X_n.~ Z
\turn \{e\}$. By consistency and singularity, $\emptyset \turn Y$ for
any $Y\subseteq X$ with $|Y| \neq 1$. Hence $X\in S(\eE)$.
\item[Ad 2, 3, 8 and 9.~] Let $A\neq\emptyset$ and pick a $y$ from $A$.
Let $X_n = y_n \cap \bigcap A$ for $n\!\in\!\IN$. Then $\bigcap A =
\bigcup_{n=0}^\infty X_n$.  Moreover, $X_0=\emptyset$ and \mbox{$X_n
\subseteq X_{n+1}$} for $n\in \IN$.  Now let $Y \subseteq X_{n+1}$ for
some $n\!\in\!\IN$. Then $Y \subseteq y_{n+1}$ and $Y \subseteq x$ for
$x\!\in\!A$. Hence there is a $Z\subseteq y_n$ with $Z\turn Y$. Moreover,
as $\eE$ is secure, for $x\in A$ we have $x \in L(\eE)$, so there must
be a $Z_x \subseteq x$ with $Z_x \turn Y$. Now by the conjunctivity of
$\eE$, or by consistency and the local conjunctivity of $\eE$, we
obtain that $Z\cap\bigcap_{x\in A} Z_x \turn Y$, with
$Z\cap\bigcap_{x\in A} Z_x \subseteq X_n$.  Hence $X\in S(\eE)$.
\hfill$\Box$
\end{list}
\end{trivlist}
By \lem{finite conflict secure}, the security requirement
$S(\eE)\subseteq L(\eE)$ holds trivially in case $\eE$ has finite conflict,
i.e., in Claims 4--9 of \thm{EtoC-secured}; it is not needed for
Claims 0 and 1 and used in the proof of claims 2 and 3. The question
whether this requirement is needed there is open.
The following example shows that Claims 4 and 5 fail for general event
structures that are not reachably pure:

\begin{example}{no hyperreachable conflict}
Let $\eE:=\tuple{\IN,\turn\;}$ be given $\{j\} \turn \{i,j\}$
for $i<j$ and $\emptyset\turn X$ when $|X|\neq 2$. Then
$S(\eE)=R_f(\eE)=\pow_{\it fin}(\IN)$ but
$S((\fS(\eE))^f) = \fL(\eE) = \pow(\IN)$.  The infinite
configurations are not secured, because events can happen only in
decreasing order. Nevertheless, each finite set of events is
(contained in) a secured configuration.  It follows that $\fS(\eE)$
does not have hyperreachable finite (or binary) conflict, even though
$\eE$ does have finite (even binary) conflict.
\end{example}
There does not appear to be an obvious way around this example, as the
above event structure has the same secured configurations as one with
$X\turn Y$ iff $X=\emptyset$ and $Y$ finite, which is a prototypical
example of an otherwise trivial event structure with infinite
conflict. Happily, our goal is to deal with secure
event structures anyway, as those fit the computational interpretation
of configuration structures.

In order to establish the completeness of the characterisations of
Table~\ref{correspondence secured} we first show that some crucial
properties of \df{properties} are preserved under closure under finite
or binary conflict.

\begin{lemma}{closure}
Let $\eC$ be a configuration structure.
\begin{itemise}
\item If $\eC$ is closed under \plat{\bbigcup^f} then so is $\eC^f$.
\item If $\eC$ is closed under \plat{\nbbigcap^f} then so is $\eC^f$.
\item If $\eC$ is closed under \plat{\bbigcup^2} then so is $\eC^b$.
\item If $\eC$ is closed under \plat{\nbbigcap^2} then so is $\eC^b$.
\item If $\eC$ is closed under $\nbigcap$ then so are $\eC^f$ and $\eC^b$.
\item If $\eC$ is rooted then so are $\eC^f$ and $\eC^b$.
\end{itemise}
\end{lemma}

\begin{proof}
Let $\eC=\tuple{E,C}$.  Note that, for each $X \subseteq E$,
$$\forall Y\!\subseteq_{\it fin}\! X.~\exists z\!\in\!C^f\!\!.~Y\!\subseteq\!z
\Leftrightarrow
\forall Y\!\subseteq_{\it fin}\! X.~\exists z\!\in\!C.~Y\!\subseteq\!z,$$
i.e., $X$ is finitely consistent w.r.t.\ $\eC^f$ iff it is
finitely consistent w.r.t.\ $\eC$---hence we can write $\fCn{X}$
without indicating whether it is w.r.t.\ $\eC^f$ or $\eC$.

Suppose first that $\eC$ is closed under $\bbigcup^f$.  Let
$A \subseteq C^f$ be a family of configurations of $\eC^f$ such that
$\fCn{\bigcup A}$.  We wish to show that $\bigcup A \in C^f$. Suppose
that $Y \subseteq_{\it fin} \bigcup A$.  Then $Y$ has the form
$\bigcup_{i = 1}^n Y_i$ for some $n \geq 0$ where $Y_i\subseteq_{\it
fin} x_i$ for some $x_i\in A$ for $i = 1,...,n$. So there are $z_i \in
C$ such that $Y_i\subseteq z_i \subseteq x_i$.  We then have that
$Z\defeq \bigcup_{i=1}^n z_i \subseteq \bigcup A$.  Since
$\fCn{\bigcup A}$ we have $\fCn{Z}$.  As $\eC$ is closed under
\plat{\bbigcup^f} it follows that $Z\in C$.  As we also have that
$Y \subseteq Z \subseteq \bigcup A$, it follows that $\bigcup A$ is a
configuration of $\eC^f$, as required.

Suppose instead that $\eC$ is closed under $\nbbigcap^f$.  Let
$\emptyset \neq A \subseteq C^f$, such that $\fCn{\bigcup A}$.  We wish
to show that $\bigcap A \in C^f$.  Suppose that $Y \subseteq_{\it fin}
\bigcap A$.  Then $Y\subseteq_{\it fin} x$ for each $x\!\in\!A$ and so
there are $z_x\!\in C$ with $Y\subseteq z_x \subseteq x$. Since
$\fCn{\bigcup_{x \in A} x}$ we have $\fCn{\bigcup_{x \in A} z_x}$.
As $\eC$ is closed under $\nbbigcap^f$ it follows that
$z \defeq \bigcap_{x\in A} z_x \in C$.  As moreover
$Y \subseteq z\subseteq X$, it follows that $\bigcap A$ is a
configuration of $\eC^f$, as required.

The claims about binary conflict are proved just like the ones about
finite conflict, and the claims about closure under $\nbigcap$ are
obtained as simplifications of the arguments about
\raisebox{-1pt}[0pt]{$\nbbigcap^f$} above. The claims about rootedness are trivial.
\end{proof}

\begin{theorem}{CtoE-secured}
A hyperconnected configuration structure $\eC$ has any package of
properties from the second column of Table~\ref{correspondence
secured} iff there is a (pure and) secure event structure $\eE$ with
the corresponding properties such that $\fS(\eE) = \eC$.
\end{theorem}

\begin{proof}
``If'' follows from \thm{EtoC-secured} and \cor{ES analogies secure}.
For ``only if'', let \plat{\eC^*\defeq\eC^b} in case the package
contains hyperreachable binary conflict; if that does not apply,
\plat{C^*\defeq\eC^f} in case the package contains hyperreachable
finite conflict, and \plat{\eC^*\defeq\eC} otherwise.  Now
$\fS(\eC^*)=\eC$ and, by \lem{closure} and \cor{conflict closure},
$\eC^*$ has the same package of properties as $\eC$ but dropping the
adjective ``hyperreachable''. Thus, using \thm{CtoE}, there exists a
pure event structure $\eE$ with the corresponding properties such that
$\fL(\eE)=\eC^*$. Using \pr{pure ES commute},
$\fS(\eE)=\fS(\fL(\eE))=\fS(\eC^*)=\eC$. By \pr{hyperconnected
secure}, the event structure $\eE$ is secure.
\end{proof}
Trivially, an event structure $\eE$ is $\fS$-ir\-re\-dun\-dant iff the
configuration structure $\fS(\eE)$ is irredundant; thus
Theorems~\ref{th-EtoC-secured} and~\ref{th-CtoE-secured} can be
upgraded by adding $\fS$-ir\-re\-dun\-dan\-cy and irredundancy to
Table~\ref{correspondence secured}.  $\fL$-ir\-re\-dun\-dan\-cy and
cycle-freeness are not particularly interesting properties when
studying secured configurations. For the property finite causes we
have correspondence results only for singular event structures:

\begin{definition}{axiom of finiteness}
A configuration structure is said to satisfy the \phrase{axiom of
finiteness} \cite{Wi87a,Wi89} if any configuration is the union of its
finite subconfigurations.
\end{definition}

\begin{proposition}{finite causes}
If $\eE$ is a singular event structure with finite causes, then
$\fS(\eE)$ satisfies the axiom of finiteness (and is closed under
$\bbigcup$). Conversely, if $\eC$ is a hyperconnected configuration
structure satisfying the axiom of finiteness and any package of
properties from the second column of Table~\ref{correspondence
secured} including closure under $\bbigcup$, then there is a pure and
secure event structure $\eE$ with finite causes and the corresponding
properties of Table~\ref{correspondence secured}, such that $\fS(\eE) = \eC$.
\end{proposition}

\begin{proof}
The first claim has been established in the first statement of
\pr{infinite secured configurations}, of which direction
``$\Rightarrow$'' only requires singularity and finite causes.

For ``conversely'', first of all note that if $\eC$ satisfies the axiom of
finiteness, then so do $\eC^f$ and $\eC^b$. Now note that in the
proof of \thm{CtoE}, which is called in the proof of \thm{CtoE-secured},
one may replace the definition of $\turn_1$ by
\begin{center}
$X \turn_1 Y$ iff $|Y| = 1$, $X \cap Y = \emptyset$ and $X \cup Y\in F(\eC)$
\end{center}
because any $Y$-event occurring in a configuration occurs in a finite
subconfiguration and whenever $X\turn Y$ all enablings $X'\turn Y$
with $X'\supseteq X$ may be dropped. By construction, the resulting event
structure has finite causes.
\end{proof}
For configuration structures satisfying the axiom of finiteness we can
reformulate the condition of being closed under
\raisebox{-1.5pt}[0pt]{$\bbigcup^f$}.

\begin{proposition}{finite-completeness}
Let $\eC$ be a configuration structure satisfying the axiom of finiteness.
Then $\eC$ is closed under \raisebox{-2pt}[0pt]{$\bbigcup^f$} iff it
is closed under \raisebox{-1.5pt}[0pt]{$\bbigcup^{\it fc}$}.
\end{proposition}
\pf
``Only if'' is trivial, so suppose $\eC=\tuple{E,C}$ is closed under 
$\bbigcup^{\it fc}$. Let $A\subseteq C$ with $\fCn{\bigcup A}$. We
have to show that $\bigcup A \in C$. Let $B$ be the set of all finite
configurations included in members of $A$. Then for all $F
\subseteq_{\it fin} B$ we have that $\bigcup F \subseteq_{\it fin}
\bigcup A$ and hence $\Cn{\bigcup F}$.
By the axiom of finiteness, $\bigcup A = \bigcup B \in C$.%
\\[1em]
Moreover, for configuration structures satisfying the axiom of
finiteness and closed under \raisebox{-1.5pt}[0pt]{$\bbigcup^f$} we
reformulate the condition of being hyperconnected.

\begin{proposition}{coincidence-freeness-hyperconnectedness}
Let $\eC$ be a configuration structure closed under
\raisebox{-1.5pt}[0pt]{$\bbigcup^f$} and satisfying the axiom of
finiteness.  Then $\eC$ is hyperconnected iff it is coincidence-free.
\end{proposition}

\begin{proof}
``Only if'' is trivial, so suppose $\eC=\tuple{E,C}$ is
coincidence-free. Closure under \raisebox{-2pt}[0pt]{$\bbigcup^f$}
immediately implies that $S(\eC)\subseteq C$, so it remains to be
shows that $C\subseteq S(\eC)$.  Let $x \in C$. For any $e\in x$ say
that $e$ can happen at stage $n$ if $n$ is the smallest cardinality of
a subconfiguration of $x$ containing $e$. By the axiom of finiteness,
this cardinality is always finite. Let $X_n$ be the set of all events
in $x$ that can happen at stage $\leq\! n$. Then $X_0 =\emptyset$,
$X_n \subseteq X_{n+1}$ for $n\!\in\!\IN$ and $\bigcup_{n=0}^\infty
X_n = x$.  As $X_n$ is the union of all subconfigurations of $x$ of
size $\leq\!n$ and $\eC$ is closed under $\bbigcup$, we have
$X_n\!\in\! C$ for $n\!\in\!\IN$.\linebreak[3] Let $X_n \subseteq Y
\subseteq X_{n+1}$ for some $n\!\in\!\IN$.  It suffices to show that
$Y \in C$. For any \mbox{$e\in Y-X_n$} pick a subconfiguration $y_e$
of $x$ of $n+1$ elements, containing $e$.  Given that $y_e$ does not
have a proper subconfiguration containing $e$, for any $d\neq e$ in
$y_e$, by coincidence-freeness, there must be subconfiguration $z_d$
of $y_e$ with $d\in z_d \subseteq y_e-\{e\}$, showing that $d\in X_n
\subseteq Y$. It follows that $y_e\subseteq Y$.  Hence $Y = X_n \cup
\bigcup_{e\in Y-X_n} y_e$ and as $\eC$ is closed under \plat{\bbigcup}
we have $Y\in C$.
\end{proof}
We now apply the results of this section to characterise the secured
configuration structures associated to the various event structures of
{\sc Winskel} \cite{Wi87a,Wi89}.

\begin{corollary}{characterisation-Wi87a}
A configuration structure arises as the family of (secured)
configurations of an event structure of \cite{Wi87a} iff it satisfies
the axioms of rootedness, finiteness, coincidence-freeness and
finite-completeness.

A configuration structure arises as the family of (secured)
configurations of a stable event structure of \cite{Wi87a} iff it
moreover is closed under $\nbbigcap$.
\hfill $\Box$
\end{corollary}
These characterisations were obtained earlier in \cite{Wi87a}.
However, the following one seems to be new.

\begin{corollary}{characterisation-prime-Wi87a}
A configuration structure arises as the family of configurations of a
prime event structure of \cite{Wi87a} iff it satisfies
the axioms of rootedness, finiteness, coincidence-freeness,
finite-completeness, irredundancy and closure under $\nbigcap$.
\hfill $\Box$
\end{corollary}
Recall that for these structures the left-closed and secured
configurations are the same.

\begin{corollary}{characterisation-Wi89}
A configuration structure arises as the family of (secured)
configurations of an event structure of \cite{Wi89} iff it satisfies
the axioms of rootedness, finiteness, coincidence-freeness and
closure under \plat{\,\bbigcup^2}.

A configuration structure arises as the family of (secured)
configurations of a stable event structure of \cite{Wi89} iff it
moreover is closed under $\nbbigcap$.
\hfill $\Box$
\end{corollary}
In \cite{Wi87a} the characterisations above were claimed, but using
coherence (cf.\ \df{compatibility}.7) instead of closure under
\plat{\bbigcup^2}.  Arend Rensink [personal communication, around
1996] provided the following counterexample against that characterisation.

\begin{example}{Rensink}
Let $\eC\mathbin{=}\tuple{E,C}$ be given by $E\mathbin{:=}\{a,b,c\}$ and\vspace{-1.4ex}
$$C:=\{\emptyset, \{a\}, \{b\}, \{a,b\}, \{a,c\}, \{b,c\}\}.$$
Then $\eC$ satisfies the axioms of rootedness, finiteness,
coincidence-freeness, closure under $\nbbigcap$ and coherence, but it
is not closed under \plat{\bbigcup^2} (cf.\ \df{properties}.7). By
\cor{characterisation-Wi89} it therefore cannot arise as the family of
configurations of an event structure of \cite{Wi89}.
\end{example}
We now propose the property of closure under \plat{\bbigcup^2} as the
replacement for coherence in this theorem. Using
\pr{directed-unions} we can replace closure under \plat{\bbigcup^2} (and
rootedness) by coherence if we have closure under $\nbigcap$.
This gives the following, apparently novel, characterisation.

\begin{corollary}{characterisation-prime-Wi89}
A configuration structure arises as the family of configurations of a
prime event structure of \cite{Wi89} iff it satisfies the axioms of
finiteness, coincidence-freeness, coherence, irredundancy and closure
under $\nbigcap$.
\hfill $\Box$
\end{corollary}

\subsubsection*{Propositional theories}

We do not have axiomatic characterisations of properties like
connectedness or hyperconnectedness, and therefore we cannot offer a
third column for Table~\ref{correspondence secured} such that for
$\eC$ a hyperconnected configuration structure satisfying a package of
properties, a suitably axiomatised $\eT$ theory can be found for which
$\fM(\eT)=\eC$. 
As best we could work up to reachable equivalence, and be content with
a theory $\eT$ such that $\fS(\fM(\eT)) = \eC$.
In this context we directly inherit the third column of
Table~\ref{correspondence}---however, only for theories that are
\emph{secure}, a property for which we have no axiomatic characterisation.
\begin{definition}{secure-PT}
A configuration structure $\fC=\tuple{E,C}$ is \phrase{secure} if
$S(\eC)\subseteq C$, and a propositional theory $\eT$ is \emph{secure}
if $\fM(\fT)$ is.
\end{definition}
Note that if $\eC$ is secure, then so is any pure event structure
$\eE$ with $\fL(\eE)=\eC$.

\begin{corollary}{CtoT-secured}
A hyperconnected configuration structure $\eC$ has any package of
properties from the second column of Table~\ref{correspondence secured} iff
there is a secure propositional theory $\eT$ whose formulae are of
one of the combinations of the forms found on the corresponding lines
of Table~\ref{correspondence}, such that $\fS(\fM(\eT)) = \eC$.
\end{corollary}

\begin{proof}
Given a hyperconnected configuration structure $\eC$ satisfying a
package of properties from the second column of
Table~\ref{correspondence secured}, \thm{CtoE-secured} yields a pure
and secure event structure with the corresponding properties such that
$\fS(\eE)=\eC$. By \thm{EtoT}, there is a theory $\eT$ whose formulae are of
one of the combinations of the forms found on the corresponding lines
of Table~\ref{correspondence}, such that $\fM(\eT) = \fL(\eE)$.
As $\eE$ is secure, so is $\fL(\eE)$ and hence $\eT$.
Using \pr{pure ES commute} we find $\fS(\fM(\eT)) = \fS(\fL(\eE))
=\fS(\eE)= \eC$.

Conversely, given a package of properties from the second column of
Table~\ref{correspondence secured}, let $\eT$ be a secure theory whose
formulae are of one of the combinations of the forms found on the
corresponding lines of Table~\ref{correspondence}.
Then \thm{TtoC} yields that $\fM(\eT)$ has the corresponding package of
properties from  the second column of Table~\ref{correspondence}
(i.e., skipping ``hyperreachable''), so by \thm{CtoE} there is a pure
event structure $\eE$ with the corresponding properties such that
$\fL(\eE)=\fM(\eT)$. As noted above, $\eE$ is secure, and by
\thm{EtoC-secured} $\fS(\eE)$ has the given package of properties. Furthermore,
$\fS(\eE)=\fS(\fL(\eE))= \fS(\fM(\eT))$.
\end{proof}
In case of a package of properties including finite conflict, or
excluding (local) conjunctivity, the requirement that $\eT$ be secure
may be dropped. This follows from the remarks following \thm{EtoC-secured}.

We do not have an axiomatic characterisation of irredundancy, nor of
the axiom of finiteness, and hence neither of the event structures
from \cite{Wi87a,Wi89}.

\subsubsection*{Reachable configuration structures}

We were unable to find correspondences of the form of
Table~\ref{correspondence secured} using reachable
configurations instead of secured ones. The problem we encountered is
that the set of reachable configurations of a singular event structure
need not be closed under bounded unions.

\begin{example}{no bounded unions}
Let $\eE\mathbin{=}\tuple{E,\turn\;}$ be given by $\eE\mathbin{:=}\IN\dcup\{e\}$,
$\{n\}\turn\{n+1\}$ and $\{e\}\turn\{n+1\}$ for $n\in\IN$
and \mbox{$\emptyset\turn X$} for $X$ not of the form $\{n+1\}$.
This event structure is rooted and singular and has finite conflict.
Its reachable configurations include $\{0\ldots,n\}$ for all numbers $n$,
together with all sets containing $e$.
In particular the set of all events is a reachable configuration,
because after $e$ all other events can happen in one step.
However, $\IN$ is not a reachable configuration.
Therefore, $\fR(\eE)$ fails to be closed under bounded intersections.
\end{example}

\subsection{Two finitary comparisons}\label{finitary comparisons}

In this section we characterise the configuration structures that
arise by taking the finite left-closed configurations of the various
classes of event structures.
We also characterise the corresponding propositional
theories, but working up to finitary equivalence only. Thus, given a
finitary configuration structure $\eC$ satisfying some relevant closure
properties, we seek a proposition theory $\eT$ of a particular form such that
$\fF(\fM(\eT))=\eC$; we do not seek a theory $\eT$ with $\fM(\eT)=\eC$.
Subsequently, we do the same for the finite \emph{reachable} configurations
of the various classes of event structures.

We can put any event structure $\eE$ into a ``finitary'' form $\eE_f$
by removing all causal relations $Y \turn X$ with $X$ or $Y$ infinite
and then adding all $\emptyset \turn X$ for $X$ infinite.  Clearly
$\eE_f$ has finite causes and finite conflict, and $\fF(\fL(\eE_f)) =
\fF(\fL(\eE))$. Thus, by \thm{CtoE}, any finitary configuration
structures arises as $\fF(\fL(\eE))$ for an event structure $\eE$ with
finite causes and finite conflict.\linebreak[2]
Next, since every clause of the form $Y \implies X$ with $Y$ infinite
is satisfied by any finite configuration, up to finitary equivalence
any configuration structure has an axiomatisation by formulae of the
form (fin, any).  Thus, at the level of finitary equivalence, we have
a general correspondence between (pure) event structures (with finite
causes and finite conflict), finitary configuration structures and
this class of propositional theories.

For particular correspondences we again consider the relevant
properties of event structures and their correspondences in
configuration structures and prop\-ositional theories.  We consider
pureness, rootedness, singularity, conjunctivity, local conjunctivity
and binary conflict, as finite conflict is already built in.  For
finitary configuration structures,
the distinctions between $\bbigcup$ and $\fbbigcup$, and between
$\nbbigcap$ and $\nbbigcap^f$, disappear, and indeed
we are left with closure conditions,
$\bsmallcupf$, $\cap$, $\bsmallcap$, $\bbsmallcupf$ and $\bbsmallcap$,
meaning, respectively: closure under finite bounded unions, binary
intersections, bounded binary intersections, finite pairwise
consistent unions and pairwise consistent binary intersections.

\begin{observation}{finite closure conditions}
A finitary configuration structure
\begin{itemise3}
\item[\bf --] is closed under $\bbigcup$ iff it is closed under $\bsmallcupf$;
\item[\bf --] is closed under $\nbigcap$ iff it is closed under $\cap$;
\item[\bf --] is closed under $\nbbigcap$ iff it is closed under $\bsmallcap$;
\item[\bf --] is closed under $\bbbigcup$ iff it is closed under $\bbsmallcupf$;
\item[\bf --] is closed under $\nbbigcap^2$ iff it is closed under $\bbsmallcap$.
\end{itemise3}
\end{observation}
Note that a configuration structure is closed under $\bsmallcupf$ iff
it is closed under $\overline\cup$ and is either rooted or empty. We say
that a configuration structure $\eC$ has \phrase{finite binary
conflict} iff for every finite set of events $X$
$$[\forall Y \!\!\subseteq\! X.~(|Y| \leq\!2 \Rightarrow \exists z
\!\in\! C.~ Y \!\!\subseteq\! z \!\subseteq\! X)] \Rightarrow X \!\!\in\! C.$$
Note that a finitary configuration structure $\eC$ has finite binary
conflict iff $\eC = \fF(\eC^b)$ with $\eC^b$ its closure under binary
conflict, as introduced in \df{conflict closure}.

\begin{observation}{finitary closure conditions}
If a configuration structure $\eC$ is closed under $\bbigcup$,
$\nbigcap$, $\nbbigcap$, $\bbbigcup$ or $\nbbigcap^2$, then so is $\fF(\eC)$.
Furthermore, if $\eC$ has binary conflict then $\fF(\eC)$ has finite
binary conflict.
\end{observation}
For propositional theories used for comparison up to finitary equivalence
we replace ``ddc'' and ``bddc'' by new forms ``ddfc''  and ``bddfc'', meaning
\emph{finite} conjunctions. The interpretation of the resulting forms $(L,R)$ 
should be clear; as before, 
they are combined by taking meets in the left and right lattices.
We get the correspondences summarised by Table~\ref{correspondence_finite}.
\begin{table}[htb]
\begin{tabular}{@{}|@{~}l@{\,}|@{~}l@{~}|@{~}l@{~}|@{}}
\hline
Event           & Configuration                 & Propositional         \\
structures      & structures                    & theories              \\
\hline\hline
rooted          & rooted                        & (nef, any)            \\
singular        & closed under $\bsmallcupf$    & (1, any), (fin., 0)   \\
conjunctive     & closed under $\cap$           & (fin., $\leq\!1$)     \\
locally conj.   & closed under $\bsmallcap$     & (fin., ddfc)          \\
binary conflict & fin.\ bin.\ conflict          & ($\leq\!2$, any)      \\
\hline
\raisebox{0pt}[12pt]{sing}.\ \& bin.\ con.& closed under $\bbsmallcupf$  & (1, any), ($\leq\!2$, 0)\\
loc.\ conj.\ \& b.c.& closed under $\bbsmallcap$& ($\leq\!2$, bddfc)      \\
\hline
\end{tabular}
\caption{Corresponding properties for finite parts \label{correspondence_finite}}
\end{table}

We define a package of properties of configuration structures from the 
table analogously to before. We call a set of properties from the second column of
Table~\ref{correspondence_finite} a {package }if
\begin{itemise2}
\itemsep 0pt
\item it contains the property
``closed under $\bbsmallcupf$'' iff  it contains the properties
``closed under $\bsmallcupf$'' and ``having finite binary conflict'',
and
\item it contains the property
``closed under $\bbsmallcap$'' iff it contains the properties
``closed under $\bsmallcap$'' and ``having finite binary conflict.''
\vspace{2pt}
\end{itemise2}
We can now formulate the correspondences explicitly as:
\begin{theorem}{finite}~
\begin{enumerate}
\item Let $\eE$ be a (pure)  event structure
satisfying any collection of properties
from Table~\ref{correspondence_finite}. Then there is a (pure) propositional
theory $\eT$ whose axioms
have forms which are combinations of the forms corresponding to the event structure 
properties, such that $\fF(\fM(\eT)) = \fF(\fL(\eE))$.
\item Let $\eT$ be a propositional theory whose axioms
have forms which are combinations of forms from a given collection of rows of 
Table~\ref{correspondence_finite}. Then $\fF(\fM(\eT))$ has the corresponding 
collection of properties of configuration structures.
\item Let $\eC$ be a finitary configuration structure satisfying a
given package of properties from Table~\ref{correspondence_finite}.
Then there is a pure event structure $\eE$ with finite causes and
finite conflict such that $\fF(\fL(\eE)) = \eC$ and with the
corresponding collection of properties of event structures.
\end{enumerate}
\end{theorem}
\pf
\begin{enumerate}
\item 
Given a (pure) event structure $\eE$
satisfying a given collection of properties of the table, 
Theorem~\ref{th-EtoT} yields an axiomatisation of 
$\fL(\eE)$ by a (pure) propositional theory $\eT$ whose axioms have 
the form of a combination of the forms corresponding to
the  properties given by Table~\ref{correspondence}.

Now we can remove any formulae of the form $(X, - )$ with $X$ infinite
from the axiomatisation as they are automatically true in finite
interpretations (i.e., the finite subsets of $E$).  (Alternatively, we
could have obtained these forms by requiring $\eE$, without limitation
of generality, to be with finite conflict.)  Next, to any formula of
the form $(-,\mbox{ddc})$ one can associate a formula of the form
$(-,\mbox{ddfc})$ by removing all infinite disjuncts, and the
associated formula is true in a finite interpretation iff the original
one is; the same holds for $(-,\mbox{bddc})$ and $(-,\mbox{bddfc})$
formulae. Making these replacements as necessary, one arrives at the
required (pure) propositional theory

\item 
Given any propositional theory $\eT$ whose axioms have the form of
combinations of forms given in rows of the table, then, by \thm{TtoC},
$\fM(\eT)$ satisfies the corresponding properties of
Table~\ref{correspondence} and so, by Observations~\ref{obs-finitary
closure conditions} and~\ref{obs-finite closure conditions},
$\fF(\fM(\eT))$ satisfies the corresponding properties of
Table~\ref{correspondence_finite}.

\item 
Let $\eC$ be a finitary configuration structure with a given package
of properties of the table, not including finite binary conflict.
Then,  by \ob{finite closure conditions}, $\eC$ satisfies the
corresponding package of properties of Table~\ref{correspondence},
and so, by \thm{CtoE}, there is a pure  event structure $\eE$ satisfying 
the corresponding properties for event structures such that 
$\fL(\eE) = \eC$.
Now $\eE_f$ also satisfies these properties, and
$\fF(\fL(\eE_f)) = \fF(\fL(\eE)) = \fF(\eC) = \eC$.

In the case of a package of properties which does include finite
binary conflict, by \lem{closure} and \cor{conflict closure}, $\eC^b$
has the same package of properties but with binary conflict instead of
finite binary conflict. Thus, by \thm{CtoE}, there is a pure event
structure $\eE$ with the corresponding properties such that
$\fL(\eE)=\eC^b$.  Now $\eE_f$ is pure and also satisfies these properties, and
$\fF(\fL(\eE_f)) = \fF(\fL(\eE)) = \fF(\eC^b) = \eC$.
\hfill$\Box$
\end{enumerate}

\subsubsection*{Comparisons via finite reachable parts}

We now turn to comparisons via finite reachable parts. 
A similar obstacle as in Section~\ref{EvsC-secured} presents itself:
an event structure $\eE$ may have binary conflict even though
$\fR_f(\eE)$ does not have finite binary conflict.
\begin{example}{pentagon}
Let $\eC$ be the configuration structure with events $\{a_0,\ldots,a_4\}$
and with configurations:
$$\emptyset, \{a_i\}, \{a_i,a_{i+1}\},\{a_i,a_{i+1}, a_{i+2}\}$$
and
$$\{a_0,\ldots,a_4\}$$
where the counting is done mod 5. Then $\eC$ is finitary and has (finite)
binary conflict, but its reachable part has not, as $\{a_0,\ldots,a_4\}$ is not reachable.
Furthermore, $\eC$ can be given by a pure rooted event structure with
finite causes and binary conflict, namely the one with the enablings
$$\emptyset \turn a_i,\;\; \emptyset \turn a_i,a_{i+1} \;\mbox{ and }\; a_{i+1}\turn a_i,a_{i+2}$$
again counting mod 5 (and omitting explicit set parentheses),
plus those needed for rootedness and binary conflict.
\end{example}
Since our primary interest is in characterising natural properties
of event structures we find a suitable weakening of this property of
configuration structures, and proceed analogously to Section~\ref{EvsC-secured}.
\begin{definition}{sbc}
A configuration structure $\eC$ has \phrase{finite reachable binary
conflict} iff $\eC = \fR(\fF(\eC^b))$.
\end{definition}
In other words, a configuration structure $\eC=\tuple{E,C}$ has
finite reachable binary conflict iff $X\in C$ exactly
when $X$ can be written as  $\{e_1, \ldots, e_n\}$  
so that for every $j \leq n$ and $Y \subseteq \{e_1, \ldots, e_j\}$ with $|Y| \leq 2$ there 
is a configuration $z\in C$ such that $Y \subseteq z  \subseteq\{e_1, \ldots, e_j\}$.

We then obtain the correspondences summarised by Table~\ref{correspondence_finite_reachable}.

\begin{table}[htb]
\begin{tabular}{@{}|@{~}l@{\hspace{3pt}}|@{~}l@{\hspace{3pt}}|@{~}l@{\hspace{3pt}}|@{}}
\hline
Event           & Configuration                 & Propositional         \\
structures      & structures                    & theories              \\
\hline\hline
rooted          & rooted                        & (nef, any)            \\
singular        & closed under $\bsmallcupf$    & (1, any), (fin., 0)   \\
conjunctive     & closed under $\cap$       & (fin., $\leq\!1$)         \\
locally conj.   & closed under $\bsmallcap$      & (fin., ddfc)         \\
binary conflict & fin.\ reach.\ b.c.             & ($\leq\!2$, any)     \\
\hline
\raisebox{0pt}[12pt]{sing}.\ \& bin.\ con.& closed under $\bbsmallcupf$  & (1, any), ($\leq\!2$, 0)\\
loc.\ conj.\ \& b.c.& closed under $\bbsmallcap$& ($\leq\!2$, bddfc)       \\
\hline
\end{tabular}
\caption{Corresponding properties for finite reachable parts}\label{correspondence_finite_reachable}
\end{table}

\begin{lemma}{G3r} Let $\eE$ be a pure event structure
with the properties given in one of the rows of
Table~\ref{correspondence_finite_reachable}. Then $\fR_f(\eE)$ has the
corresponding property, as given in the table.
\end{lemma}
\begin{proof} The event structure $\eE_f$ has the same properties as
$\eE$ and in addition has finite conflict. Clearly $\fR_f(\eE)=\fR_f(\eE_f)$,
and by \pr{finite reachable ES} we have $\fR_f(\eE_f) = \fF(\fS(\eE_f))$.
By \lem{finite conflict secure} $S(\eE_f)\subseteq L(\eE_f)$. Hence, by
\thm{EtoC-secured}, $\fS(\eE_f)$ has the corresponding property given in
Table~\ref{correspondence secured}.  In case the row we started with
was not that of binary conflict, by Observations~\ref{obs-finitary
closure conditions} and~\ref{obs-finite closure conditions}
$\fF(\fS(\eE_f))$ has the corresponding property of
Table~\ref{correspondence_finite_reachable}.  In case the row we
started with \emph{was} that of binary conflict, by expanding
\df{hyperreachable conflict} we find that $\fS(\eE_f) =
\fS(\fS(\eE_f)^b)$. Now observe that $\fF(\eC^b)=\fF(\fF(\eC)^b)$
for any configuration structure $\eC$.  Applying
Propositions~\ref{pr-finite reachable ES} and~\ref{pr-idempotence}
this yields $$\begin{array}{r@{~=~}l}
\fF(\fS(\eE_f)) & \fF(\fS(\fS(\eE_f)^b))\\
& \fF(\fR(\fS(\eE_f)^b))\\
& \fR(\fF(\fS(\eE_f)^b))\\
& \fR(\fF(\fF(\fS(\eE_f))^b))~.
\end{array}$$
Hence $\fR_f(\eE)=\fF(\fS(\eE_f))$ has finite reachable binary conflict.
\end{proof}
Note that the purity requirement in \lem{G3r} can be weakened to
reachable purity, and is only needed for the
binary conflict row, namely in the application of \thm{EtoC-secured}.
The following example shows that this requirement cannot be omitted.

\begin{example}{impure finite reachable binary conflict}
Let $\eE=\tuple{E,\vdash\;}$ be the event structure with
$E:=\{a_1,a_2,a_3\}$ and the enablings $\emptyset \vdash X$ when
$|X|\neq 2$, as well as $$a_i \vdash a_i,a_{i+1}$$ where the counting
is done mod 3. Then $\eE$ has binary conflict, $\fL(\eE)=\pow(E)$ and
$\fR_f(\eE)=\pow(E)-\{E\}$. Hence $\fR_f(\eE)$ does not have finite
reachable binary conflict.
\end{example}

We can now establish the correspondences of Table~\ref{correspondence_finite_reachable}. 
We define packages of properties of configuration structures from the
table just as we did for finitary equivalence, substituting finite reachable
binary conflict for finite binary conflict; and we keep the same form
lattices and their interpretation as just used for finitary
equivalence.
\begin{theorem}{finite_reachable}~
\begin{enumerate}
\item Let $\eE$ be a (pure) event structure
satisfying any collection of properties
from Table~\ref{correspondence_finite_reachable}. Then there is a (pure) propositional
theory $\eT$ whose axioms
have forms which are combinations of the forms corresponding to the event structure 
properties such that $\fR(\fF(\fM(\eT))) = \fR(\fF(\fL(\eE)))$.
\item Let $\eT$ be a propositional theory whose axioms
have forms which are combinations of forms from a given collection of rows of 
Table~\ref{correspondence_finite_reachable}. Then $\fR(\fF(\fM(\eT)))$
has the corresponding collection of properties of configuration structures.
\item Let $\eC$ be a finitary connected configuration structure
satisfying a given package of properties from
Table~\ref{correspondence_finite_reachable}. Then there is a pure
event structure $\eE$ with finite causes and finite conflict such that
$\fR_f(\eE) = \eC$ and with the corresponding collection of properties
of event structures.
\end{enumerate}
\end{theorem}
\pf
\begin{enumerate}
\item This is immediate from part 1 of Theorem~\ref{th-finite}.
\item  Let $\eT$ be such a theory. It follows from Theorem~\ref{th-finite}
that there is a pure event structure $\eE$ satisfying the corresponding properties 
from the table such that  $\fF(\fL(\eE)) = \fF(\fM(\eT))$. The
conclusion then follows from \lem{G3r} and \pr{pure ES commute}.
\item This follows just as in the proof of \thm{finite}:
Let $\eC$ be a finitary connected configuration structure with a given package
of properties of the table, not including finite reachable binary conflict.
Then, by \ob{finite closure conditions}, $\eC$ satisfies the
corresponding package of properties of Table~\ref{correspondence},
and so, by \thm{CtoE}, there is a pure  event structure $\eE$ satisfying 
the corresponding properties for event structures such that 
$\fL(\eE) = \eC$.
Now $\eE_f$ is pure and also satisfies these properties, so \pr{pure ES
commute} yields
$\fR_f(\eE_f) = \fR(\fF(\fL(\eE_f))) = \fR(\fF(\fL(\eE))) = \fR(\fF(\eC)) = \eC$.

In the case of a package of properties which does include finite reachable
binary conflict, by \lem{closure} and \cor{conflict closure}, $\eC^b$
has the same package of properties but with binary conflict instead of
finite reachable binary conflict. Thus, by \thm{CtoE}, there is a pure event
structure $\eE$ with the corresponding properties such that
$\fL(\eE)=\eC^b$.  Now $\eE_f$ also satisfies these properties, and
$\fR_f(\eE_f) = \fR(\fF(\fL(\eE_f))) = \fR(\fF(\fL(\eE))) =
\fR(\fF(\eC^b)) = \eC$.
\hfill$\Box$
\end{enumerate}


\noindent
We now apply the results of this section to characterise the finitary
configuration structures associated to the various event structures of
{\sc Winskel} \cite{Wi87a,Wi89}.

\begin{corollary}{characterisation-Winskel finitary}
A configuration structure arises as the family of finite left-closed
configurations of an event structure of \cite{Wi87a} iff it is
finitary, rooted and closed under $\overline\cup$. It arises as the
family of finite left-closed configurations of a stable event
structure of \cite{Wi87a} iff it moreover is closed under $\bsmallcap$.

A configuration structure arises as the family of finite left-closed
configurations of an event structure of \cite{Wi89} iff it is
finitary, rooted and closed under $\overline\cup^2$.  It arises as the
family of finite left-closed configurations of a stable event
structure of \cite{Wi89} iff it moreover is closed under $\bsmallcap$.
\hfill $\Box$
\end{corollary}
The four classes of configuration structures mentioned in the above
corollary arise as the finite models of propositional
theories whose axioms have the forms
\begin{center}
\begin{tabular}{c}
(1, any), (nef, 0)\\
(1, ddfc), (nef, 0)\\
(1, any), (2, 0)\\
(1, bddfc), (2, 0),
\end{tabular}
\end{center}
respectively.

When dealing with finite \emph{reachable} configurations, the same
characterisations as in \cor{characterisation-Winskel finitary} are
obtained, but now the resulting configuration structures are
additionally connected.

\begin{proposition}{coincidence-freeness-connectedness}
Let $\eC$ be a finitary configuration structure closed under
$\bbigcup$.  Then $\eC$ is connected iff it is coincidence-free.
\end{proposition}

\begin{proof}
Similar to the proof of \pr{coincidence-freeness-hyperconnectedness}.
\pagebreak[3]
\end{proof}

\begin{corollary}{characterisation-Winskel finitary reachable}
A configuration structure arises as the family of finite reachable
configurations of an event structure of \cite{Wi87a} iff it is
finitary, rooted, coincidence-free and closed under $\overline\cup$.  It
arises as the family of finite reachable configurations of a stable
event structure of \cite{Wi87a} iff it moreover is closed under
$\bsmallcap$.

A configuration structure arises as the family of finite reachable
configurations of an event structure of \cite{Wi89} iff it is
finitary, rooted, coincidence-free and closed under $\overline\cup^2$.
It arises as the family of finite reachable configurations of a
stable event structure of \cite{Wi89} iff it moreover is closed under
$\bsmallcap$.
\hfill $\Box$
\end{corollary}
The first class of configuration structures in this corollary was the
class of configuration structures originally considered in \cite{GG90}.

We have no characterisation of the finitary configuration structures
associated to the event structures from \cite{NPW81}; in particular,
the property of $\fL$-irredundancy appears hard to express in terms of
finite configurations. As for the prime event structures from
\cite{Wi87a,Wi89}, recall that their finite left-closed configurations
are the same as their finite reachable or finite secured configurations.

\begin{corollary}{characterisation-prime-Winskel finitary}
A configuration structure arises as the family of finite
configurations of a prime event structure of \cite{Wi87a} iff it is
finitary, rooted, coincidence-free, irredundant and closed under
$\overline\cup$ and $\cap$.

A configuration structure arises as the family of finite
configurations of a prime event structure of \cite{Wi89} iff it is
finitary, coherent, coincidence-free, irredundant and closed under
$\cap$.
\hfill $\Box$
\end{corollary}
The two classes of configuration structures mentioned above
arise as the finite reachable models of propositional
theories whose axioms have the forms
\begin{center}
\begin{tabular}{c}
(1, 1), (nef, 0)\\
(1, 1), (2, 0)
\end{tabular}
\end{center}
respectively. We do not have axiomatic characterisations of
connectedness or coincidence-freedom; that lack is circumvented in the
above characterisations by talking about \emph{reachable} models.

\subsection{Tying-in Petri nets}\label{Tie-nets}

The third columns of Tables~\ref{correspondence},~\ref{correspondence_finite},%
~\ref{correspondence_finite_reachable} also provide characterisations
of classes of Petri nets corresponding to various combinations of
properties of event structures or configuration structures.  The pure
1-occurrence nets correspond up to finite configuration equivalence to
event and configuration structures that are rooted and with finite
conflict. We do not have a structural characterisation of the subclass
of those pure 1-occurrence nets corresponding to locally conjunctive
event structures. However, for each combination of the properties
singular, conjunctive, and binary conflict, the forms $(L,R)$ that
characterise the associated propositional theory also provide a
structural characterisation of the associated subclass of pure
1-occurrence nets.  Here $L$ restricts the cardinality of the set of
posttransitions of any given place, and $R$ restricts the cardinality of
its set of pretransitions; we say that the place \emph{has} the form $(L,R)$.
For example, rooted singular pure event structures correspond to the
pure 1-occurrence nets each of whose places have either no pretransitions
or exactly one posttransition. The proof of the correspondences goes
via the characterisations of the associated propositional theories.

\begin{theorem}{Petri-correspondence}
Let $\eT$ be a (pure) rooted propositional theory in conjunctive
normal form, whose clauses are combinations
of the forms found in lines 2, 3 and 5 of Table~\ref{correspondence}.
Then $\fN(\eT)$ is a (pure) 1-occurrence net whose places have the
corresponding combinations of forms, as well as (nef, any).

Similarly, if $\eN$ is a (pure) 1-occurrence net whose places have
combinations of the forms found in lines 2, 3 and 5 of
Table~\ref{correspondence}, then $\fT(\eN)$, as defined
\hyperref[NtoT]{at the end} \hyperref[NtoT]{of Section~\ref*{PN}}, is
a (pure) rooted propositional theory axiomatised by clauses obeying
these forms, as well as (nef, any).
\end{theorem}

\begin{proof}
The first statement follows immediately from the construction in \df{TtoN}.
For the second statement, recall that $\fT(\eN)$ consists of the formulae
$\bigwedge Y\Rightarrow\bigvee_{X\in\,^{^\bullet Y(s)\raisebox{-1pt}{-}I(s)}\!s} \bigwedge X$
for $s\in S$ and $Y \subseteq_{\it fin} s^\bullet$. When converting such
formulae to conjunctive normal form, one obtains clauses $Y \implies Z$
with $Y \subseteq_{\it fin} s^\bullet$ for some place $s$. 
As remarked at the end of Section~\ref*{PN},
one can omit any clauses for $Y=\emptyset$, or more generally for
which $^\bullet Y(s){-}I(s) \leq 0$,
as then $\emptyset \mathbin\in \mathord{^{^\bullet Y(s){-}I(s)}\!s}$.
Hence all clauses obey the restriction
\mbox{(nef, any)} and $\fT(\eN)$ is rooted. By construction, $\fT(\eN)$ is
pure when $\eN$ is.  If $s$ has the form (1, any) or \mbox{($\leq\!2$, any)}, then so do the associated clauses.  Furthermore, if
$s$ has no pretransitions, then the associated clauses
have the form $Y\implies\emptyset$, and if $s$ has one pretransition
$t$, then the associated clauses have the form $Y\implies\emptyset$
for $Y\subseteq_{\it fin} s^\bullet$ with 
$^\bullet Y(s)- I(s)>t^\bullet(s)$, 
and $Y\implies t$ for $Y\subseteq_{\it fin} s^\bullet$ with
$1 \leq {^\bullet Y(s)- I(s)} \leq t^\bullet(s)$. Thus, if $s$ has the form
(any, 0) or \mbox{(any, $\leq\!1$)}, then so do the associated clauses.
\end{proof}
This theorem also holds when using the third columns of
Tables~\ref{correspondence_finite} or~\ref{correspondence_finite_reachable}
(which are the same) instead of the one of Table~\ref{correspondence}.
For these columns are obtained by additionally imposing the condition
(finite, any), a condition that is implied by (nef, any). Furthermore,
the theorem remains true if any place $s$ with $n$ incoming arcs and $k$
initial tokens is deemed to additionally have the form
``($\leq\! k\!+\!n$, $\leq\!n$) or ($k\!+\!n\!+\!1$, 0)''.
Namely if $Y \subseteq_{\it fin} s^\bullet$ and $|Y|>k+n$ then the
transitions in $Y$ cannot all happen, so we obtain the clause $\bigwedge
Y \implies \emptyset$. Among such clauses one only needs to retain the
ones with $|Y|$ minimal, that is, with $|Y|=k+n+1$.
Finally, places without posttransitions may be ignored.\linebreak
Thus, for example, pure 1-occurrence nets whose places either have
$\leq\!1$ posttransition, or one incoming arc and no initial
tokens, or no incoming arcs and $\leq\!1$ initial token correspond to
pure singular event structures with binary conflict.

This theorem, together with \thm{PtoNtoC} and \pr{NtoT}, yields a
bijection up to finitary equivalence between the stated subclasses of
pure rooted propositional theories and the corresponding subclasses of
pure 1-occurrence nets. As the nets are pure, these bijections
also hold up to finitary reachable equivalence.


\section{Related Work}\label{related work}

The notion of a configuration structure as a model of concurrency in
its own right stems from {\sc Winskel} \cite{Wi82}; our configuration
structures are obtained by dropping the requirements imposed in \cite{Wi82}:
coherence, stability, coincidence-freeness and the axiom of finiteness.
The term \emph{configuration structures} stems from \cite{GG90}; their
configuration structures obeyed the requirements of finitariness, rootedness,
coincidence-freeness and closure under $\overline\cup$, that together
ensured that these structures were exactly the families
of finite configurations of Winskel's event structures \cite{Wi87a}.
Two further partial generalisations of this model were previously proposed by
{\sc Pinna \& Poign\'e} \cite{PP95} and {\sc Hoogers, Kleijn \& Thiagarajan}
\cite{HKT96}. The \phrase{event automata} of \cite{PP95} are rooted
finitary configuration structures together with a transition relation
between the configurations; each transition extends a configuration
with exactly one event. The \phrase{local event structures} of
\cite{HKT96} are rooted, finitary, connected configuration structures
together with a step transition relation $\rightarrow$ between the
configurations that satisfies
\begin{itemise}
\item $X \rightarrow X$,
\item $X \rightarrow Y$ implies $X\subseteq Y$, and
\item $X \rightarrow Z$ and $X \subseteq Y\subseteq Z$ implies $X \rightarrow Y \rightarrow Z$.
\end{itemise}
In \cite{HKT96} $X \rightarrow Y$ is denoted $X \vdash (Y-X)$, so that
their notation $X \vdash Y$ implies $X \cap Y = \emptyset$ and
translates to $X \rightarrow X\cup Y$.

Our configuration structures are, up to isomorphism, the
\emph{extensional} \phrase{Chu spaces} of {\sc Gupta \& Pratt}
\cite{GP93a,Gup94,Pr94a}. It was in their work that the idea arose of
using the full generality of such structures in modelling
concurrency. Also the propositional representation of configuration
structures stems from \cite{GP93a,Pr94a}.  It should be noted however
that the computational interpretation in \cite{GP93a,Gup94,Pr94a}
differs somewhat from that in \cite{Wi87a,GG90,PP95,HKT96} and the
current work.  In particular, in \cite{GP93a,Gup94,Pr94a} unreachable
configurations may be semantically relevant, as witnessed by the
notions of \phrase{causality} and \phrase{internal choice} in
\cite{GP93a,Pr94a} and that of \phrase{history preserving
bisimulation} in \cite{Gup94}.

{\sc Gunawardena} proposes \phrase{causal automata} in \cite{Gun92b} and
\phrase{geometric automata} in \cite{Gun91}. The first are given by a
set of events with, for each event $e$, a boolean expression $\rho(e)$
over the set of events. Each event occurrence in $\rho(e)$ is
interpreted as the proposition that it happened, and $e$ is enabled
when $\rho(e)$ evaluates to true.  In geometric automata, a more
complicated infinitary logic is used, and the boolean expression is
replaced by two positive logical expressions, one of which must
evaluate to true, and the other to false, in order for the associated
event to be enabled.  Both models can be interpreted in a natural way
in terms of event automata; plain configuration structures are not
sufficient here.

Our event structures are directly inspired by, and generalise, the
ones of {\sc Winskel} \cite{NPW81,Wi87a,Wi89}. Many other variants
of these event structures have been proposed in the literature.

A \phrase{bundle event structure}, as studied in {\sc Langerak}
\cite{Lk92}, is given as a tuple $(E,\#,\mapsto,l)$ with $E$ a set of
events, $\#$ an irreflexive, symmetric conflict relation,
$\mathord{\mapsto} \subseteq \pow(E) \times E$, the \phrase{bundle
relation}, and $l:E\rightarrow Act$ a \emph{labelling function},
labelling events with actions from a given alphabet $Act$. When $X
\mapsto e$, the events in $X$ should be pairwise in conflict; in this
case $e$ can happen only if one of the events in $X$ occurred
earlier. Ignoring the labelling function, a bundle event structure can
in our framework best be understood as a propositional theory, namely
one whose formulae have the forms \mbox{(2, 0)} and (1, dds). Here ``dds''
stands for \phrase{disjoint disjunction of singletons}; in the right
form lattice of Figure~\ref{form-lattices} it can be positioned right
below ``bddc'', or right below ``bddfc'' of Section~\ref{finitary comparisons}.
The configurations used in
\cite{Lk92} are in our terminology finite reachable configurations.
Using the translations of Section~\ref{ComparingModels}, preserving
finitary reachable equivalence, the bundle event structures map to a
subclass of rooted, singular, locally conjunctive event structures
with binary conflict, and hence to a subclass of stable event
structures as defined in \cite{Wi89} that contains the class of prime
event structures of \cite{Wi89}.\vspace{0pt plus 1pt}

Langerak's notion of an \phrase{extended bundle event structure} on the
other hand does not correspond to an event structure as in
\cite{Wi87a,Wi89}.  Here the symmetric binary conflict relation $\#$
is replaced by an asymmetric counterpart $\leadsto$\index{asymmetric
conflict}, a relation that was considered independently in
\cite{PP95}, writing $e \not\leadsto d$ for $d \leadsto e$.
When $d \leadsto e$, the event $e$ can happen either initially or after $d$;
however, as soon as $e$ happens, $d$ is blocked. When both $d$ and $e$
happen, $d$ causally precedes $e$.  Asymmetric conflict 
$d \leadsto e$ can be translated into our framework as $\{d\} \turn
\{d,e\}$, where it is important that $\{d\}$ is the \emph{only} set of
events enabling $\{d,e\}$. The absence of both $d \leadsto e$
and $e \leadsto d$ translates to $\emptyset \vdash \{d,e\}$, and the
conjunction of $d \leadsto e$ and $e \leadsto d$ is simply $d \# e$
and translates to the absence of any $X$ with $X\vdash \{d,e\}$.
Under this translation, the configurations of extended bundle event
structures defined in \cite{Lk92} are exactly our finite reachable
configurations of \df{reachable ES}.  Thus, the class of extended
bundle event structures can be regarded as a subclass of our rooted,
locally conjunctive event structures with binary conflict.  However,
they are not pure, and cannot be faithfully represented by
configuration structures as studied in this paper.\vspace{0pt plus 1pt}
The relationship between event structures with asymmetric conflict,
Petri nets, and domains, is studied in \cite{BCM01}.

A \phrase{dual event structure}, as studied in {\sc Katoen} \cite{Ka96},
is like an extended bundle event structure, but without the
requirement that when $X \mapsto e$ the events in $X$ should be
pairwise in conflict. This amounts to generalising the formulae of the
form (1, dds) to \mbox{(1, any)}. They correspond to a subclass of our
rooted event structures with binary conflict.
The same can be said for the \phrase{extended dual event structures} of
\cite{Ka96}. Here the new feature is the irreflexive and symmetric
\phrase{interleaving relation} $\rightleftharpoons$, modelling mutual 
exclusion of events, i.e., disallowing them to overlap in time. As for
the event structure M in Section~\ref{ES computational},
$d \rightleftharpoons e$ can in our framework be modelled as
$~~~\{d\} \turn \{d,e\}~~~\{e\} \turn \{d,e\}$.\vspace{0pt plus 1pt}

As remarked in the introduction, behaviour preserving translations
from safe Petri nets to a class of event structures, and from there to
configuration structures, are defined in \cite{NPW81}.  In
Section~\ref{prime-bc} we saw that the event structures of
\cite{NPW81} can be seen as a subclass of our event structures, in the
sense that there are translations back and forth that respect the
identify of events and the sets of associated configurations. The
translation in \cite{NPW81} from safe nets to event structures
proceeds in two steps: an \phrase{unfolding} turns every safe net into
an \phrase{occurrence net}---a particular kind of pure safe 1-occurrence
net---and a mapping $\xi$ takes occurrence nets to event structures.
The transitions of an occurrence net $N$ become the events of the
event structure $\xi(N)$, and the finite configurations of $N$, as defined in
this paper, equal the finite configurations of $\xi(N)$ as defined in
\cite{NPW81}. This follows directly from the definitions. Hence the
translation $\xi$ preserves finitary configuration equivalence.  It is not hard
to check that the unfolding of a safe 1-occurrence net preserves finitary
configuration equivalence as well. Thus, restricted to safe
1-occurrence nets, the translations of \cite{NPW81} are entirely in agreement with ours.

This agreement extends to pure safe nets that are not
1-occurrence nets. However, this cannot be stated in the terminology
of this paper, for the unfolding may make multiple copies of a single
transition, namely one for every possible way in which it can be
fired.  Since the identify of events is thereby not preserved, this
unfolding does not respect the equivalences of this paper.

Define a \phrase{1-reachable-occurrence net} to be a net in which every
\emph{reachable} configuration is a set. This notion is a slight
generalisation of a 1-occurrence net. When working up to reachable
equivalence, all our work generalises without change from 1-occurrence
nets to 1-reachable-occurrence nets. Similarly, define a
1-reachable-occurrence net $\eN$ to be \phrase{semantically pure} if
there exists a pure net $\hat\eN$ with the same places and
transitions, and possibly less arcs and less initial tokens, that has
the same reachable configurations and the same step transition
relation between those configurations.  When working up to
reachable equivalence, also preserving the transitions between
reachable configurations, our connections between pure
1-reachable-occurrence nets and pure event structures evidently
generalise to semantically pure 1-reachable-occurrence nets---just as
they did to reachably pure event structures.

{\sc Boudol} \cite{Bo90} provides translations between a class of
1-reachable-occurrence nets, the \phrase{flow nets}, and a class of {\em
flow event structures} that have expressive power strictly between
the bundle event structures of \cite{Lk92} and the stable event
structures of \cite{Wi89}.  His correspondence extends the
correspondence due to \cite{NPW81} between occurrence nets and prime
event structures with binary conflict.  Flow nets are defined to have
the property that transitions that can occur in the same firing
sequence do not share a preplace. This implies that the reachable
configurations of a flow net $\eN$, as well as the transition relation
$\goto{}_\eN$ between them, are unaffected by omitting the arcs from a
transition $e$ to a place $s$ for which there also is an arc from $s$
to $e$. Any flow net can thereby be transformed, in a behaviour
preserving way, into a pure 1-reachable-occurrence net. Hence flow
nets are semantically pure.

As Boudol's translations preserve the notions of event (= transition)
and finite reachable configuration, they are consistent with our
approach. Our translations can thus be regarded as an extension of the
work of \cite{Bo90} to a more general class of Petri nets and event
structures.

Another translation between Petri nets and a model of event
structures has been provided in {\sc Hoogers, Kleijn \& Thiagarajan}
\cite{HKT96}, albeit only for systems without autoconcurrency.  As
mentioned, their event structures are families of configurations with
a step transition relation between them. The translations of
\cite{HKT96} are quite different from ours: even on 1-occurrence nets
an individual transition may correspond to multiple events in the
associated event structure.  We conjecture that the two approaches are
equivalent under a suitable notion of history preserving bisimulation.

\subsubsection*{Future research}

As we have seen, both event structures and Petri nets
have naturally associated transition relations. In the pure case these
transition relations can be derived from their associated sets of
configurations, but this fails more generally. A natural line of future work
is therefore to go beyond the pure case, looking for a suitable notion of
configuration structure equipped with a transition relation and, perhaps, a
suitable notion of propositional theory.

We would also like to connect our models with appropriate versions of
higher dimensional automata \cite{Pr91a}. An embedding up to finitary
reachable equivalence of rooted configuration structures as well as
Petri nets into a form of higher dimensional automata called \phrase{cubical
sets} is proposed in \cite{vG06}. Another form of higher dimensional
automata called \phrase{labelled step transition systems} is
considered in \cite{vG05}.

After the initial work of~\cite{NPW81} it was natural to ask whether their
unfolding could be seen as a universal construction. This led to a
development of categories of event structures, nets and related models, and, in turn, to a
general process algebra whose constructions were natural categorically: see
\cite{Wi87a,WN95,SNW96}. In our case it would be natural to look for
categories of configuration structures and the other models of this paper, so that, for example,
the connections developed in Section 1 became functorial.  The recent work
of~\cite{Win08, WH08} on adding symmetry to structures may prove helpful here.
Proposals for a category of configuration structures can be found in
\cite{Pr94a} and \cite{BMMS98}.

In a different direction, the equivalences considered in this paper are
quite fine and it would be interesting to look at coarser ones, say
along the lines of history preserving bisimulation.  In that connection,
and also the categorical one, it may be useful to consider configuration
structures, and other models, equipped with event labellings.

\paragraph{Acknowledgement} We thank the referees for their helpful comments.


\printindex

\end{document}